\documentclass[aps,preprint,tightenlines,superscriptaddress,showpacs,byrevtex]{revtex4}

\usepackage{graphicx} 
\usepackage{dcolumn}  

\newcolumntype{B}{D{B}{\;}{-1}} 

\def\preprintA{\hbox{\hfil Belle Preprint 2005-23}}
\def\preprintB{\hbox{\hfil KEK   Preprint 2005-34}}
\def\preprintC{}

\renewcommand{\thefootnote}{\fnsymbol{footnote}}
\newcommand{\dpprim}{\ensuremath{D^+_{prim}}}
\newcommand{\dzprim}{\ensuremath{D^0_{prim}}}
\newcommand{\parj}{\ensuremath{\text{\textbf{PARJ(13)}}}}
\newcommand{\Ys}{\ensuremath{\mathrm{\Upsilon(4S)}}}
\newcommand{\Zn}{\ensuremath{\mathrm{Z}}}
\newcommand{\fb}{\ensuremath{\mathrm{fb^{-1}}}}

\newcommand{\pbc}{\ensuremath{\mathrm{pb}}}
\newcommand{\nb}{\ensuremath{\mathrm{nb}}}
\newcommand{\DZ}{\ensuremath{D^0}}
\newcommand{\DP}{\ensuremath{D^+}}
\newcommand{\DSZ}{\ensuremath{D^{*0}}}
\newcommand{\DSP}{\ensuremath{D^{*+}}}
\newcommand{\Ds}{\ensuremath{D_s^+}}

\newcommand{\LC}{\ensuremath{\Lambda_c^+}}

\newcommand{\KP}{\ensuremath{K^+}}
\newcommand{\KM}{\ensuremath{K^-}}
\newcommand{\PP}{\ensuremath{\pi^+}}

\newcommand{\PZ}{\ensuremath{\pi^0}}
\newcommand{\PHI}{\ensuremath{\phi}}
\newcommand{\pr}{\ensuremath{p^+}}
\newcommand{\TO}{\ensuremath{\to}}
\newcommand{\xP}{\ensuremath{\mathrm{x_P}}}
\newcommand{\xE}{\ensuremath{\mathrm{x_E}}}
\newcommand{\GeV}{\ensuremath{\mathrm{GeV}}}
\newcommand{\MeV}{\ensuremath{\mathrm{MeV}}}
\newcommand{\epem}{\ensuremath{e^+e^-}}
\newcommand{\MC}{MC}
\newcommand{\qqbar}{\ensuremath{q\bar q}}

\newcommand{\ccbar}{\ensuremath{c\bar c}}
\newcommand{\frf}  {fragmentation function}

\newcommand{\ZPC}[3] {Z.~Phys.\ {\bf C#1}, #3 (#2)}

\newcommand{\PLB}[3] {Phys.~Lett.\ {\bf B#1}, #3 (#2)}

\newcommand{\PRD}[3] {Phys.~Rev.\ {\bf D#1}, #3 (#2)}

\newcommand{\JPG}[3] {J.~Phys.\ {\bf G#1}, #3 (#2)}

\begin{document}

\preprint{\vbox{
  \preprintA
  \preprintB
  \preprintC
}}

\title{ \quad\\[0.5cm]
Charm Hadrons from Fragmentation and $B$ decays in\\
\epem\ Annihilation at $\sqrt{s}=$10.6~\GeV}

\affiliation{Budker Institute of Nuclear Physics, Novosibirsk}
\affiliation{Chiba University, Chiba}
\affiliation{Chonnam National University, Kwangju}
\affiliation{University of Cincinnati, Cincinnati, Ohio 45221}
\affiliation{University of Frankfurt, Frankfurt}
\affiliation{University of Hawaii, Honolulu, Hawaii 96822}
\affiliation{High Energy Accelerator Research Organization (KEK), Tsukuba}
\affiliation{Hiroshima Institute of Technology, Hiroshima}
\affiliation{Institute of High Energy Physics, Chinese Academy of Sciences, Beijing}
\affiliation{Institute of High Energy Physics, Vienna}
\affiliation{Institute for Theoretical and Experimental Physics, Moscow}
\affiliation{J. Stefan Institute, Ljubljana}
\affiliation{Kanagawa University, Yokohama}
\affiliation{Korea University, Seoul}
\affiliation{Kyungpook National University, Taegu}
\affiliation{Swiss Federal Institute of Technology of Lausanne, EPFL, Lausanne}
\affiliation{University of Ljubljana, Ljubljana}
\affiliation{University of Maribor, Maribor}
\affiliation{University of Melbourne, Victoria}
\affiliation{Nagoya University, Nagoya}
\affiliation{National Central University, Chung-li}
\affiliation{National United University, Miao Li}
\affiliation{Department of Physics, National Taiwan University, Taipei}
\affiliation{H. Niewodniczanski Institute of Nuclear Physics, Krakow}
\affiliation{Nippon Dental University, Niigata}
\affiliation{Niigata University, Niigata}
\affiliation{Nova Gorica Polytechnic, Nova Gorica}
\affiliation{Osaka City University, Osaka}
\affiliation{Osaka University, Osaka}
\affiliation{Panjab University, Chandigarh}
\affiliation{Peking University, Beijing}
\affiliation{Princeton University, Princeton, New Jersey 08544}
\affiliation{University of Science and Technology of China, Hefei}
\affiliation{Sungkyunkwan University, Suwon}
\affiliation{University of Sydney, Sydney NSW}
\affiliation{Tata Institute of Fundamental Research, Bombay}
\affiliation{Toho University, Funabashi}
\affiliation{Tohoku Gakuin University, Tagajo}
\affiliation{Tohoku University, Sendai}
\affiliation{Department of Physics, University of Tokyo, Tokyo}
\affiliation{Tokyo Institute of Technology, Tokyo}
\affiliation{Tokyo University of Agriculture and Technology, Tokyo}
\affiliation{University of Tsukuba, Tsukuba}
\affiliation{Virginia Polytechnic Institute and State University, Blacksburg, Virginia 24061}
\affiliation{Yonsei University, Seoul}
   \author{R.~Seuster}\affiliation{University of Hawaii, Honolulu, Hawaii 96822} 
   \author{K.~Abe}\affiliation{Tohoku Gakuin University, Tagajo} 
   \author{H.~Aihara}\affiliation{Department of Physics, University of Tokyo, Tokyo} 
   \author{Y.~Asano}\affiliation{University of Tsukuba, Tsukuba} 
   \author{T.~Aushev}\affiliation{Institute for Theoretical and Experimental Physics, Moscow} 
   \author{A.~M.~Bakich}\affiliation{University of Sydney, Sydney NSW} 
   \author{M.~Barbero}\affiliation{University of Hawaii, Honolulu, Hawaii 96822} 
   \author{I.~Bedny}\affiliation{Budker Institute of Nuclear Physics, Novosibirsk} 
   \author{U.~Bitenc}\affiliation{J. Stefan Institute, Ljubljana} 
   \author{I.~Bizjak}\affiliation{J. Stefan Institute, Ljubljana} 
% ...
\collaboration{The Belle Collaboration}

\begin{abstract}
 We present an analysis of charm quark fragmentation at 10.6~\GeV,
 based on a data sample of 103~\fb\
 collected by the Belle detector at the
 KEKB accelerator. We consider fragmentation into the main charmed
 hadron ground states, namely \DZ, \DP, \Ds\ and \LC, as
 well as the excited states \DSZ\ and \DSP. The fragmentation
 functions are important to measure as they describe processes at a
 low energy scale, where calculations in perturbation theory lead to
 large uncertainties. Fragmentation functions can also be used as
 input distributions for Monte Carlo generators. Additionally, we
 determine the average number of these charmed hadrons produced per
 $B$ decay at the \Ys\ resonance and measure the distribution of their
 production angle in \epem\ annihilation events and in $B$ decays.
 \vskip 1cm
\end{abstract}

\pacs{13.66.Bc, 13.66.Jn, 14.20.Lq, 14.40.Lb}

\maketitle


\newpage
\renewcommand{\thefootnote}{\arabic{footnote}}
\setcounter{footnote}{0}

\section{Introduction}
 Over recent years perturbative quantum chromodynamics (pQCD) has
 shown impressive agreement with various inclusive measurements at
 \epem\ colliders at many center-of-mass energies (CME) ranging from
 14~GeV up to 206~GeV.
 These measurements utilised variables called event shapes or jet rates,
 see \cite{OPAL_PETRAqcd} for such an analysis.
 These are inclusive variables, whose values are calculated from the
 four-momenta of all particles in an event.
 
 Other properties, such as the momentum spectra of charged or neutral
 particles, have also been measured, but their prediction has proven to be
 more difficult. The necessary calculations have to cover the entire
 energy range
 from the production of the partons at the CME down to the scale of
 the hadron masses (typically $1~\GeV/c^2$), at which hadronisation occurs.
 Typically, powers of the form $\log{(Q^2/m^2)}$ arise when quark masses
 are taken into account, making pQCD calculations difficult to interpret.
 
 Attempts have been made to extend the applicable range of pQCD to
 lower scales. These attempts have to be validated, for example by comparing
 so-called fragmentation functions.
 Due to the scaling violation of QCD, a fundamental property of
 this theory, the fragmentation function for a given particle depends
 explicitly on the CME. This energy dependence must follow the
 Dokshitzer-Gribov-Lipatov-Altarelli-Parisi (DGLAP) \cite{DGLAP}
 evolution equations.
 
 Thus, the fragmentation functions have to be properly evolved.
 Monte Carlo (MC) generators which include this scaling can be used
 instead of analytical evolution.
 Common \MC\ generators which include this scaling are JETSET
 \cite{JETSET}, (its variant) PYTHIA \cite{PYTHIA} and
 HERWIG \cite{HERWIG}.
 
 These \MC\ generators are also needed to model 
 hadronisation, the transition of partons into hadrons, which
 cannot be calculated from first principles within QCD.
 Various models are implemented in MC generators. These
 can be distinguished by comparing the (identified) heavy
 hadron momentum spectra predicted by each model to the spectra seen
 in data.
 
 Fragmentation functions for heavy quarks are attractive both
 experimentally and theoretically. Concerning theory, mass effects in
 the matrix elements only have to be considered for the heavy quark;
 in the limit of $m_{light}\rightarrow0$, a pQCD calculation based 
 on an effective Lagrangian reduces the complexity of the calculation
 compared to the case of light quark fragmentation.
 
 Experimentally, it is important to measure heavy quark
 fragmentation functions as their shapes are different
 from the corresponding functions for light quarks; such a
 measurement is furthermore straightforward, as very often hadrons
 containing heavy quarks can easily be identified.
 Since the production
 of heavy quarks is strongly suppressed in both the perturbative splitting
 of one parton to many partons (the so-called ``parton shower'') and in
 hadronisation, a heavy quark found in an event will most likely be
 produced in the primary interaction.
 
 At LEP and SLD, $b$ quark fragmentation functions have been measured
 with high precision \cite{ALEPHb,DELPHIb,OPALb,SLDb}. 
 These measurements found that these fragmentation functions are in fact
 close to the ones of light quarks, suggesting that one combined model for all
 five flavours might describe the measured momentum spectra better
 than functions which have been introduced for heavy quarks alone.
 These collaborations have also published measurements of $c$ quark
 fragmentation functions \cite{ALEPHc,OPALc}, but with large
 statistical uncertainties due to the small product of the branching fraction
 and reconstruction efficiency for the various final states.
 Some commonly used \frf s are described by the models of Peterson {\it et
 al.} \cite{Peterson}, of Kartvelishvili {\it et al.}
 \cite{Kartvelishvili} and of
 Collins and Spiller \cite{ColSpi}, as well as by the models of the Lund
 group \cite{Lundsymm} and one of its variants by Bowler \cite{Bowler}.
 
 For charm quark \frf s at lower energies, the most recent published
 results for \DZ, \DP, \DSZ\ and \DSP\ are those of CLEO \cite{CLEOprel}.
 The analysis presented here has better statistical precision as the
 data sample is five times larger.
 Other measurements are more than 10 years old \cite{CLEOc,ARGUSc};
 their data sample is over three orders of magnitude smaller than that
 used in this analysis.
 The systematic uncertainties are reduced significantly and are
 comparable to those in \cite{CLEOprel}.
 For a recent review of \frf\ measurements and theory, see \cite{Biebel}.
 
 A measurement of \DZ\ and \DP\ performed by the same
 experiment on the same data set allows for an easy comparison of
 charged meson production rates and momentum spectra, as well as a
 comparison of the momentum-dependent production of secondary- to
 primary-produced mesons. 
 The measurement of the excited states \DSZ\ and \DSP\ allows the
 determination of the feed-down contribution to the ground states \DZ\ and
 \DP\ and also a momentum-dependent determination of $V/(V+P)$,
 the ratio of the production rates of vector and
 the sum of vector and pseudo-scalar mesons.
 A comparison between \Ds\ production, and the production of \DZ\ and
 \DP, can be used to determine the fraction of $s$ quark production in
 hadronisation. Comparing the results for the \LC\ to those of the $D$ mesons
 makes a study of the baryon production mechanism possible.
 
 In addition to charm fragmentation in the \epem\TO\ccbar\
 continuum, charmed hadrons in \epem\ annihilation events can be produced in
 decays of b-hadrons. The dataset for this analysis includes events
 above the production threshold for $B\bar{B}$ pairs,
 at the \Ys\ resonance, so the lower momentum hadrons include
 contributions from $B^0$ and $B^+$ decays. This allows a measurement
 of the production rate of charmed hadrons in B-meson decay.
 
\section{Data Sample and Event Selection}
 This analysis uses data recorded at the Belle detector at the KEKB
 accelerator.
 The KEKB \epem\ collider is a pair of storage rings for electrons and
 positrons with asymmetric energies,
 $8.0\,\mathrm{GeV}$ ($e^-$) and $3.5\,\mathrm{GeV}$
 ($e^+$), and a single intersection point with a 22 mrad
 crossing angle. 
 The beam energies are tuned to produce an available CME of $\sqrt{s}=10.58~\GeV$,
 corresponding to the mass of the $\Upsilon(\mathrm{4S})$.
 A detailed description can be found in \cite{KEKB}.
 
 The Belle detector covers a solid angle of almost $4\pi$. Closest to
 the interaction point is a high resolution silicon micro-vertex
 detector (SVD). It is surrounded by the central drift chamber
 (CDC). Two dedicated particle identification systems, the aerogel
 \v{C}erenkov counter (ACC) and the time-of-flight system (TOF), are
 mounted between the CDC and the CsI(Tl) crystal electromagnetic
 calorimeter (ECL). All these sub-detectors are located inside a 
 super-conducting coil that provides a magnetic field of 1.5~T. The
 return yoke of the coil is instrumented as a $K^0_L$ and $\mu$
 detector.
 A detailed description can be found in \cite{BelleDet}.

 This analysis uses 87.7~\fb\ of \epem\ annihilation
 data taken at the \Ys\ resonance at $\sqrt{s}=10.58~\GeV$ (``on-resonance
 data''), above the production threshold for $B\bar{B}$ pairs.
 Additional 15.0~\fb\ are taken 60~\MeV\ below the
 resonance at $\sqrt{s}=10.52~\GeV$ (``continuum data''), which is also below the production
 threshold for $B\bar{B}$ pairs.
 Hadronic events are selected as described in \cite{HadronB}.
 The selection efficiency of events originating from light quarks
 \mbox{($d$, $u$ and $s$)} passing this hadronic preselection has been
 estimated to be 84.0\%, using $9.6 \times 10^6$ MC events.
 For $c$ quarks, the efficiency has been determined with $6.6 \times
 10^6$ MC events to be 93.0\%.
 The light quark sample contains almost no true candidates, reflecting
 the small rate for gluon splitting into open charm states, i.e.\ two
 mesons containing $c$ quarks.
 
 To estimate the efficiency of reconstructing charmed hadrons
 and to correct for distortions due to the finite acceptance of the
 detector, \MC\ samples of $\epem\to\ccbar$ events
 corresponding to a data luminosity of 217~\fb\ (approximately 2 1/2
 times the on-resonance data), and $\epem\to\qqbar$ ($q=u$, $d$ and $s$)
 events corresponding to 18~\fb\ (approximately 1.2 times the
 continuum data), have been studied. The \MC~samples were generated
 using the QQ98 generator \cite{QQ98} employing the Peterson 
 fragmentation function for $c$ quarks and were processed through a detailed
 detector simulation based on GEANT 3.21 \cite{GEANT}. This sample
 will be referred to as the generic sample. Special samples of several million
 $\epem\to\ccbar$ events were generated with the EvtGen
 \cite{EvtGen} generator using the Peterson as well as the
 Bowler fragmentation functions and were also run through the detector
 simulation. These samples will be referred to as reweighted samples;
 see Section \ref{MC} for details about the reweighting procedure.
 For each charmed hadron used in this analysis, a sample was
 generated where that hadron 
 was forced to decay in the same channel as later reconstructed. 
 These samples were reconstructed using the same procedures as for data.
 
 \subsection{Particle Identification}
 To minimise possible kinematic biases due to tight selection criteria
 for identified particles, only loose cuts on 
 the particle identification of the stable particles have been applied.
 All
 particles with mean lifetime longer than $100~\mathrm{ps}$ have been
 called ``stable''.
 Apart from reducing a potential kinematic bias, this increased the
 reconstruction efficiency at the cost of introducing more background,
 especially in the low momentum region.
 
 In general, the identification for each track was based on one or more
 likelihood ratios, which combined the information from the
 time-of-flight and \v{C}erenkov counters and the energy loss dE/dx in
 the drift chamber.  Pions and kaons were separated by a single
 likelihood ratio $\mathcal{L} (K)/ (\mathcal{L} (K)+
 \mathcal{L}(\pi))$.  Charged particles were identified as pions if
 this ratio was less than 0.95 and as kaons if this ratio was larger
 than 0.05.  This overlap allowed a charged particle to be identified
 as both a pion and a kaon, potentially resulting in identifying a
 mother (candidate) particle as its own anti-particle ({\it i.e.,}\ a
 $D^0\rightarrow K^- \pi^+$\ decay could be identified as a
 $\overline{D}{}^0 \rightarrow \pi^- K^+$\ decay), and therefore
 overestimating the number of candidates.  As this misidentification
 was only possible for neutral particles, an additional systematic
 uncertainty has been assigned for the $D^0$\ and $D^{*0}$; see
 section \ref{systematics} for details.

 \begin{sloppypar}
 For proton identification, similar likelihood ratios were
 required to fulfil
 \mbox{${\cal L}(p)/({\cal L}(p)+{\cal L}(\pi))>0.6$} and 
 \mbox{${\cal L}(p)/({\cal L}(p)+{\cal L}(K))>0.6$}.
 For the \PZ, photon candidates with energies above 30 MeV
 were combined to form a \PZ\ candidate.
 Under the assumption that the \PZ\ candidate decayed at the
 interaction point, it was required to have an invariant mass
 consistent with the \PZ\ mass.
 \end{sloppypar}
 
 The efficiencies $\epsilon$ and misidentification probabilities
 $f$ for tracks from
 signal candidates under these cuts have been measured in
 data, and are listed in Table \ref{eff}; in all cases except the proton,
 $\epsilon>95\%$ and $f \leq 26\%$.
 For kaons and pions the efficiencies and 
 misidentification probabilities
 have been
 estimated in bins of the particle's momentum from \DSP\ and
 subsequent \DZ\TO\KM\PP\ decays; for protons, $\mathrm{\Lambda}$
 decays have been used. The observed momentum spectra in data have
 been used to derive the listed numbers.
 \begin{table}[h]
   \caption{\label{eff} Typical efficiencies and
     misidentification probabilities for tracks from
     signal candidates used in this analysis. The 
     misidentification probabilities listed under
     $\pi^{\pm}$ means the probability of mis-identifying it as a $K^{\pm}$.}
   \begin{center}
     \begin{ruledtabular}
       \begin{tabular}{rcccccccc}
	 & $\pi^{\pm}$ & $K^{\pm}$ & $p$ \\ \hline
	 \DZ\ $(\epsilon|f)$ & $(96\%|26\%)$ & $(96\%|26\%)$ & $-$ \\
	 \DP\ $(\epsilon|f)$ & $(96\%|12\%)$ & $(97\%|24\%)$ & $-$ \\
	 \Ds\ $(\epsilon|f)$ & $(98\%|17\%)$ & $(97\%|21\%)$ & $-$ \\
	 \LC\ $(\epsilon|f)$ & $(98\%|15\%)$ & $(97\%|21\%)$ & $(81\%|7\%)$ \\
       \end{tabular}
     \end{ruledtabular}
   \end{center}
 \end{table}
 
 In addition to the requirements on the particle identification, all
 tracks had to be consistent with coming from the interaction point
 (IP). 
 For the slow pion from the \DSP\TO\DZ\PP\ decay, all track quality and
 particle identification requirements were removed to increase the
 efficiency.
 
 \subsection{Reconstruction of charmed Hadrons}
 The reconstructed hadron decay chains used in this
 analysis are the following:\\
 \hspace*{0.25cm}
 $\DZ\to\KM\PP$, $\DP\to\KM\PP\PP$,
 $\Ds\to\PHI\PP$ $(\PHI\to\KP\KM)$, $\LC\to\pr\KM\PP$, \\
 \hspace*{0.25cm}
 $\DSP\to\DZ\PP$ $(\DZ\to\KM\PP)$,
 $\DSP\to\DP\PZ$ $(\DP\to\KM\PP\PP)$ and \\
 \hspace*{0.25cm}
 $\DSZ\to\DZ\PZ$ $(\DZ\to\KM\PP)$. 
 
 The inclusion of charge-conjugate modes is implied throughout this
 paper and natural units are used throughout.
 For all charmed ground state hadrons, candidates whose masses were
 within 50 $\MeV/c^2$ of their respective nominal mass were considered.
 For the intermediate \DZ\ and \DP\ coming from  the excited states
 \DSZ\ and \DSP\, a mass window of 15 $\MeV/c^2$ around the nominal masses of
 the \DZ\ and the \DP was chosen.
 Additionally, the selection window for the two excited states was
 tightened to 15 $\MeV/c^2$ around the nominal mass difference between the
 excited meson and the \DZ\ or \DP. For the
 intermediate \PHI\ from the \Ds\ decay, the mass window was chosen to
 be 7 $\MeV/c^2$.
 Multiple candidates for each particle and anti-particle were removed
 by a best candidate selection. Most false \DSZ\ and \DSP\ candidates
 were formed from a true \DZ\ and a random slow pion. Therefore, the slow
 pions were used to determine the best candidate. For the neutral slow
 pion, the smallest $\chi^2$ of the vertex fit was used.
 For the charged slow pion, the smallest distance to the IP of all
 hits used in the reconstruction was used.
 For all other charmed mesons, the selection was based on
 the particle identification of the kaon. In the rare case that
 multiple candidates were formed with the same kaon, the first
 candidate was randomly chosen.
 
 \begin{table}[h]
  \caption{\label{pdgvalues}The values for the masses or mass
    differences and branching
    fractions for all charmed hadrons used in this analysis. The
    masses are used only to shift the mass or the mass difference
    distributions in order to center their peaks near zero, therefore no errors
    are assigned.
    The branching fractions are taken from \cite{PDG2004}.}
  \begin{center}
  \begin{ruledtabular}
   \begin{tabular}{ccccc}
    hadron & decay mode & mass [$\GeV/c^2$] & product branching fraction \\ \hline
    \DZ  & \KM\PP\    & $1.8645$ & $0.0380 \pm 0.0009$ \\
    \DP  & \KM\PP\PP\ & $1.8693$ & $0.092 \pm 0.006$   \\
    \Ds  & \PHI\PP\   & $1.9685$ & $(0.036 \pm 0.009) \cdot (0.491 \pm 0.006)$ \\
    \LC  & \pr\KM\PP\ & $2.2849$ & $0.050 \pm 0.013$   \\
    \DSP & \DZ\PP\    & $0.1455$ & $(0.677 \pm 0.005) \cdot (0.0380 \pm 0.0009)$ \\
         & \DP\PZ\    & $0.1407$ & $(0.307 \pm 0.005) \cdot (0.092 \pm 0.009)$ \\
    \DSZ & \DZ\PZ\    & $0.1422$ & $(0.619 \pm 0.029) \cdot (0.0380 \pm 0.0009)$ \\
   \end{tabular}
  \end{ruledtabular}
  \end{center}
 \end{table}

 \section{Analysis Procedure}
 \label{anaproc}
 There are
 two variables commonly used in the measurements of fragmentation
 functions. These are the scaled energy
 $\mathrm{x_E={E_{candidate}}/{E^{MAX}_{candidate}}}$ and 
 the scaled momentum
 $\mathrm{x_P={|\vec{p}_{candidate}|}/{|\vec{p}^{MAX}_{candidate}|}}$,
 where $\mathrm{E^{MAX}_{candidate}}=\sqrt{s}/2$,
 $\mathrm{|\vec{p}^{MAX}_{candidate}|=\sqrt{s/4-m_H^2}}$, and $\mathrm{m_H}$
 denotes the mass of the charmed hadron.
 For $b$ quarks at higher CMEs, the scaled energy \xE\ is often used.
 In this case, the mass of the $B$ hadron reduces only slightly the
 allowed range at small \xE. For charmed hadrons at 10.58~\GeV\
 the range of \xE\ is significantly reduced; hence \xP\ is
 prefered and will be used in this analysis.
 Unless otherwise stated, 
   all variables are given in the \epem\ rest frame, taking
   into account the different beam energies for the on-resonance and the
   continuum samples.
 
 For various bins in the range from 0.0 to 1.1
 in the scaled momentum $\mathrm{x_P}$, the signal yield has
 been determined from a fit to the mass or mass difference
 distributions of all candidates within the aforementioned selection
 windows.
 The finite momentum resolution of the detector can result in events
 being recorded in the region above the na\"{\i}ve limit of \xP=1,
 however, in the case of $D^*$, the principal contribution is
 due to the process \mbox{$\epem\TO D^*D$}. See Section
 \ref{highxp} for details.
 
 A bin width of 0.02 in \xP\ has been chosen for all particles as a
 compromise between the statistical precision in each bin and the
 momentum resolution, which is a factor of two smaller.
 Additionally, to investigate the high \xP\ region around and above the
 na\"{\i}ve limit of \xP=1, the bin width has been decreased to 0.01;
 an expanded view of the region $0.90<\xP<1.05$ with this binning will
 be discussed in Section \ref{highxp}.
 Since this decreased bin width is still larger than, but comparable
 to, the momentum resolution, an unfolding using the
 singular-value-decomposition (SVD) approach \cite{SVD} was tried
 in addition to the normal bin-by-bin correction and
 is discussed in Section \ref{highxp}.
 
 The mass or mass difference distributions were parametrized by a
 single Gaussian, except for the \DSP\TO\DZ\PP\ decay channel where a
 double Gaussian was employed. For the mass distributions, the
 background was parametrized by a quadratic function; for $\xP>0.9$
 a linear function was found to be sufficient to fit the considerably lower
 background.
 For the mass difference distributions of the excited $D$ mesons, a
 phase-space-like function \mbox{$f(\Delta m)=a(\Delta m-\Delta
 M_0)^b$} was used with $a$ and $b$ being free parameters and
 $\Delta M_0$ the nominal difference between the mass of the excited
 mother particle and that of the ground state charm meson. 
 
 For all charmed hadrons, the mean mass $m_i$ and the width $\mu_i$ of
 the signal Gaussian was fitted separately for \MC, continuum and
 on-resonance data. For these fits, \xP\ was divided into 4 bins from
 $0.2<\xP<1.0$ with a constant bin size of 0.2. 
 In a second fit, two quadratic functions $m_i(\xP)$ and
 $\mu_i(\xP)$ were fitted to the results of the first fit in these
 four bins.
 For the distributions with a bin width of 0.02 and 0.01 in \xP,
 the mean and width parameters in the fit were fixed to the values
 of the quadratic functions $m_i(\xP)$ and $\mu_i(\xP)$ for the
 appropriate \xP\ value.
 
 For the \DSP\TO\DZ\PP\ decay mode full correlations between
 the two Gaussians of the signal function were taken into account when
 determining the fit yield. 
 
 When combining the on-resonance data with the 
 continuum data, two corrections have been applied to the on-resonance
 data. After normalising
 using the integrated luminosities of the respective samples, the na\"{\i}ve
 $1/s$ dependence on the total hadronic cross section has been taken
 out by multiplying the distributions of the on-resonance sample by the
 square of the ratio of the CME's, namely by $(10.58~\GeV/10.52~\GeV)^2$.
 Second, from \MC\ an additional correction of 
 $+0.27\%$ due to different initial state radiation (ISR) at the two
 energy points has been
 applied to the on-resonance samples. This correction was  based on a
 \MC\ study of the total cross sections at these two energy points.
 
 \subsection{\xP-dependent Mass Fits}
 Fig.\ \ref{mass-xp-fita} and Fig.\ \ref{mass-xp-fitb} show 
 the mass distributions of all charmed hadrons 
 reconstructed in this analysis for two representive bins in \xP.
 The \xP\ bins shown are $0.28<\xP<0.30$ in Fig.\ \ref{mass-xp-fita}
 and $0.68<\xP<0.70$ in Fig.\ \ref{mass-xp-fitb}.
 They represent a low \xP\ bin with higher background and a bin close
 to the maximum of the \xP\ distribution with less background,
 respectively.
 All mass (mass difference) distributions have been shifted by
 their nominal mass (mass difference) to center the peaks at zero.
 See table \ref{pdgvalues} for the masses used.
 Note that the scale on the $y$-axis does not start at zero in the
 upper four plots in Fig.\ \ref{mass-xp-fita} and
 Fig.\ \ref{mass-xp-fitb}.
 
 For $0.28<\xP<0.30$ (shown in Fig.\ \ref{mass-xp-fita}), the mass
 distributions for the \DZ\ and \DP\ ground states and the mass
 difference distributions for the excited states show clear peaks at
 the expected value for signal.
 Compared to higher \xP\ values, the
 background is higher due to a larger amount of combinatorial
 background, and the signal-to-background ratio is lower. 
 At higher \xP\ values, such as those shown in Fig.\ \ref{mass-xp-fitb}
 $(0.68<\xP<0.70)$, the background is considerably reduced,
 whereas the signal yield is enhanced. This significantly increased the
 signal-to-background ratio.
 
 \begin{center}
   \pagestyle{empty}
  \begin{figure}[h]
   \includegraphics[width=0.47\textwidth]{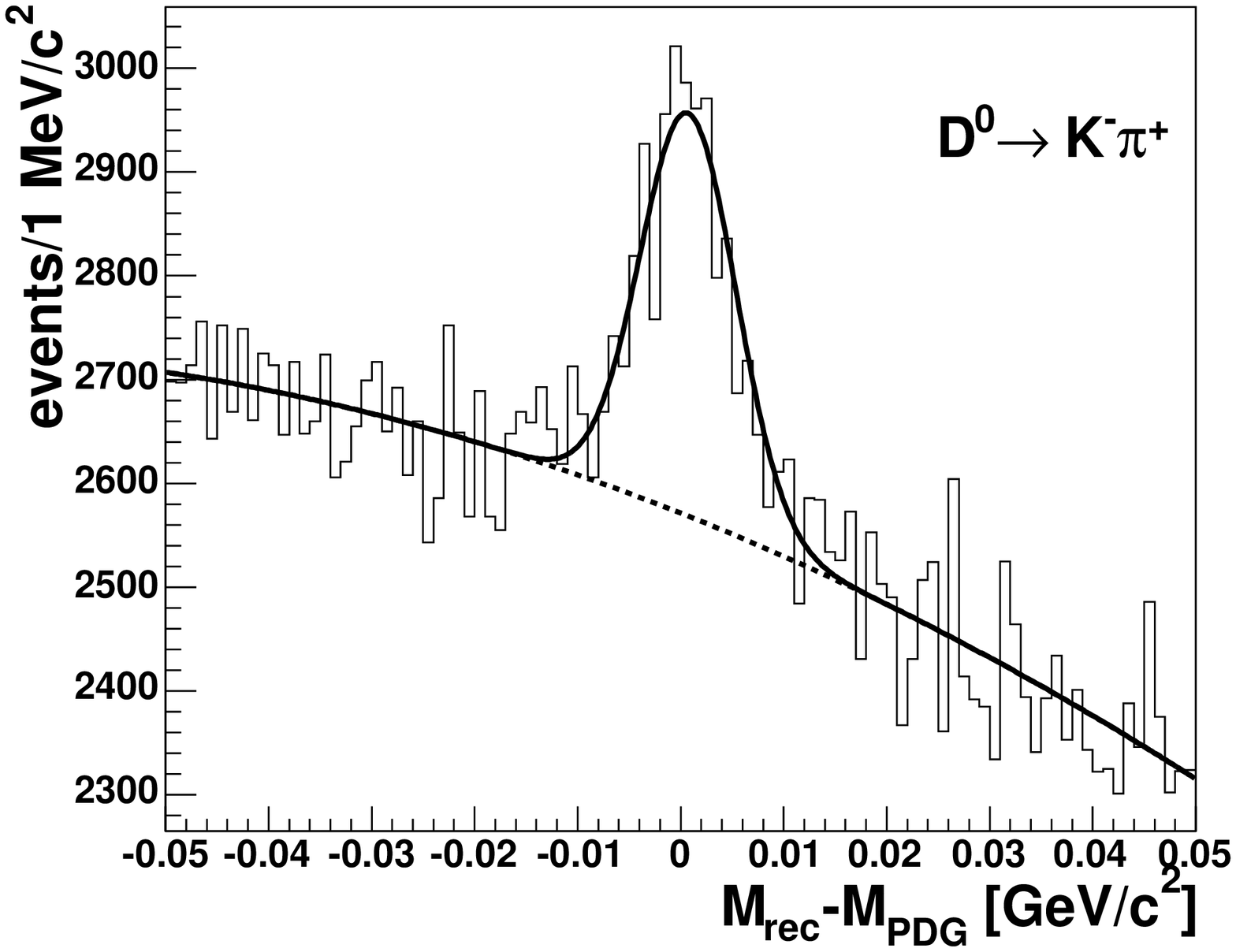}
   \includegraphics[width=0.47\textwidth]{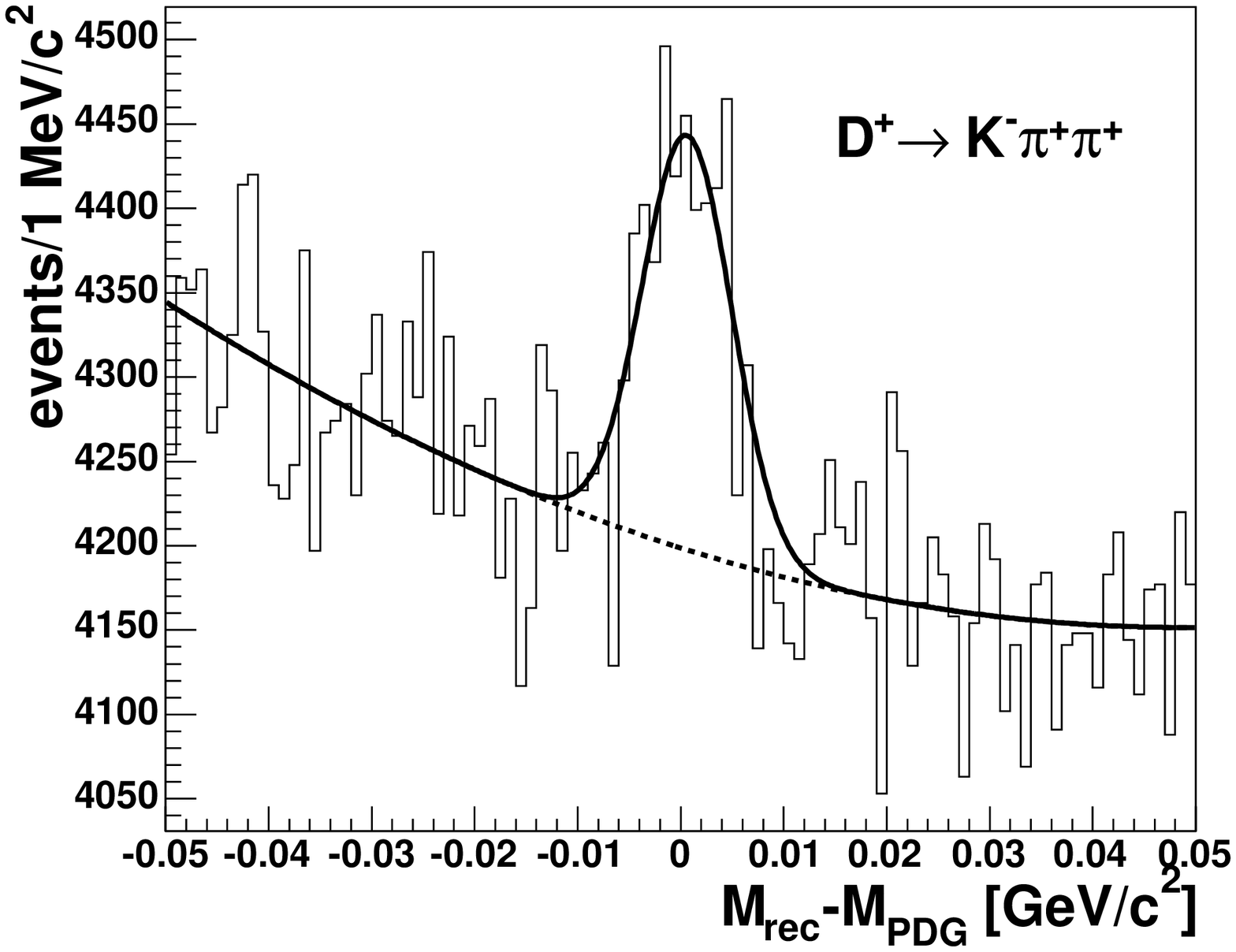}\\
   \includegraphics[width=0.47\textwidth]{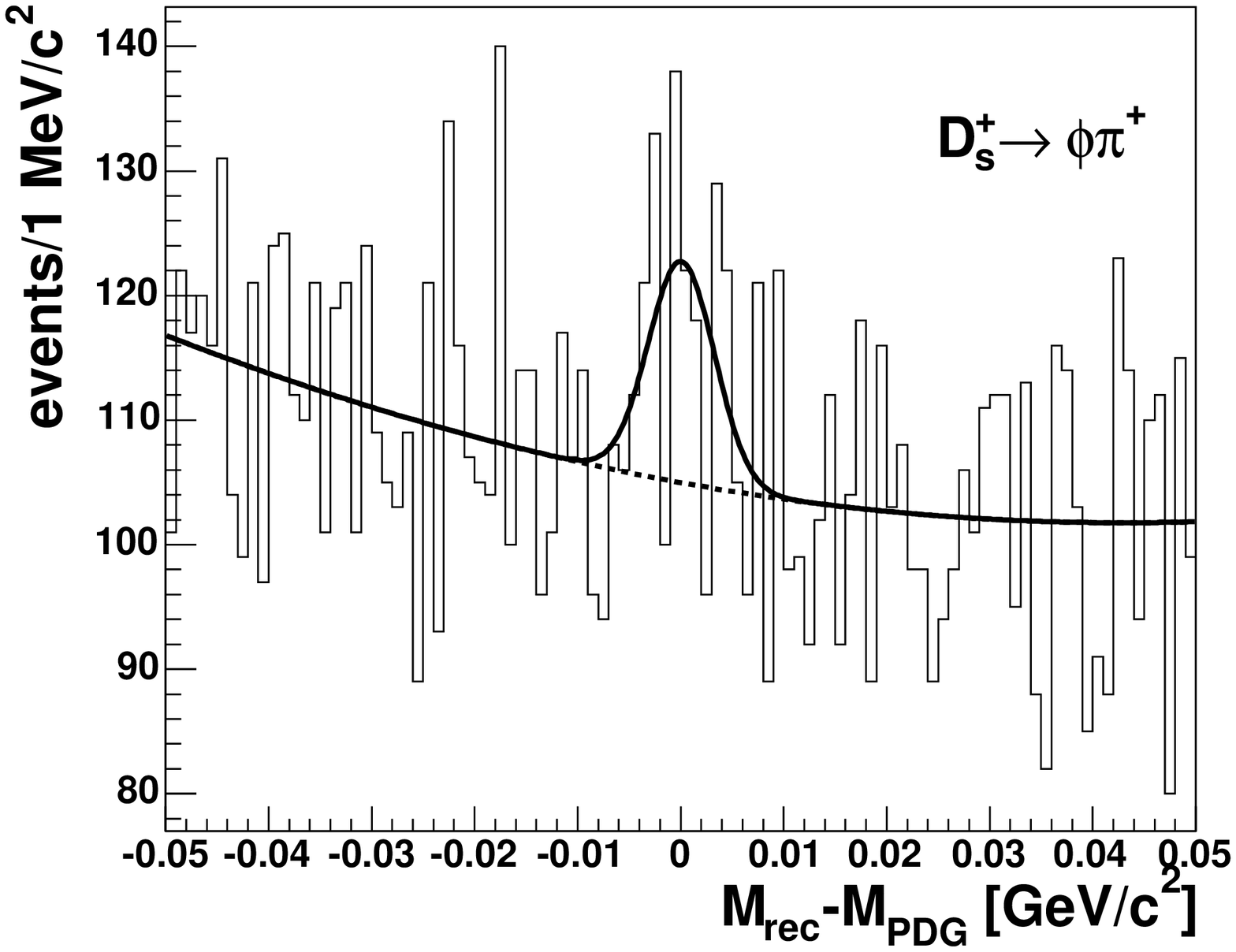}
   \includegraphics[width=0.47\textwidth]{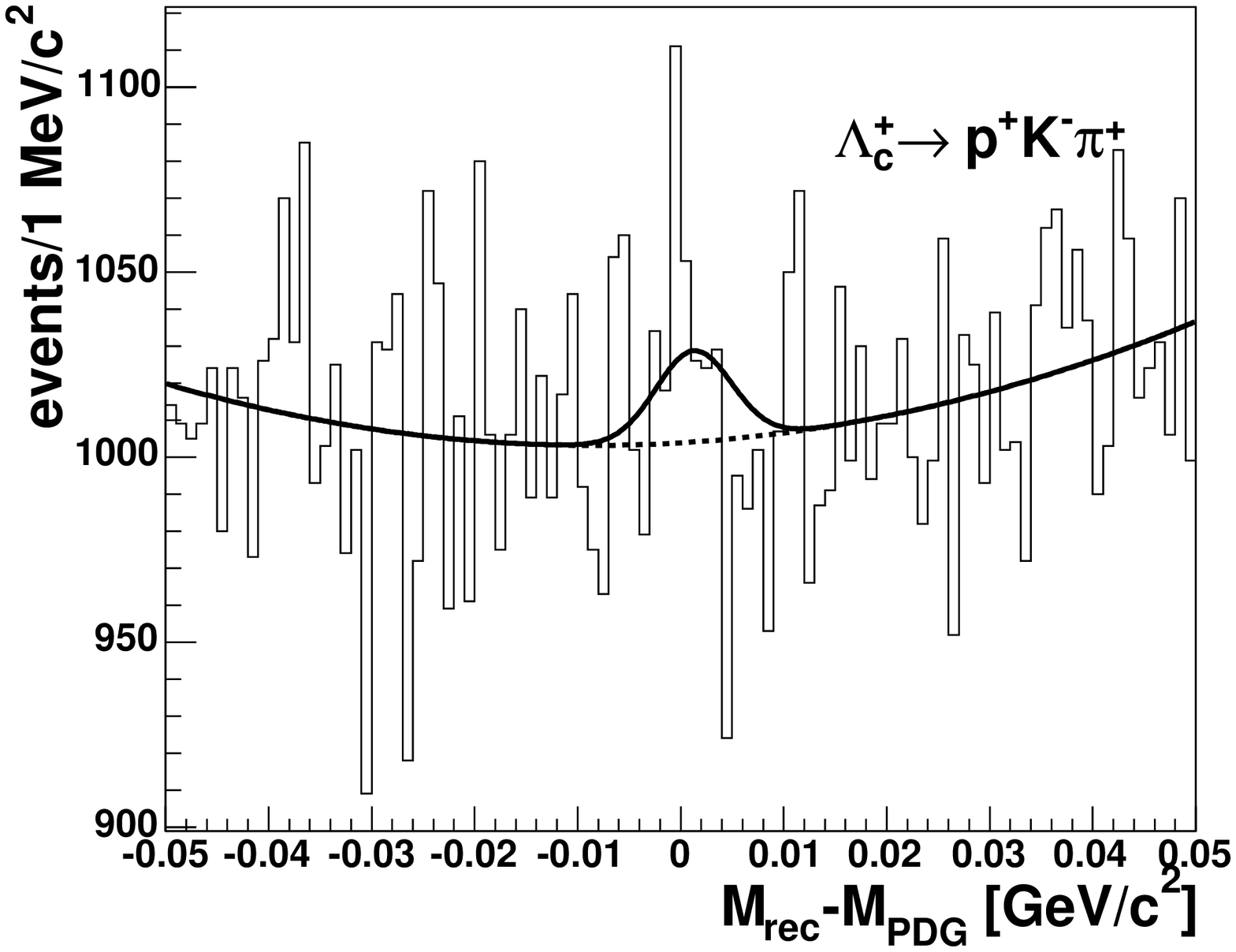}\\
   \includegraphics[width=0.31\textwidth]{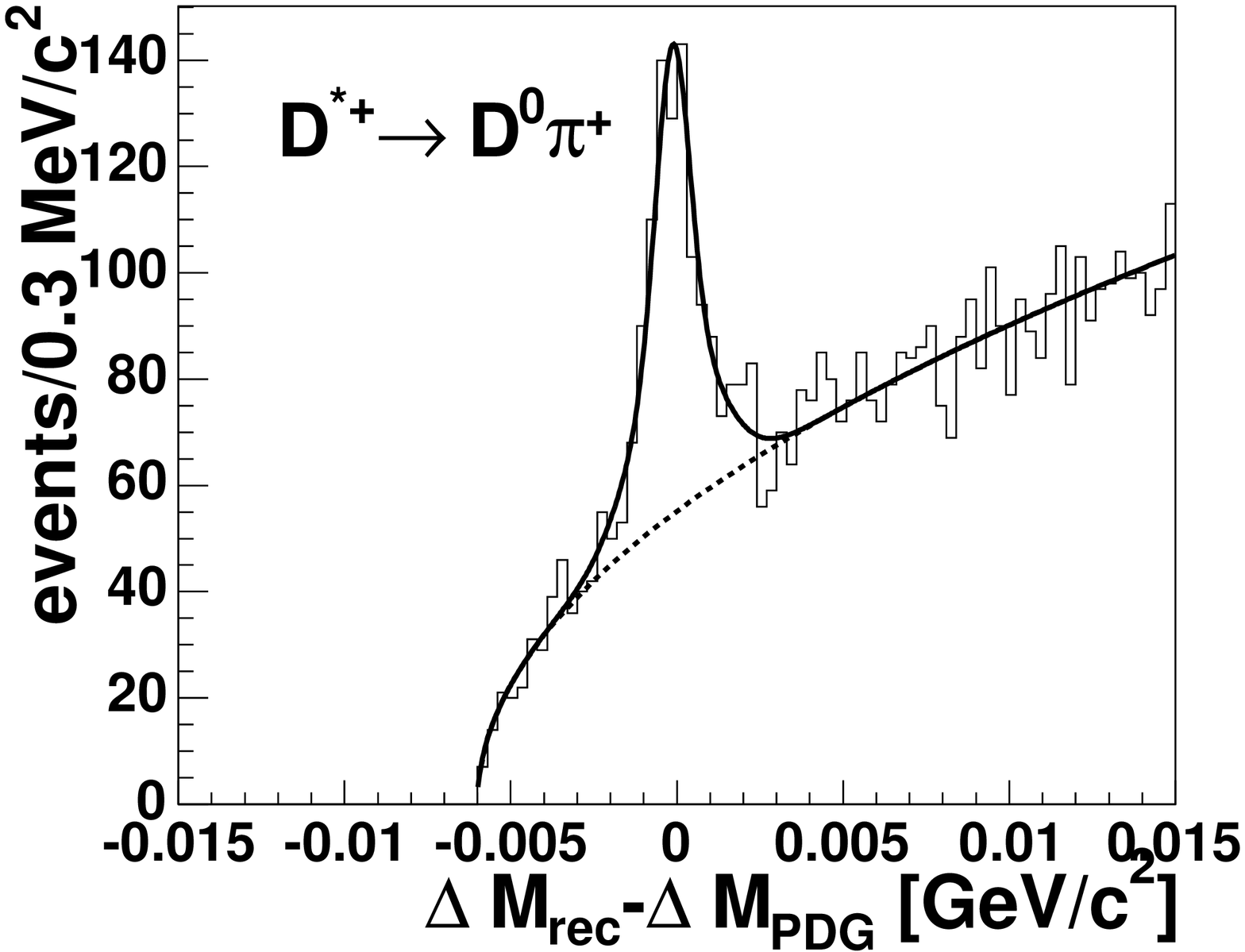}
   \includegraphics[width=0.31\textwidth]{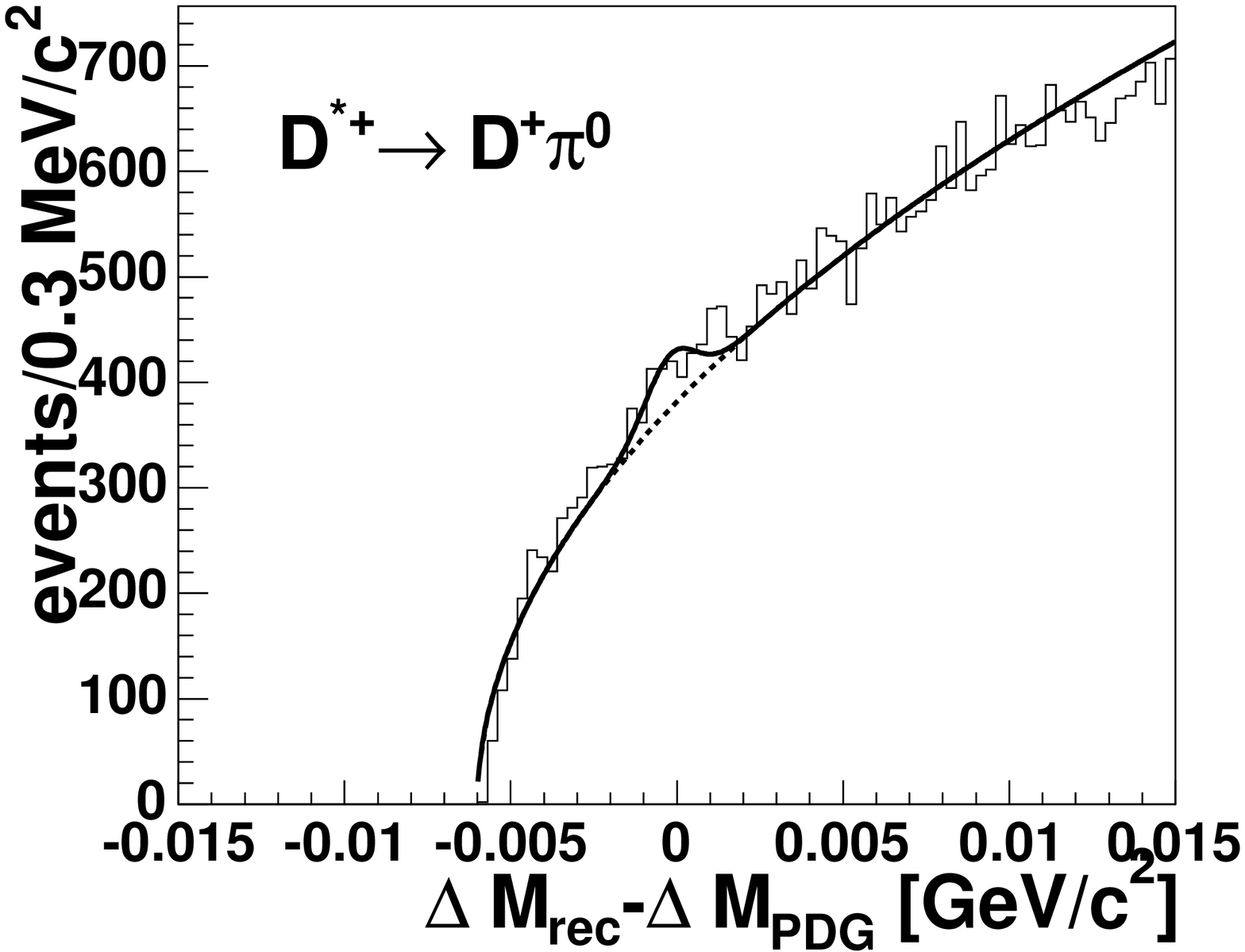}
   \includegraphics[width=0.31\textwidth]{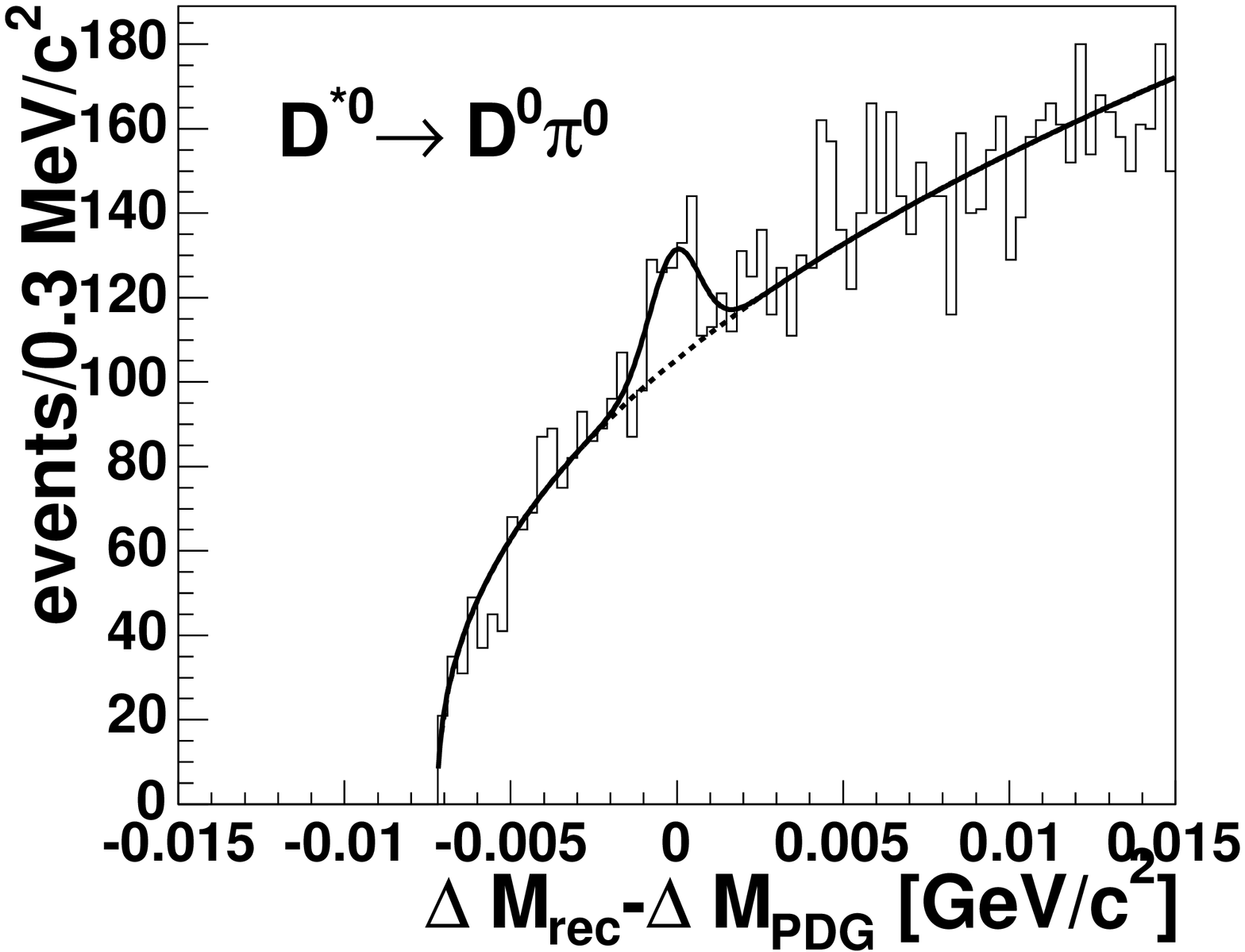}
   \caption{\label{mass-xp-fita}
     Mass and mass difference distributions for all charmed hadrons
     reconstructed in this analysis, for $0.28 < \xP < 0.30$ for the
     continuum sample. The histograms show the data, the dotted
     line describes only the background, the full line includes the
     signal.
     The top row shows the $D^0$\ (left) and the $D^+$\ (right), the
     middle shows the $D^+_s$\ (left) and the $\Lambda^+_c$\ (right),
     and the bottom row shows $D^{*+}\to D^0\pi^+$\ (left), the
     alternative decay mode $D^{*+}\to D^+\pi^0$\ (middle), and the
     $D^{*0}$ (right).
   }
  \end{figure}
 \end{center}
 
 \begin{center}
  \begin{figure}[h]
   \includegraphics[width=0.47\textwidth]{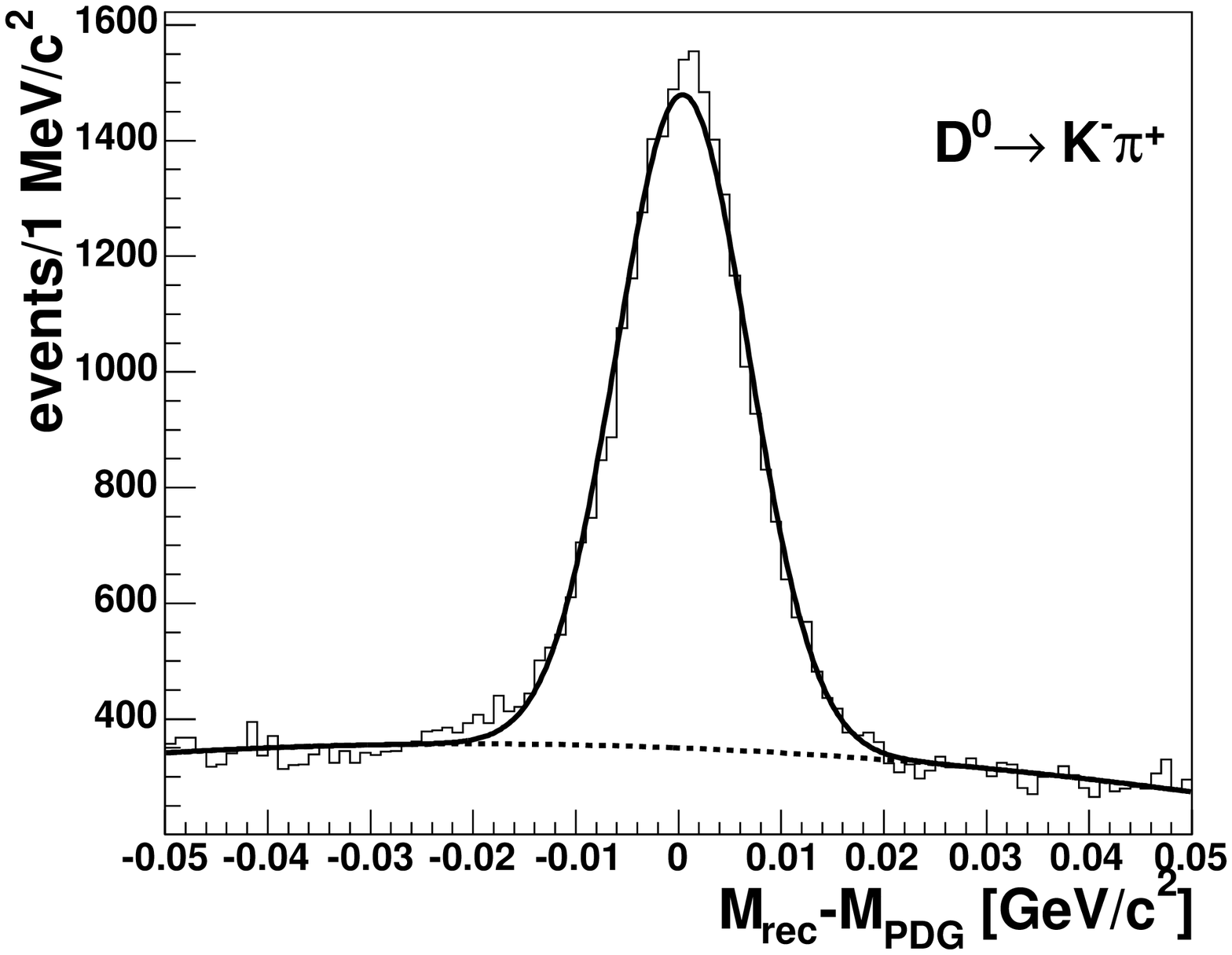}
   \includegraphics[width=0.47\textwidth]{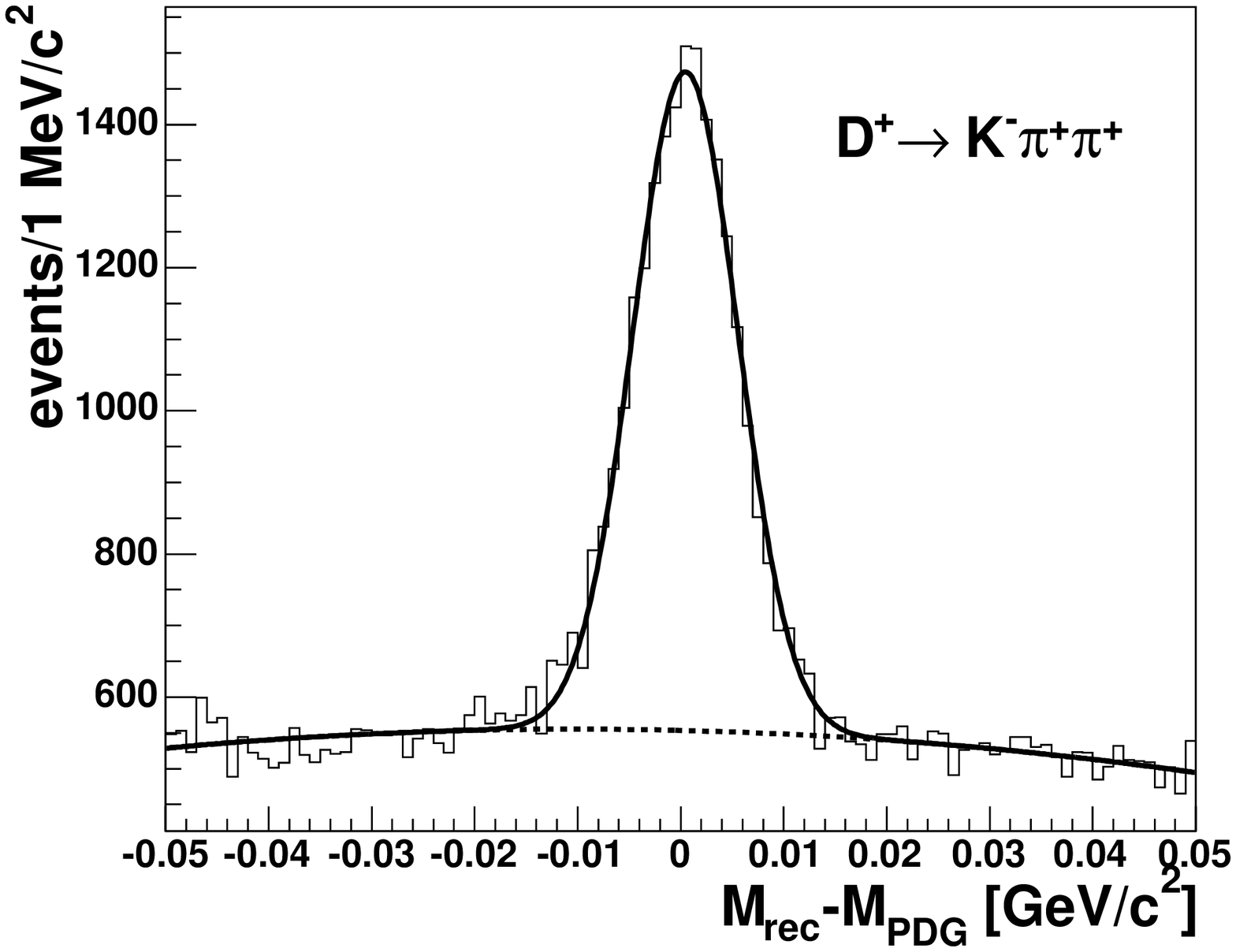}\\
   \includegraphics[width=0.47\textwidth]{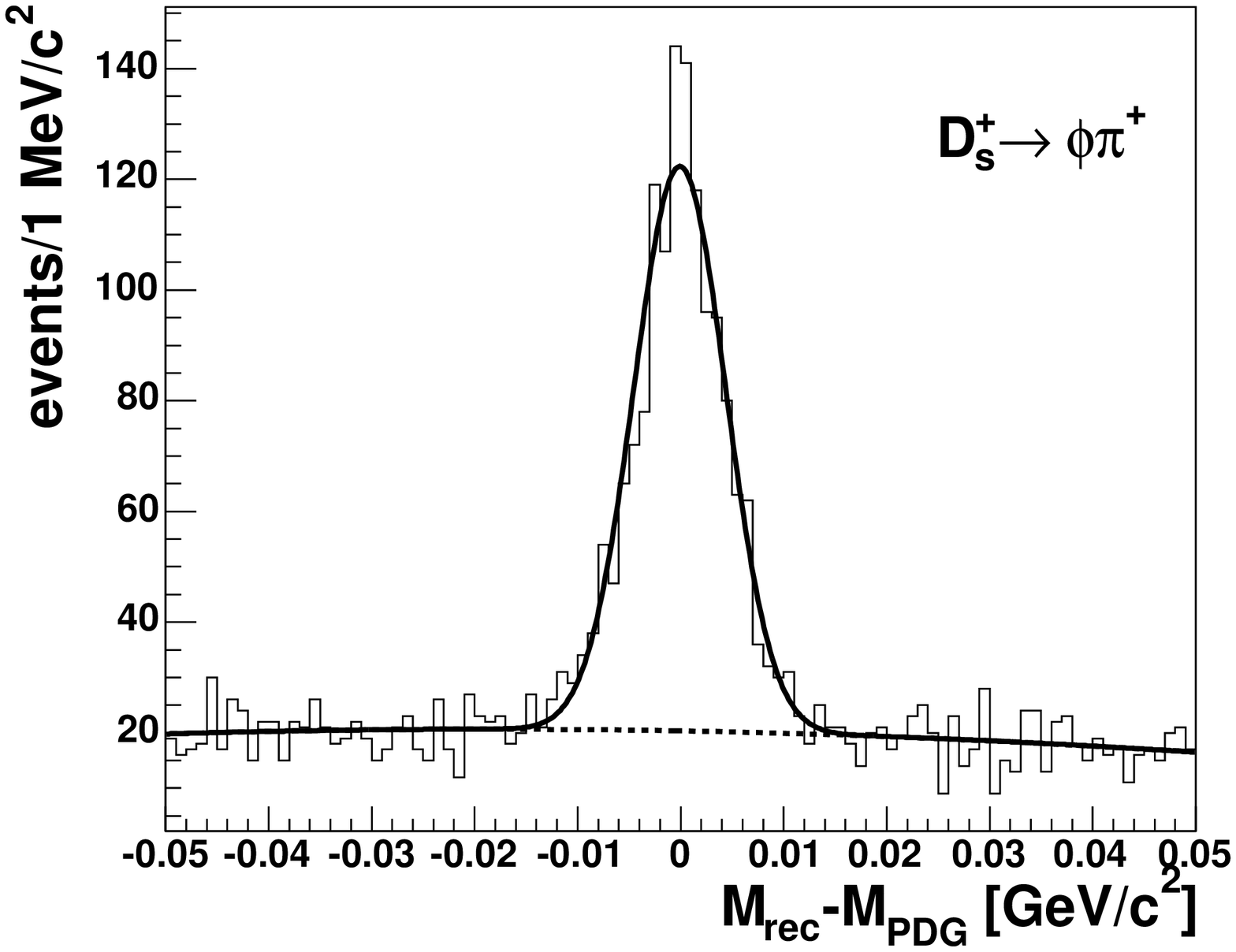}
   \includegraphics[width=0.47\textwidth]{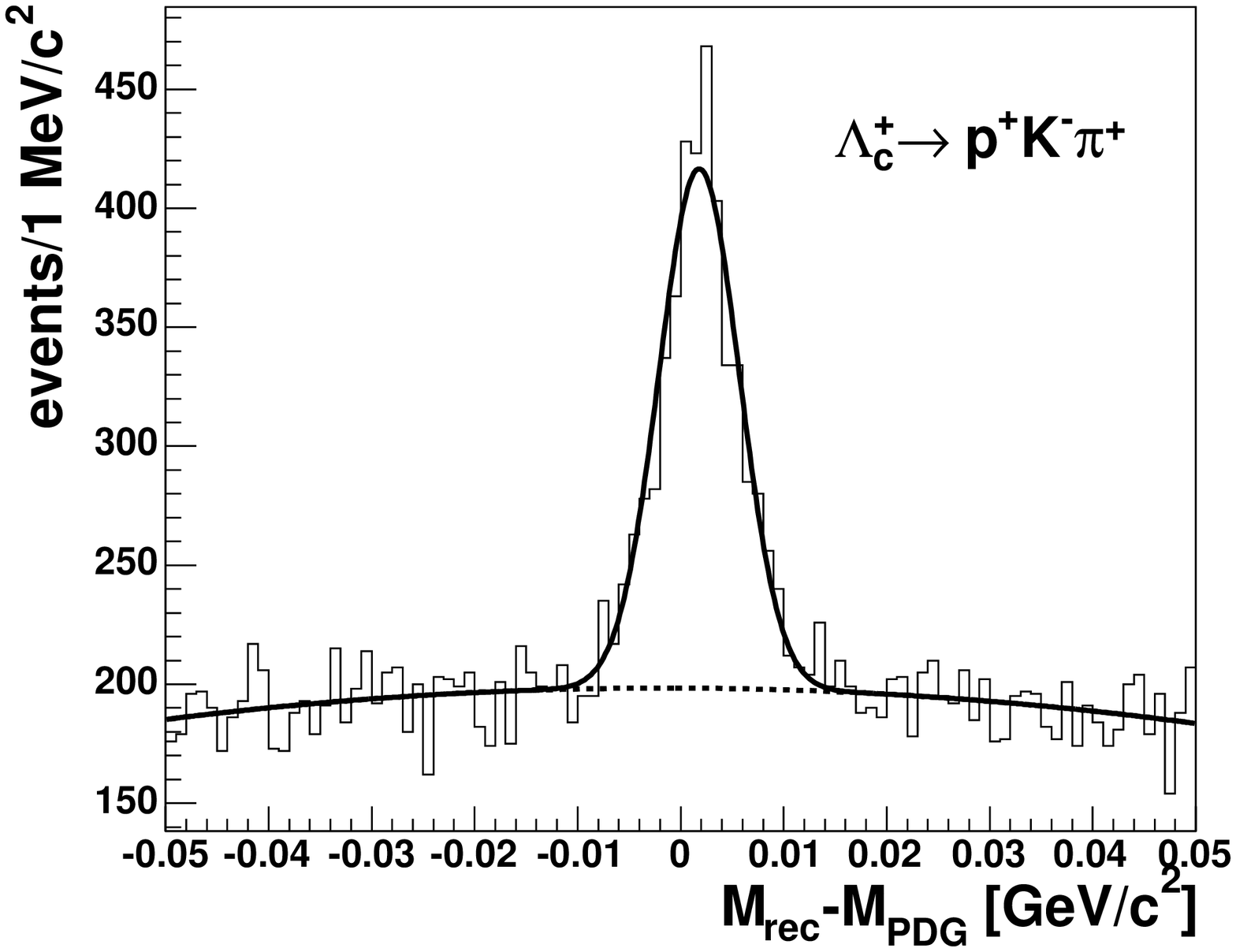}\\
   \includegraphics[width=0.31\textwidth]{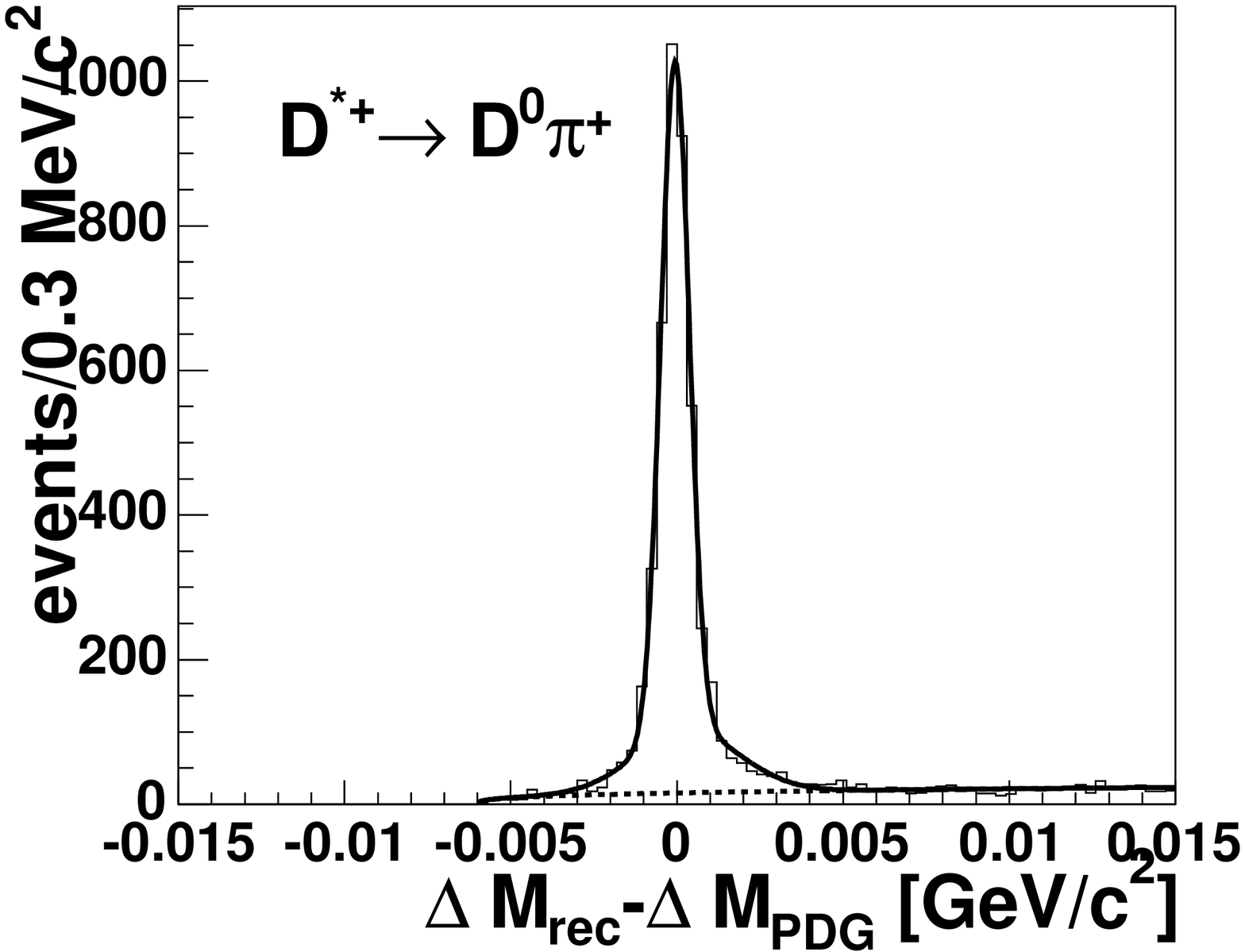}
   \includegraphics[width=0.31\textwidth]{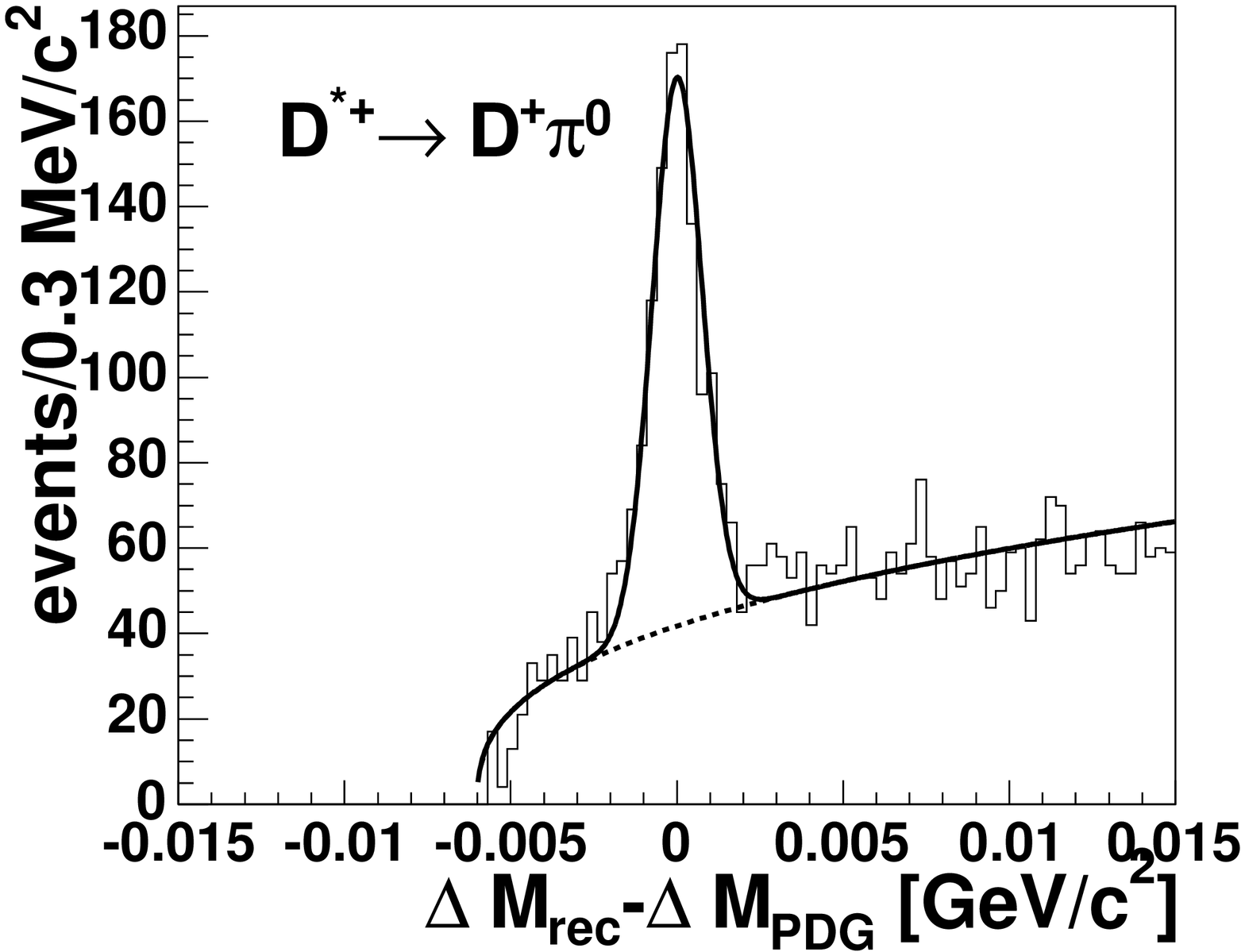}
   \includegraphics[width=0.31\textwidth]{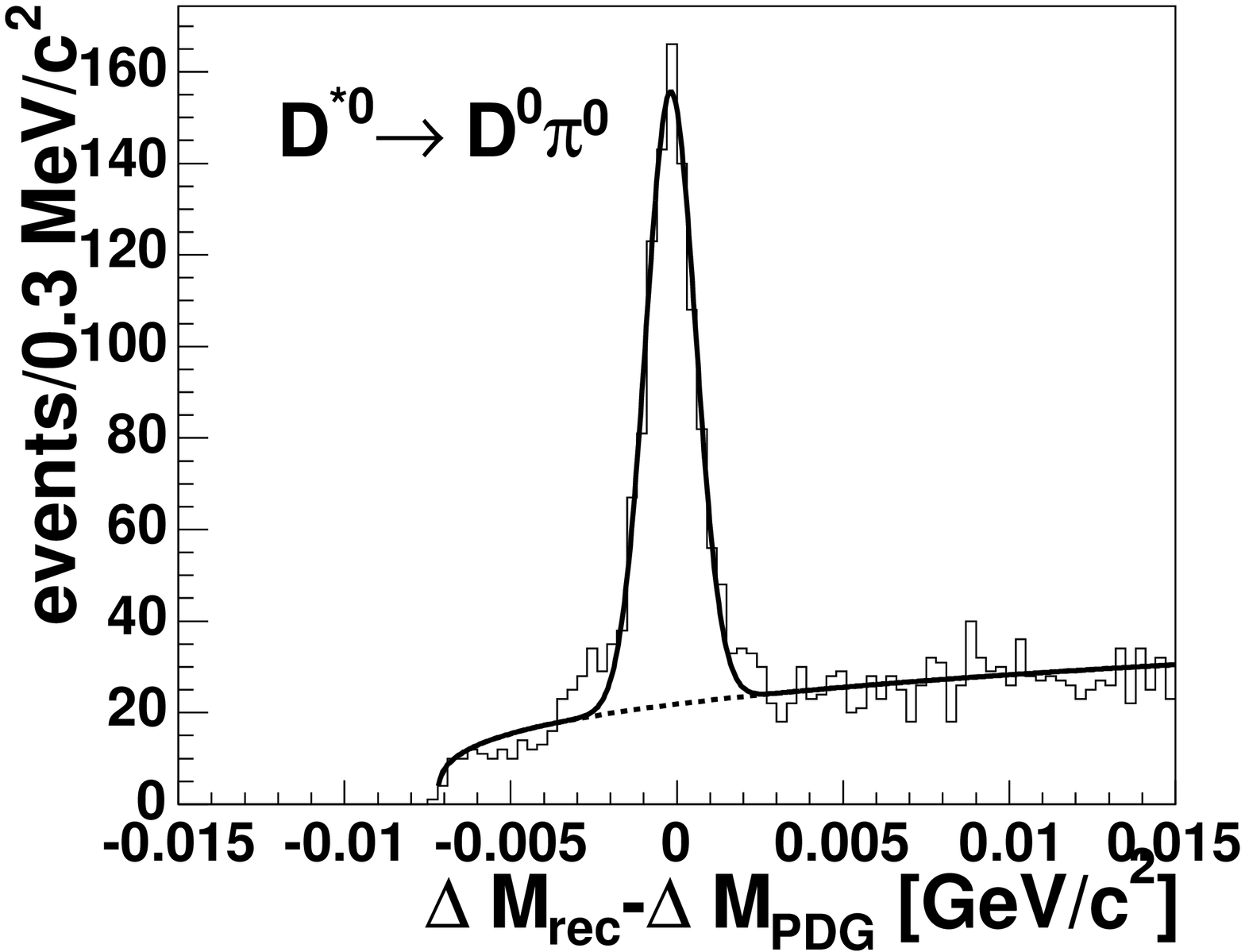}
   \caption{\label{mass-xp-fitb}
     Mass and mass difference distributions for all charmed hadrons
     reconstructed in this analysis, for $0.68<\xP<0.70$.
     The order of the plots is the same as in Fig.\
     \ref{mass-xp-fita}. As in the previous figure, 
     the histograms show the data, the dotted
     lines describe only the background, the full lines include the
     signal.
   }
  \end{figure}
 \end{center}
 
 \subsection{Raw Signal Yield}
\begin{center}
  \begin{figure}[h]
   \includegraphics[width=0.47\textwidth]{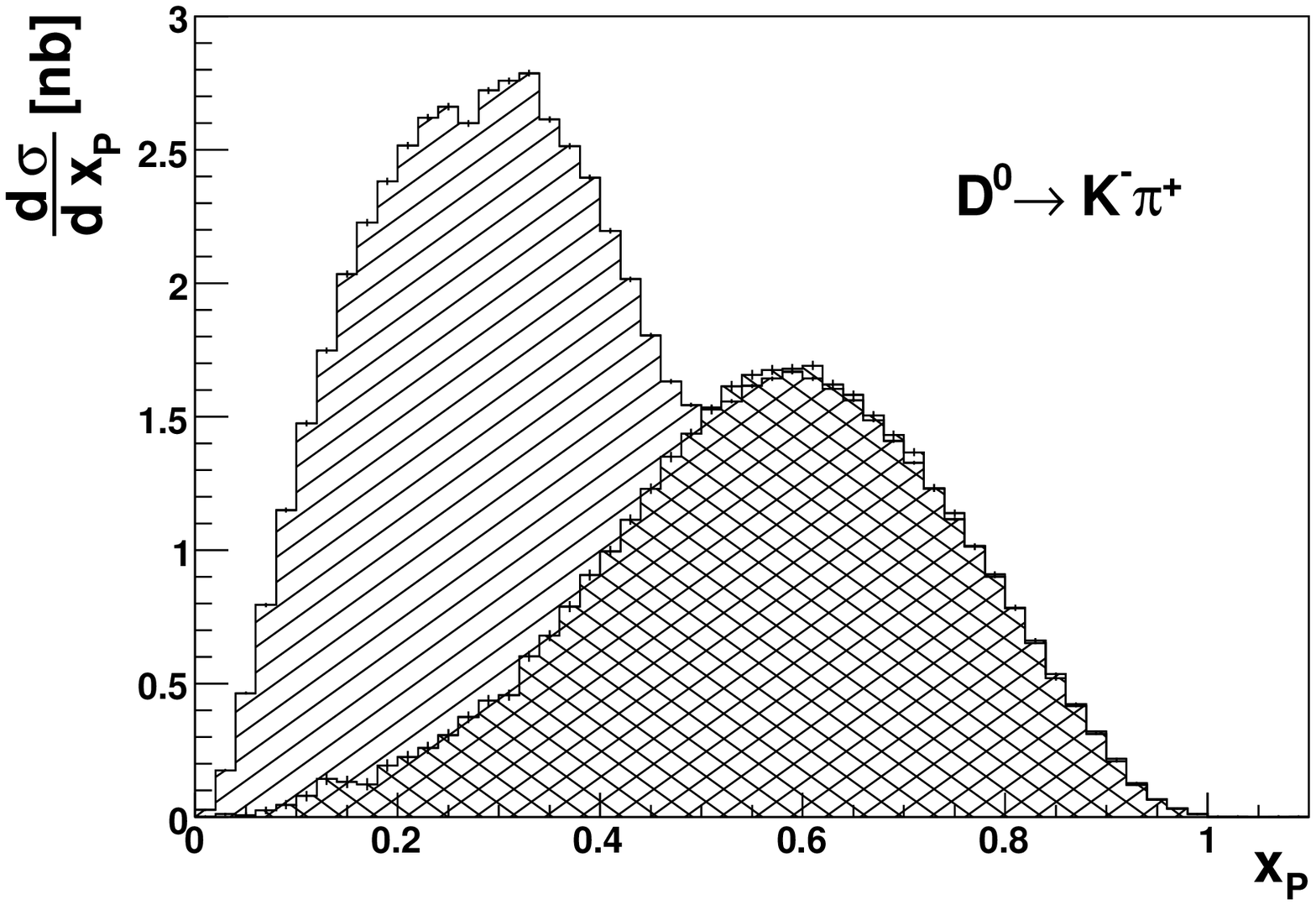}
   \includegraphics[width=0.47\textwidth]{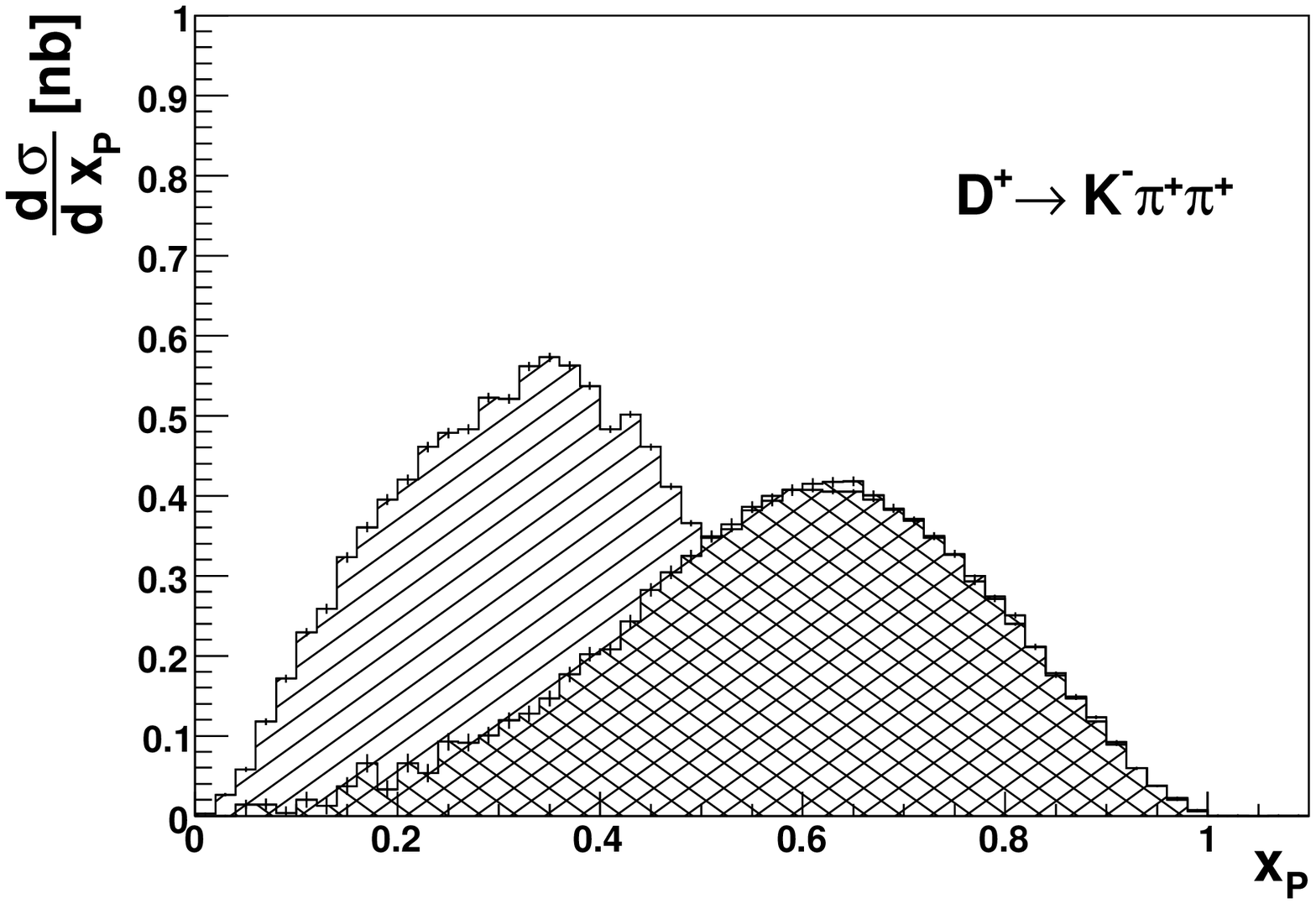}\\
   \includegraphics[width=0.47\textwidth]{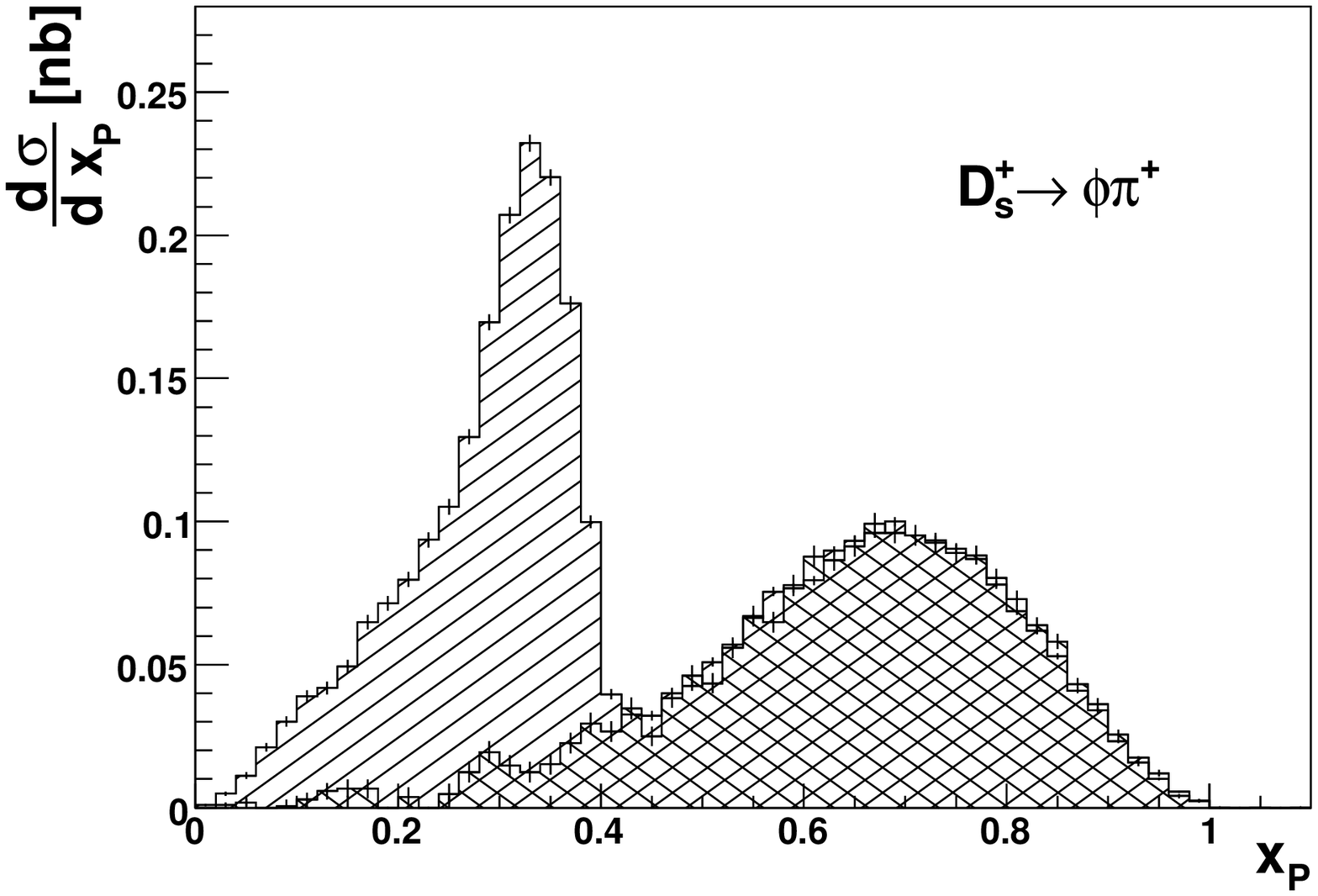}
   \includegraphics[width=0.47\textwidth]{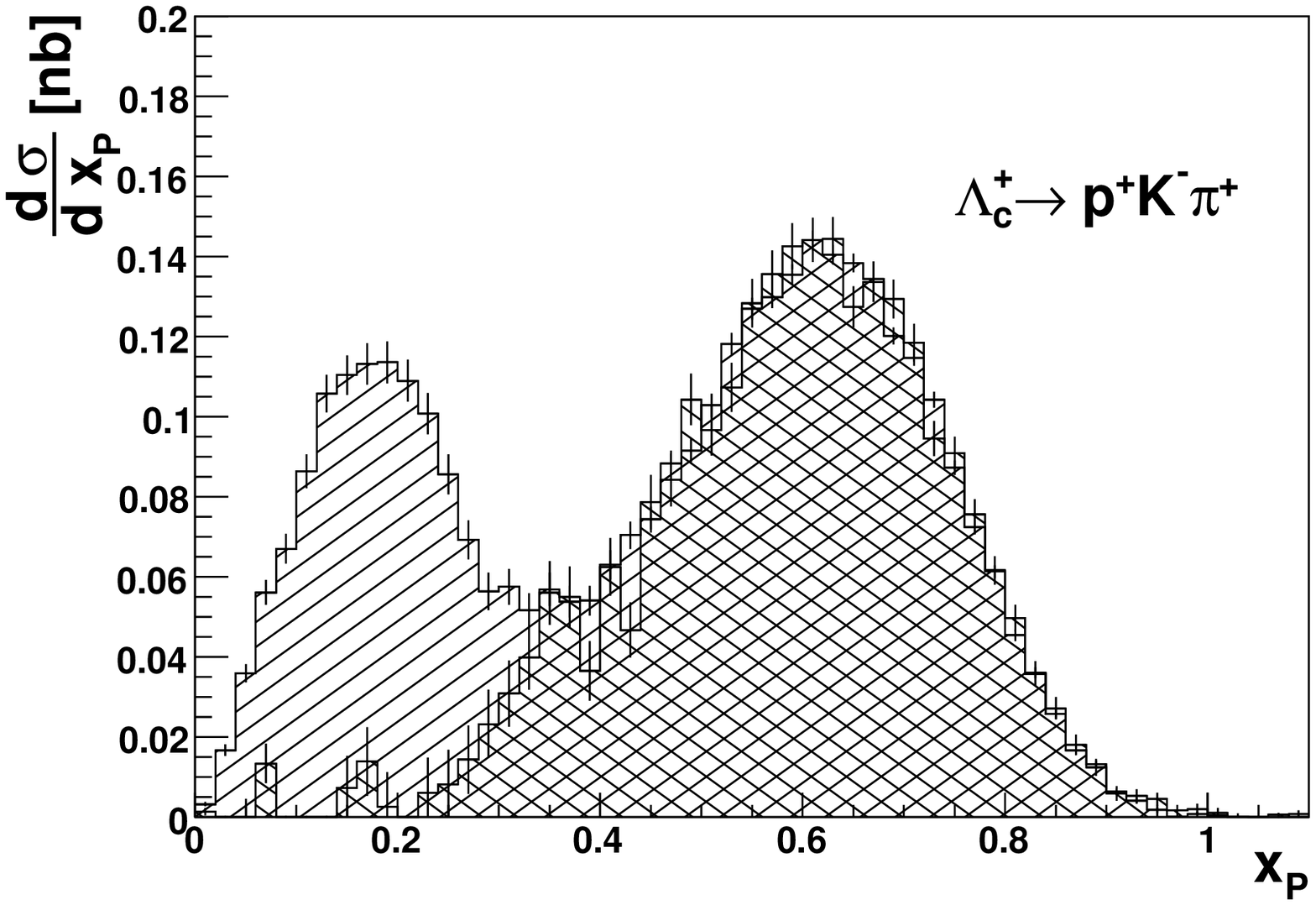}\\
   \includegraphics[width=0.3\textwidth]{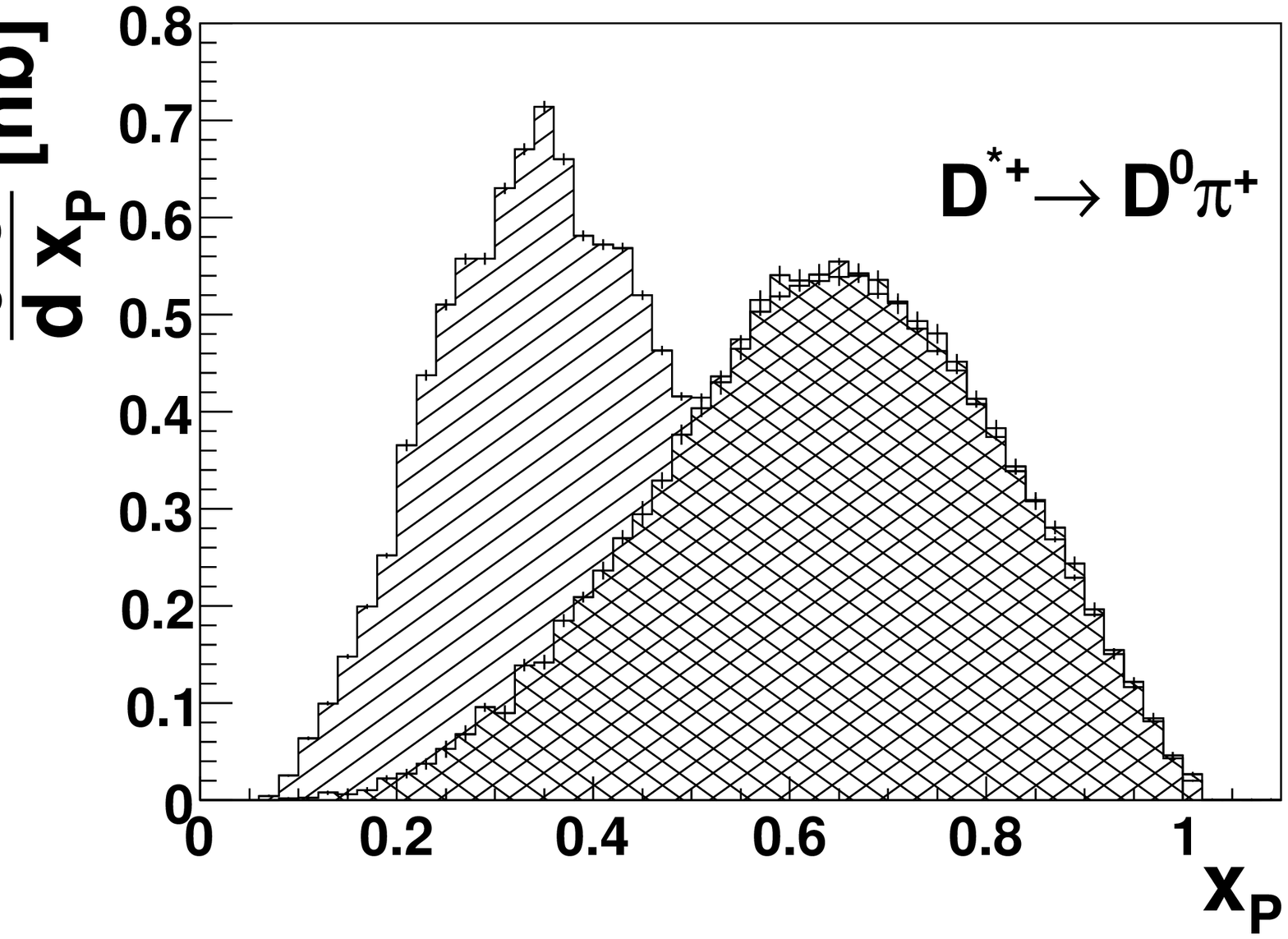}
   \includegraphics[width=0.3\textwidth]{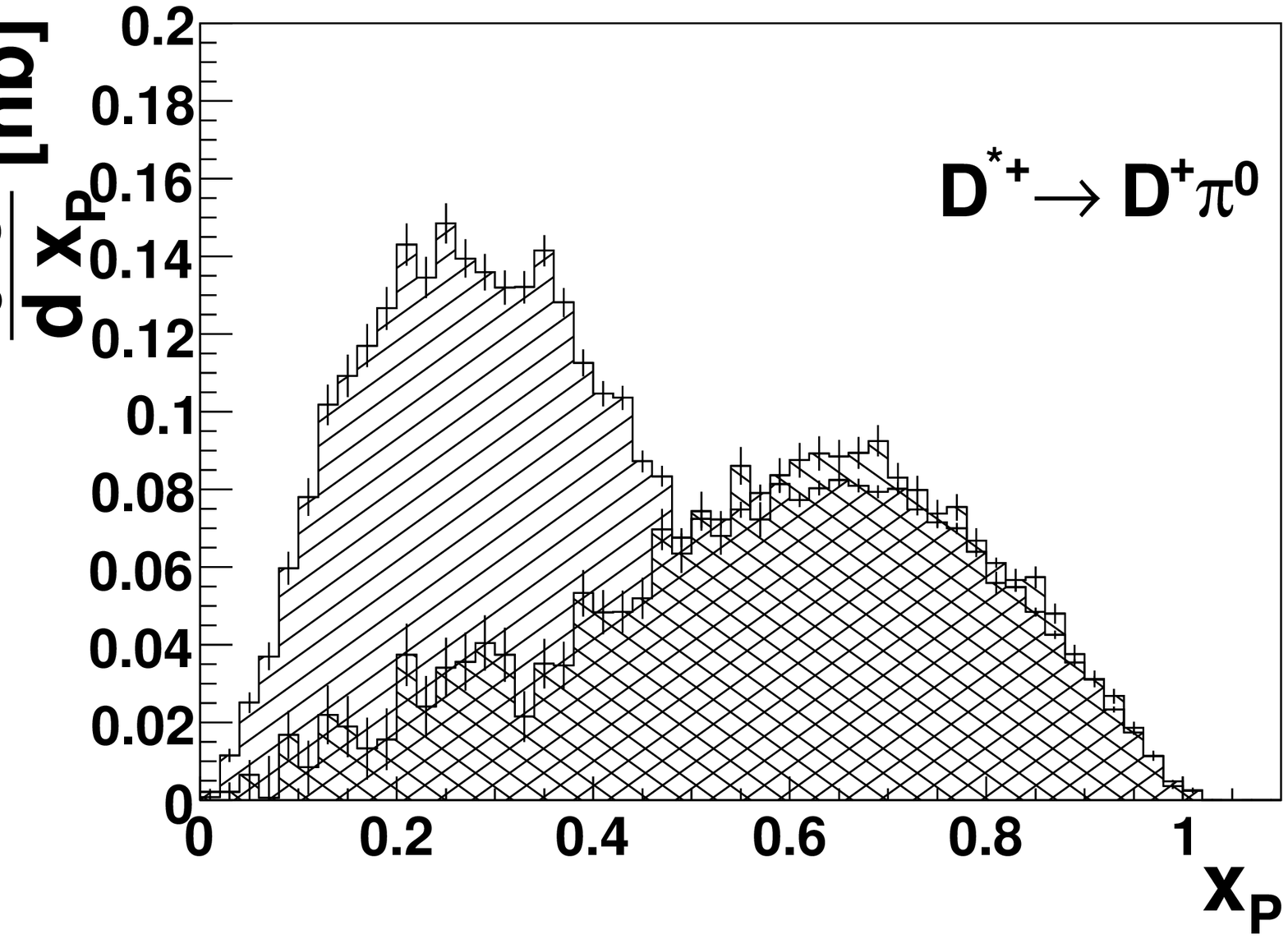}
   \includegraphics[width=0.3\textwidth]{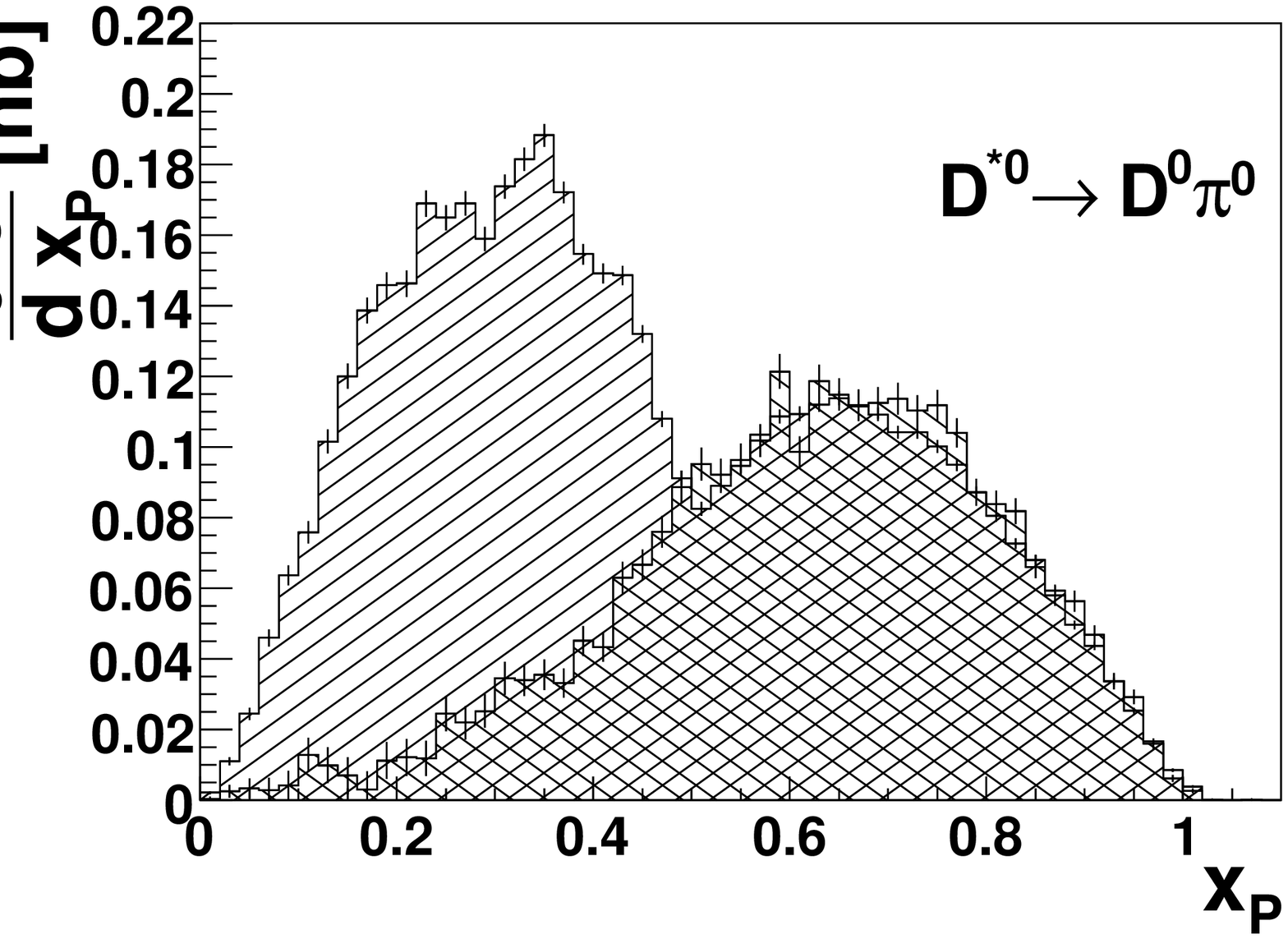}
   \caption{\label{signal-yield}The signal
     yield not corrected for efficiencies for the charmed hadrons. The order of the particles is the
     same as in Fig.\ \ref{mass-xp-fita}.
     The contribution from $B$ decays in the on-resonance samples
     (down-left hatching)
     is clearly visible in the region $\xP<0.5$.
     The error bars show the statistical uncertainties only.
   }
  \end{figure}
 \end{center}

 Fig.\ \ref{signal-yield} shows the signal yields as a function of
 \xP\ for all charmed hadrons, not corrected for the reconstruction
 efficiencies and for the branching
 fractions, denoted with ``B'' in the plots.
 For all particles, the contribution from $B$ decays is clearly
 visible in the low \xP\ range, which is $\xP<0.5$ for all charmed
 mesons containing a light quark as the spectator. For \Ds\ from
 $B$ decays, the upper bound is approximately
 $\xP\sim0.4$, reflecting the energy required to produce an additional
 strange quark. Contributions from the $b\to u$ transition,
 where the \Ds\ is formed at the upper vertex, can populate the region
 up to $\xP=0.5$, but are strongly suppressed.
 For the \LC, the only baryon reconstructed in this
 analysis, the upper bound is further decreased to approximately
 $\xP\sim0.37$, due to the production of an additional anti-baryon.
 
 All distributions peak around $\xP\sim0.6-0.7$ and show similar
 shapes.
 
 \subsection{Efficiency Correction}
\begin{center}
  \begin{figure}[h]
   \includegraphics[width=0.47\textwidth]{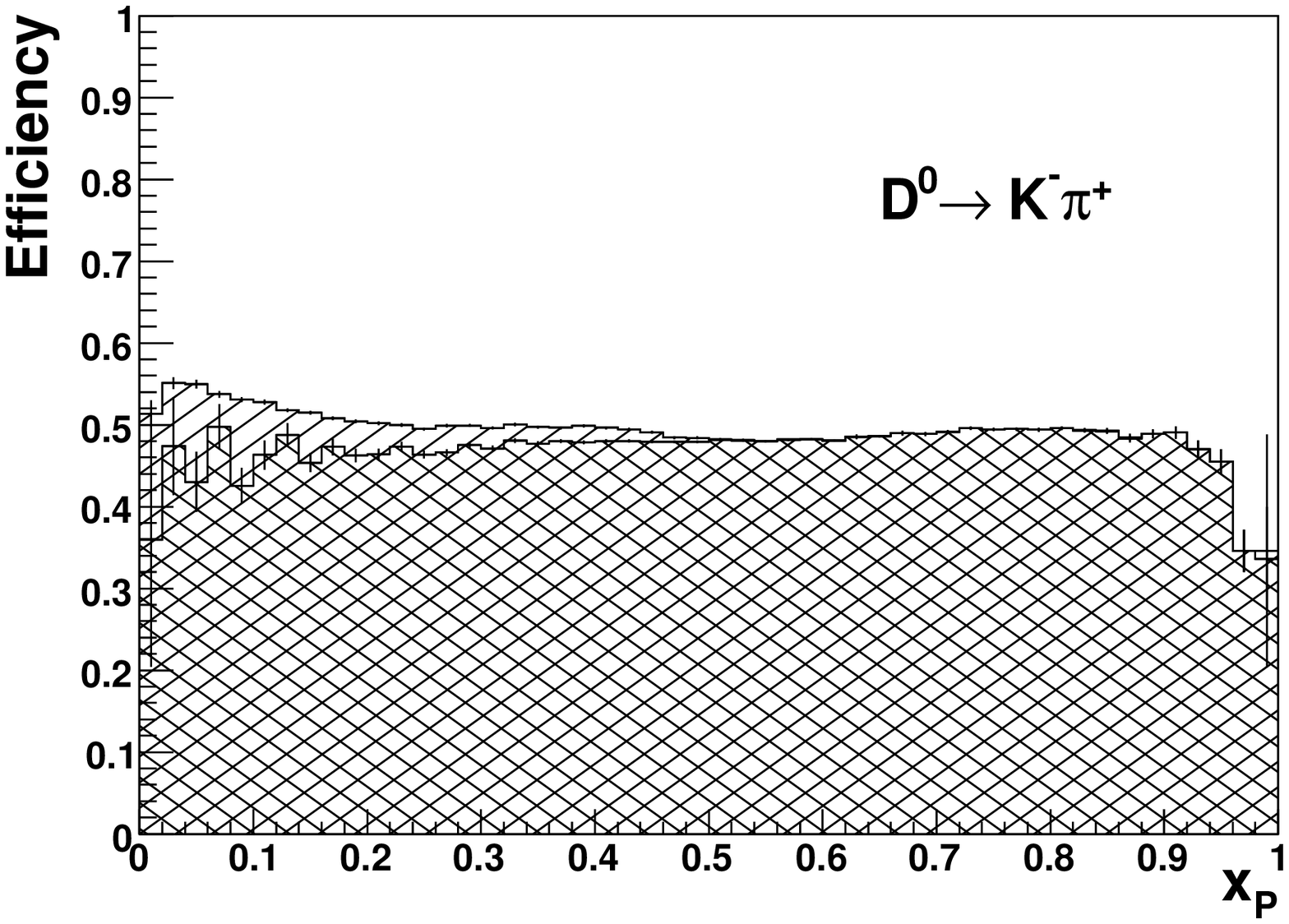}
   \includegraphics[width=0.47\textwidth]{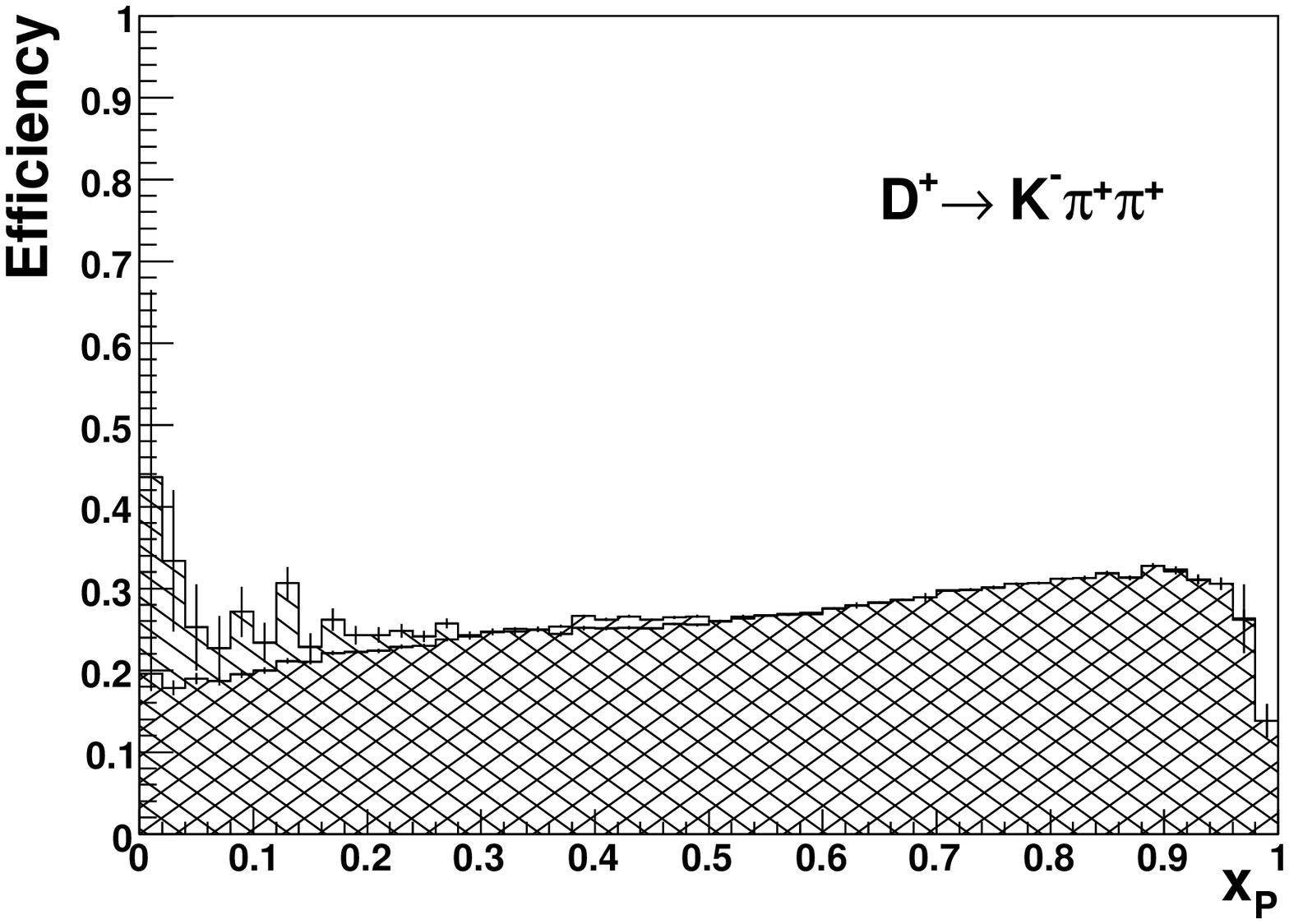}\\
   \includegraphics[width=0.47\textwidth]{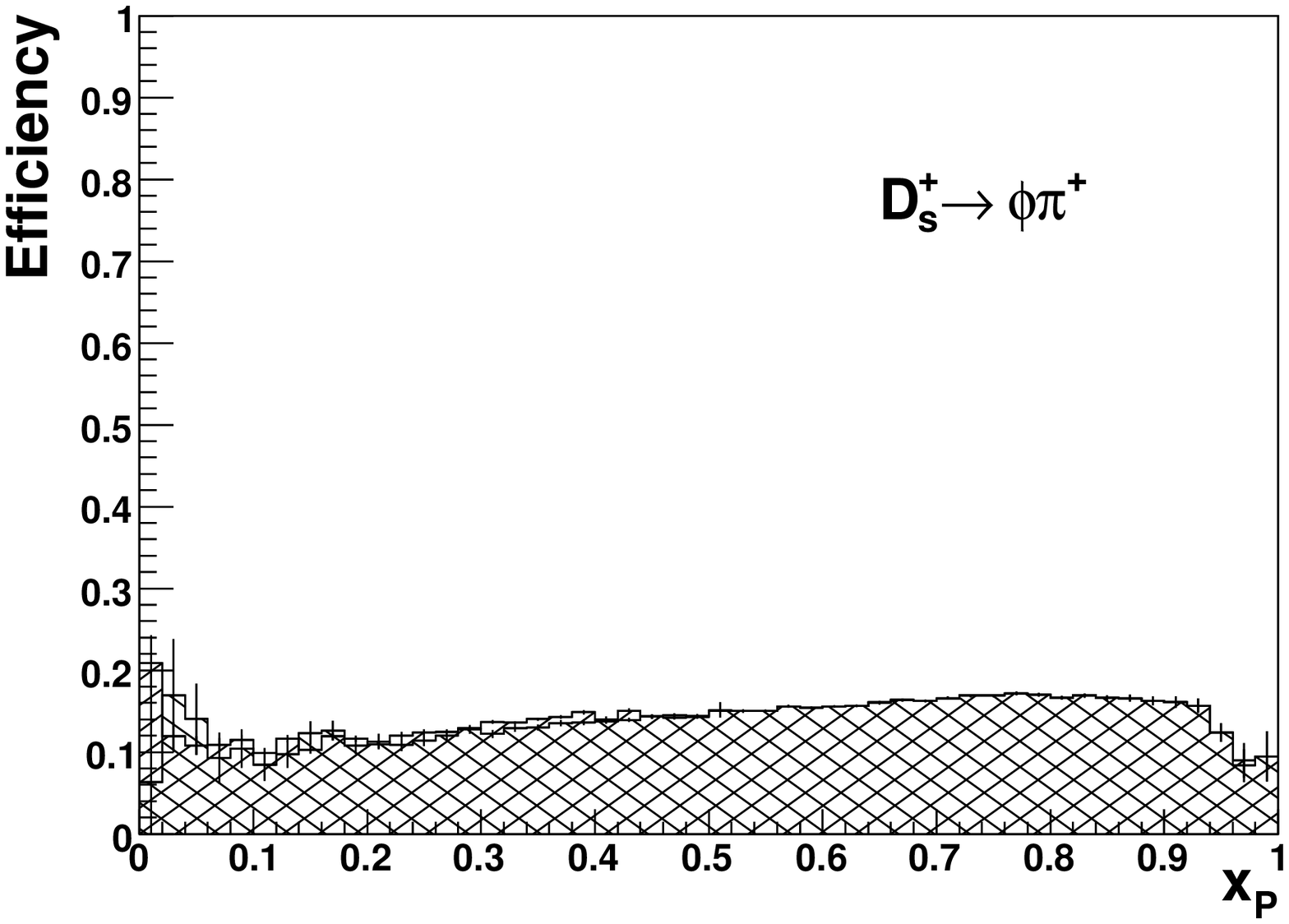}
   \includegraphics[width=0.47\textwidth]{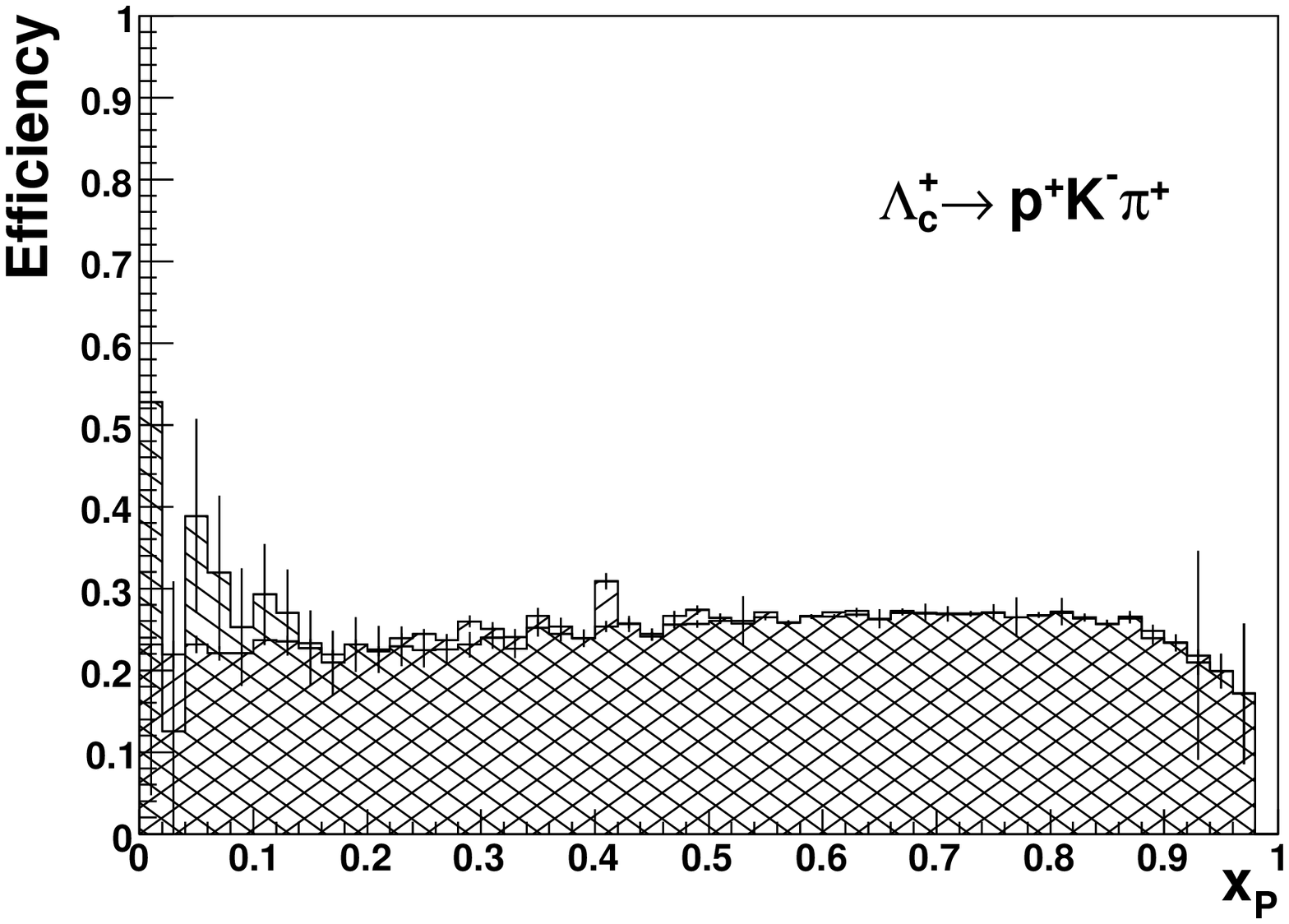}\\
   \includegraphics[width=0.3\textwidth]{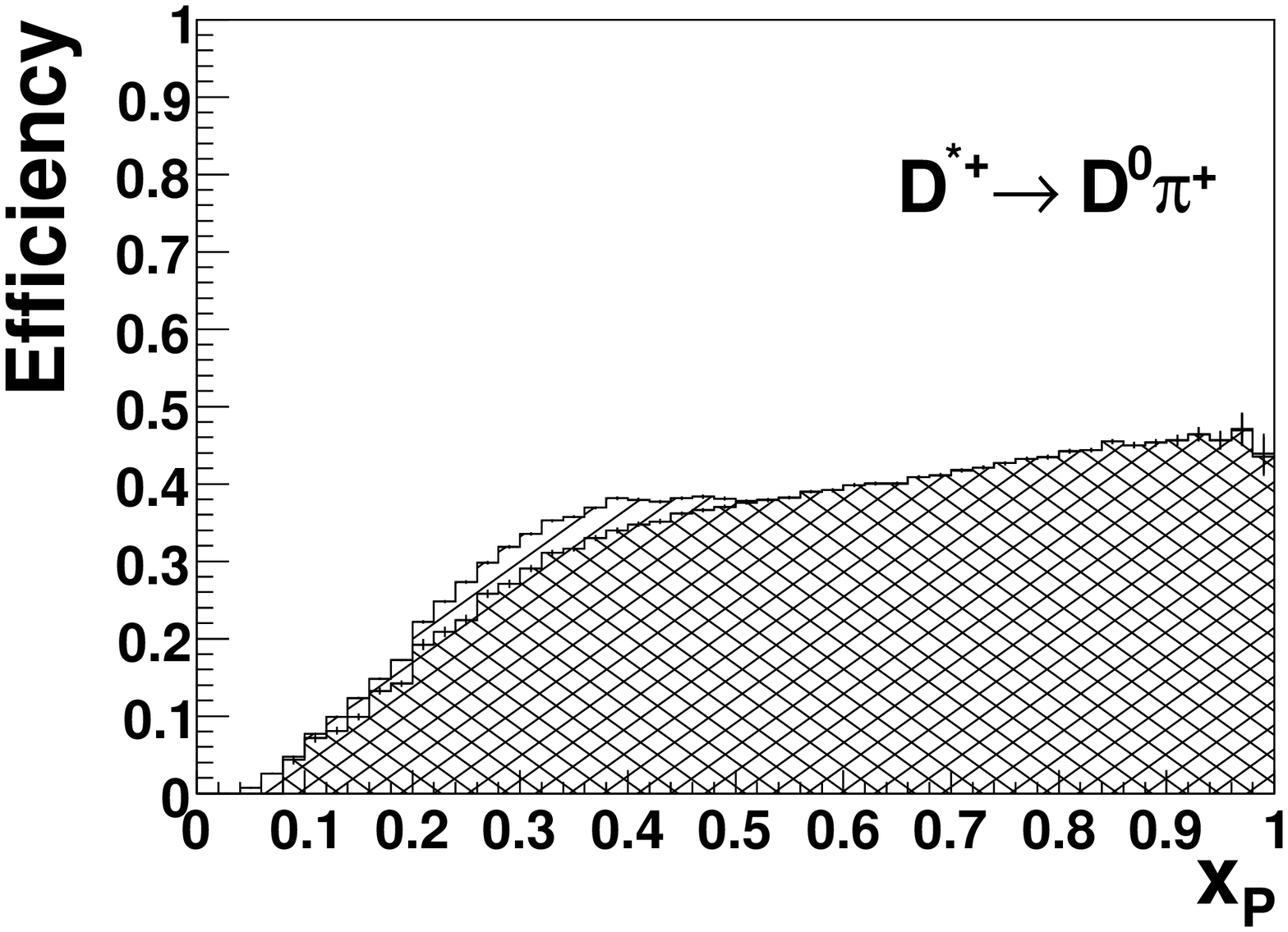}
   \includegraphics[width=0.3\textwidth]{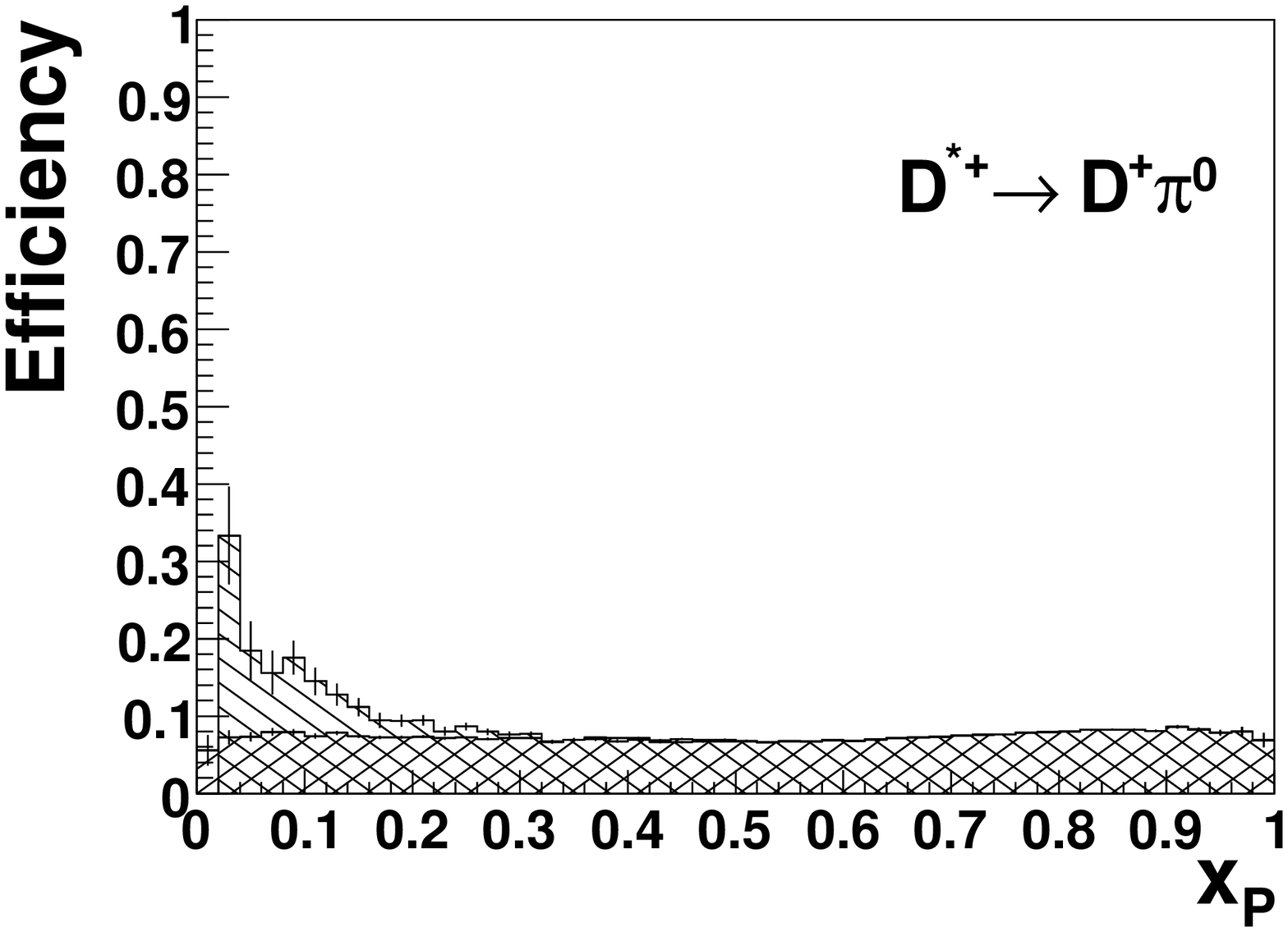}
   \includegraphics[width=0.3\textwidth]{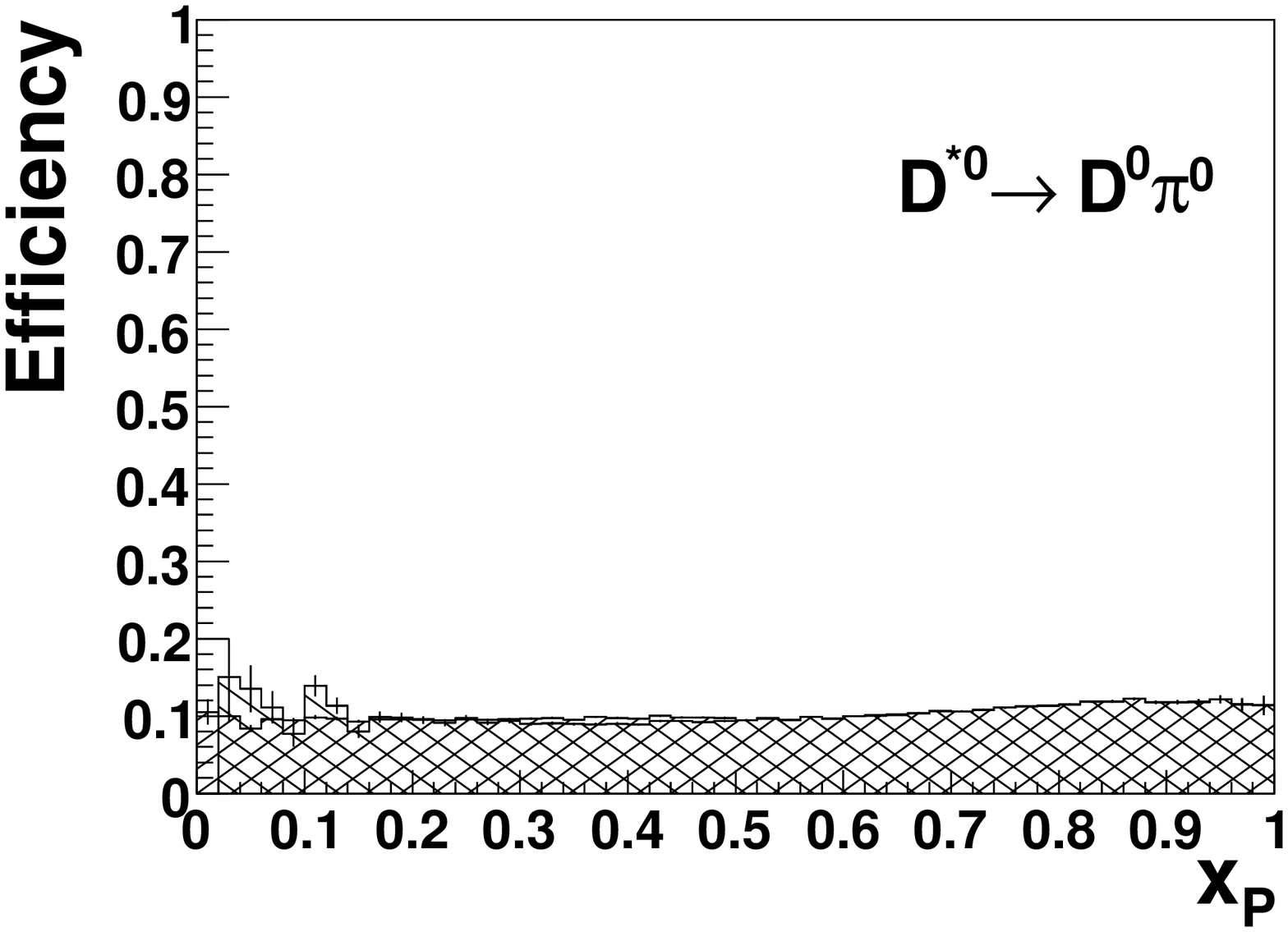}
   \caption{\label{efficiency}
     The efficiencies for the charmed hadrons used in this
     analysis. The order of the particles is the same as in
     Fig.\ \ref{mass-xp-fita}. The different production angle
     distributions for the on-resonance (down-left hatching) and the
     continuum sample (down-right hatching) result in different
     efficiencies for these samples. The error bars show the
     statistical uncertainties only.}
  \end{figure}
 \end{center}

 The efficiencies were determined from \MC\ and are defined as the
 appropriate raw signal yield (determined by the same procedure as for
 data) divided by the generated \MC\ \xP\ distribution.
 The seven histograms in Fig.\ \ref{efficiency} show the \xP-dependent
 efficiency of each charmed hadron used in this analysis for continuum
 data and on-resonance data. The \DZ\ efficiency is close to 50\% and
 almost constant over the entire \xP\ range.
 The efficiency for \DSP\TO\DZ\PP\ approaches the \DZ\ efficiency at
 high values of \xP\ and diminishes at lower values of \xP,
 reflecting the reduced  efficiency of reconstructing low-momentum
 pions. The
 two $D^{*}$ decay modes that include a neutral slow
 pion show a different behaviour: the efficiencies stay constant over
 a wide range of about $0.3<\xP<1$ and below $0.3<\xP$ the efficiency
 increases for \xP\TO0 due to the increasing
 reconstruction efficiency for slow \PZ.
 The reconstruction efficiencies for the three-particle decay modes do
 not show a strong dependence on \xP, slightly varying between 
 15\% and 20\% for the \Ds\ and remaining constant at about 30\% for the \LC.
 The decreasing efficiency for particles at values close to the
 kinematic limit is an artefact of the decreasing statistics in all
 generic \MC\ samples. The reweighted samples, which were generated
 with a different fragmentation function than the generic samples,
 contain significantly more events in the very high \xP\ region and
 do not show such behaviour. This difference between the two
 efficiency estimates was added to the systematic uncertainty.
 
 The efficiency is a function of the production angle, which
 differs for charmed hadrons from $B$ decays and from 
 continuum events. For the
 on-resonance samples, the efficiency has been determined by a
 luminosity-weighted mixture of charmed \MC\ and dedicated samples
 containing decays of charged and neutral $B$ mesons.
 For the continuum sample, only charmed \MC\ was used.
 
 In data, it was verified that \DSP\ produced in \epem\
 annihilation are unpolarised by verifying that the distribution of
 the cosine of the helicity angle is flat. The helicity angle is
 defined as the angle between
 the slow charged pion in the \DSP\ rest frame and the flight
 direction of the \DSP\ in the center of mass system of the event.
 Because the efficiency for \DSP\TO\DZ\PP\ strongly depends on the
 momentum distribution of the slow \PP, which in turn depends on the
 helicity angle, polarised \DSP\ can introduce a bias into the
 efficiency correction.
 
 \section{Systematic Uncertainties}
 \label{systematics}
 Various sources of systematic uncertainties have been
 considered: 

 Uncertainties due to tracking were estimated to be 1\% per
 track using a sample of partially reconstructed \DSP\ decays.
 As the uncertainty increased at very low momentum, the estimated
 momentum-dependent uncertainty of the slow charged pion was
 folded with the observed momentum spectrum. The
 systematic uncertainty due to the slow neutral pion detection
 efficiency was assessed by examining the differences in the shapes of
 the fragmentation function of the two $D^{*+}$\ decay modes,
 $D^{*+}\rightarrow D^0\pi^+$\ and $D^{*+}\rightarrow D^+\pi^0$.
 
 Uncertainties due to the modeling of ISR
 in the MC were determined by restricting the longitudinal momentum
 in the laboratory frame
 of all candidates to $p_z^{lab} > 0$\ only.  This cut preferentially
 removed events with ISR photons in the negative $z$ direction,
 potentially introducing an artificial asymmetry. 
 The $z$ direction is defined as being
 anti-parallel to the positron beam, which coincides up to corrections
 due to the crossing angle with the boost vector into the \epem\
 rest-frame.
 
 The cut on the likelihood ratios for kaon and proton candidates was
 tightened to 0.2 and 0.8, respectively, and the difference was taken
 into account in the systematic uncertainty.
 
 Potential differences between the actual signal shape and the fitting
 function were estimated by determining the signal yield with a
 counting method instead of using the fit.  Here, the number of entries
 in the mass (mass difference) distribution was counted in a window
 about one third the size of the total 50 $\MeV/c^2$ (15 $\MeV/c^2$) window around
 the peak position, corresponding to roughly three times the
 resolution.  The number of background events was subtracted after
 integrating the background function of the standard fit within the same
 mass window.
 
 An additional flavour assignment systematic uncertainty was taken
 into account for the neutral states $D^0$\ and $D^{*0}$.  The loose
 cuts on the charged pion and kaon particle identification allowed a
 $D^0$\ to be identified as a $\overline{D}{}^0$:  the flavour of the
 $D^0$\ from $D^{*+}$\ decays was identified by the charge of the slow
 pion (except for a small contribution from doubly-Cabibbo-suppressed
 decays).  In the MC sample, the likelihood ratio of the pion candidate
 was larger than that of the kaon candidate for 1.3\% of all $D^0$\
 candidates; the corresponding fraction was determined to be 1.1\% for
 $D^{*+}$\ decays.  The statistical uncertainties on these numbers are
 less than 0.05\%.  Accordingly, a difference of 0.2\% was assigned
 as the uncertainty of the flavour assignment due to the overlap of the
 pion and kaon likelihoods of the particle identification.
 
 The luminosity of the data sample was determined to have an uncertainty
 of about 1.4\%. A corresponding scale uncertainty of 1.4\%
 was assigned to the normalisation of the shape.
 It has been checked that the normalisation of the fragmentation
 functions of the on-resonance and continuum sample agree with each
 other, and their difference of 0.94\% is well within the scale
 uncertainty.
 
 Finally, the reconstruction efficiencies of the generic and the
 reweighted samples differed slightly. This small difference was added
 to the systematic uncertainty.
 
 All systematic uncertainties were added in quadrature to give the
 total systematic uncertainty.
 
 \section{Results}
 In this section, various results for the charmed hadrons are presented.
 
 \subsection{\xP\ Distributions}
\begin{center}
  \begin{figure}[h]
    \includegraphics[width=0.45\textwidth]{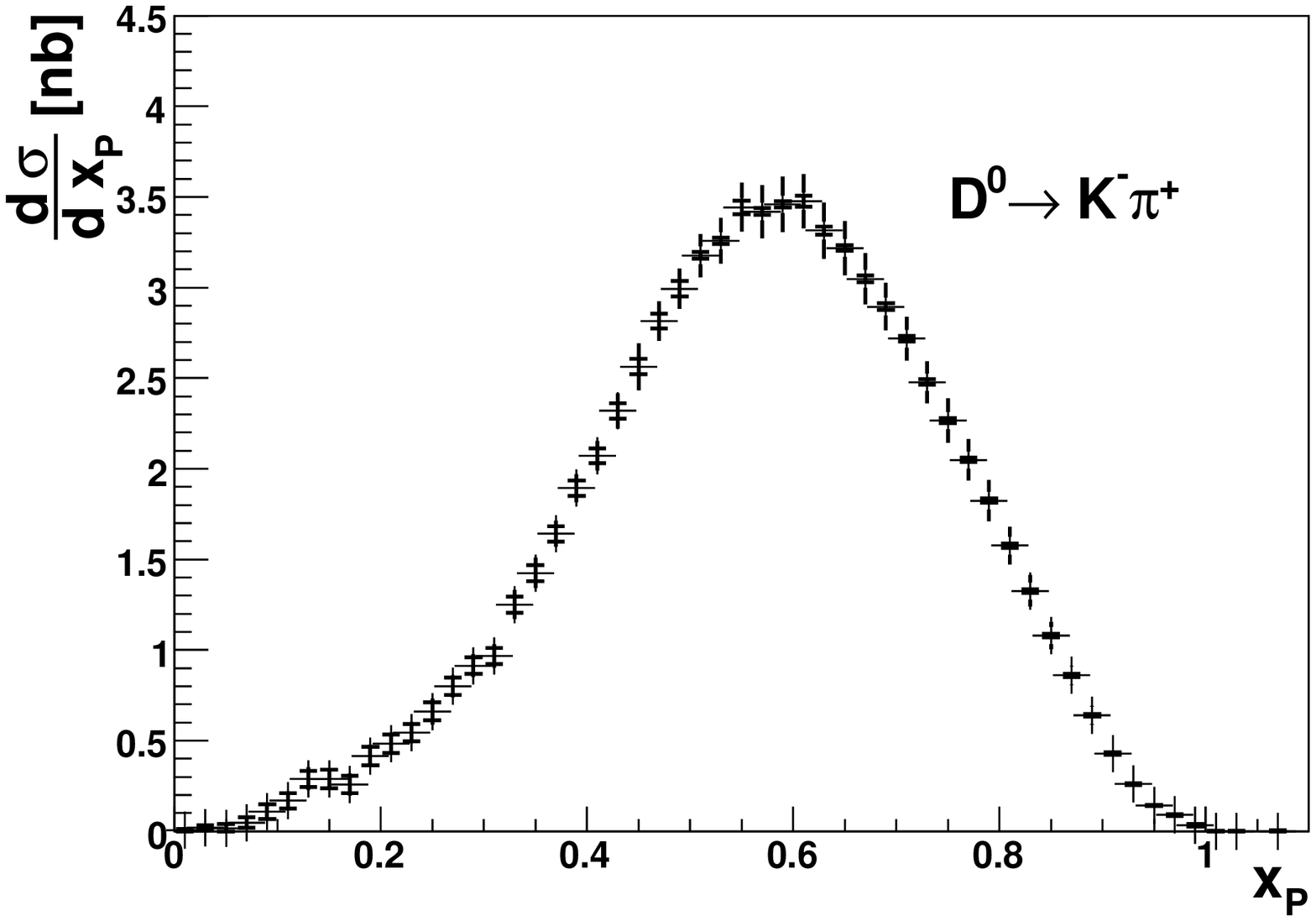}
    \includegraphics[width=0.45\textwidth]{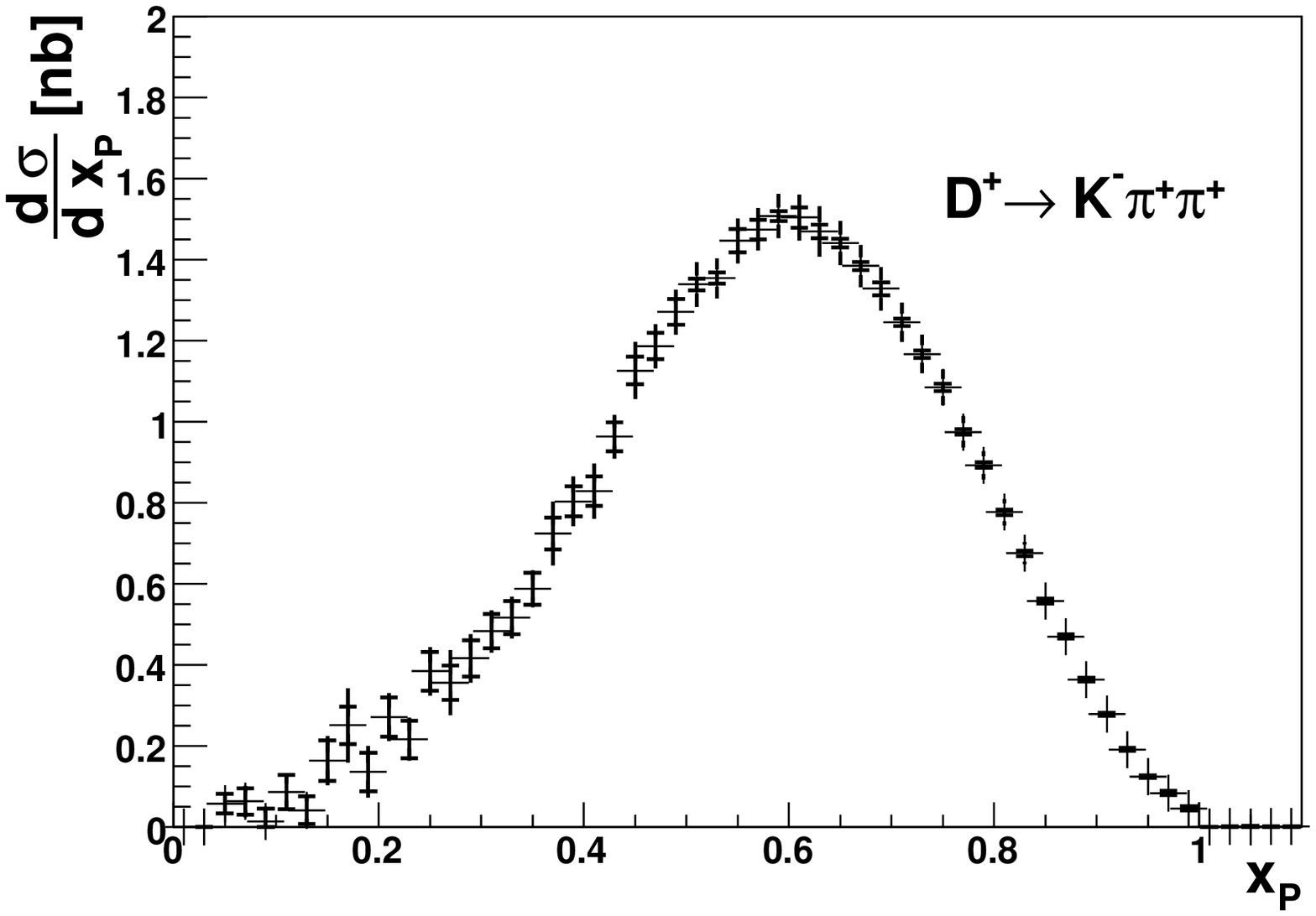} \\
    \includegraphics[width=0.45\textwidth]{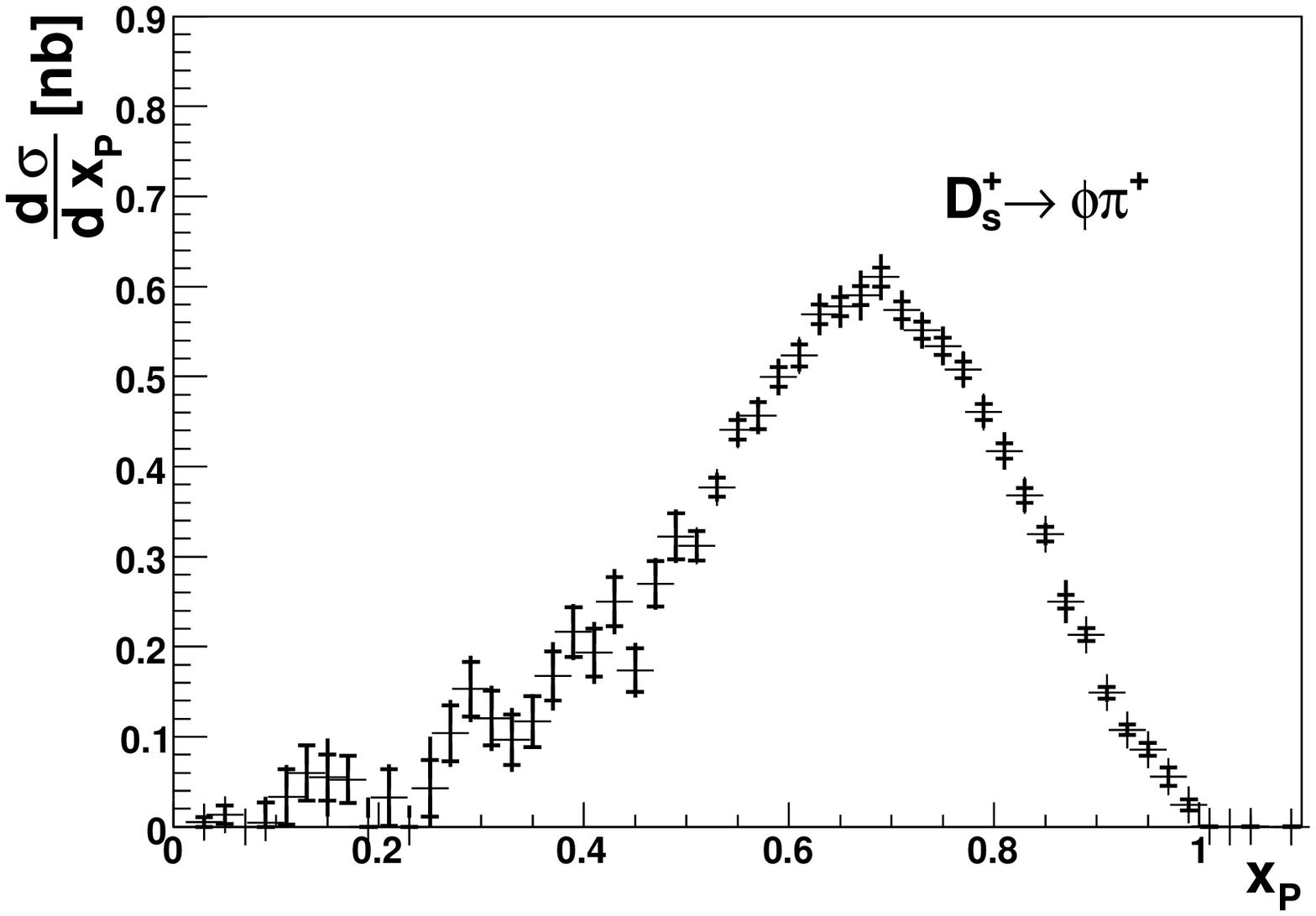}
    \includegraphics[width=0.45\textwidth]{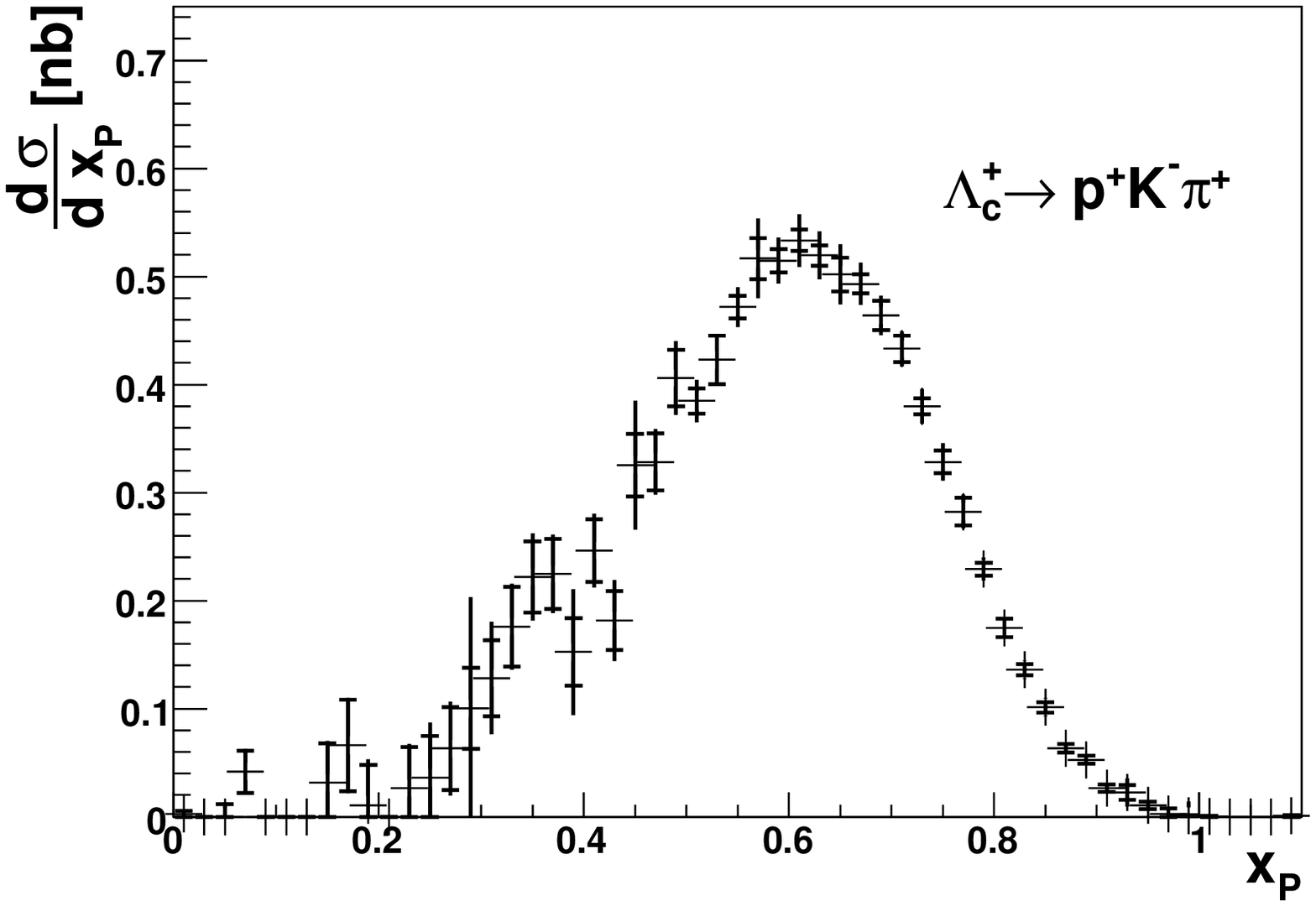} \\
    \includegraphics[width=0.3\textwidth]{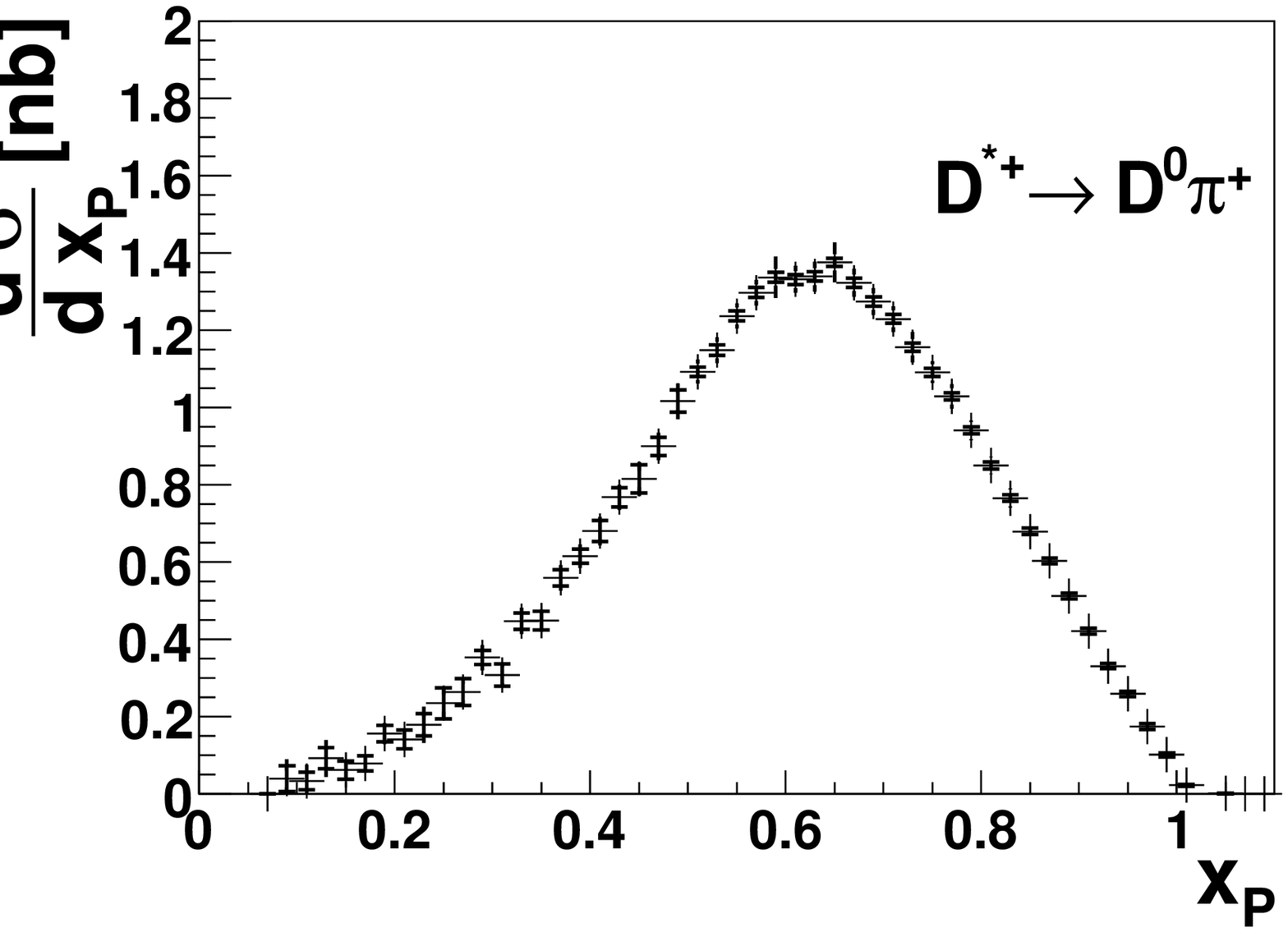}
    \includegraphics[width=0.3\textwidth]{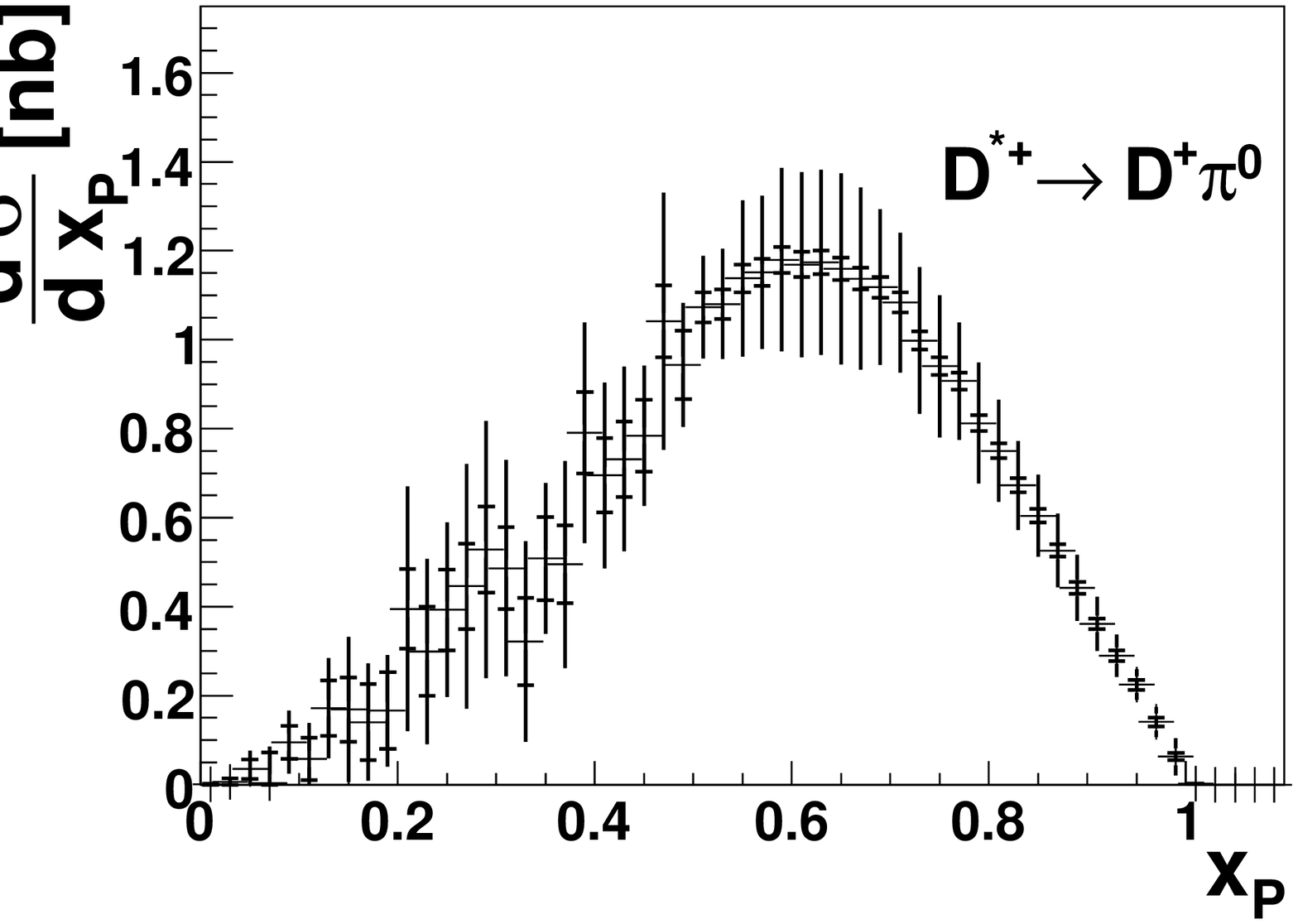}
    \includegraphics[width=0.3\textwidth]{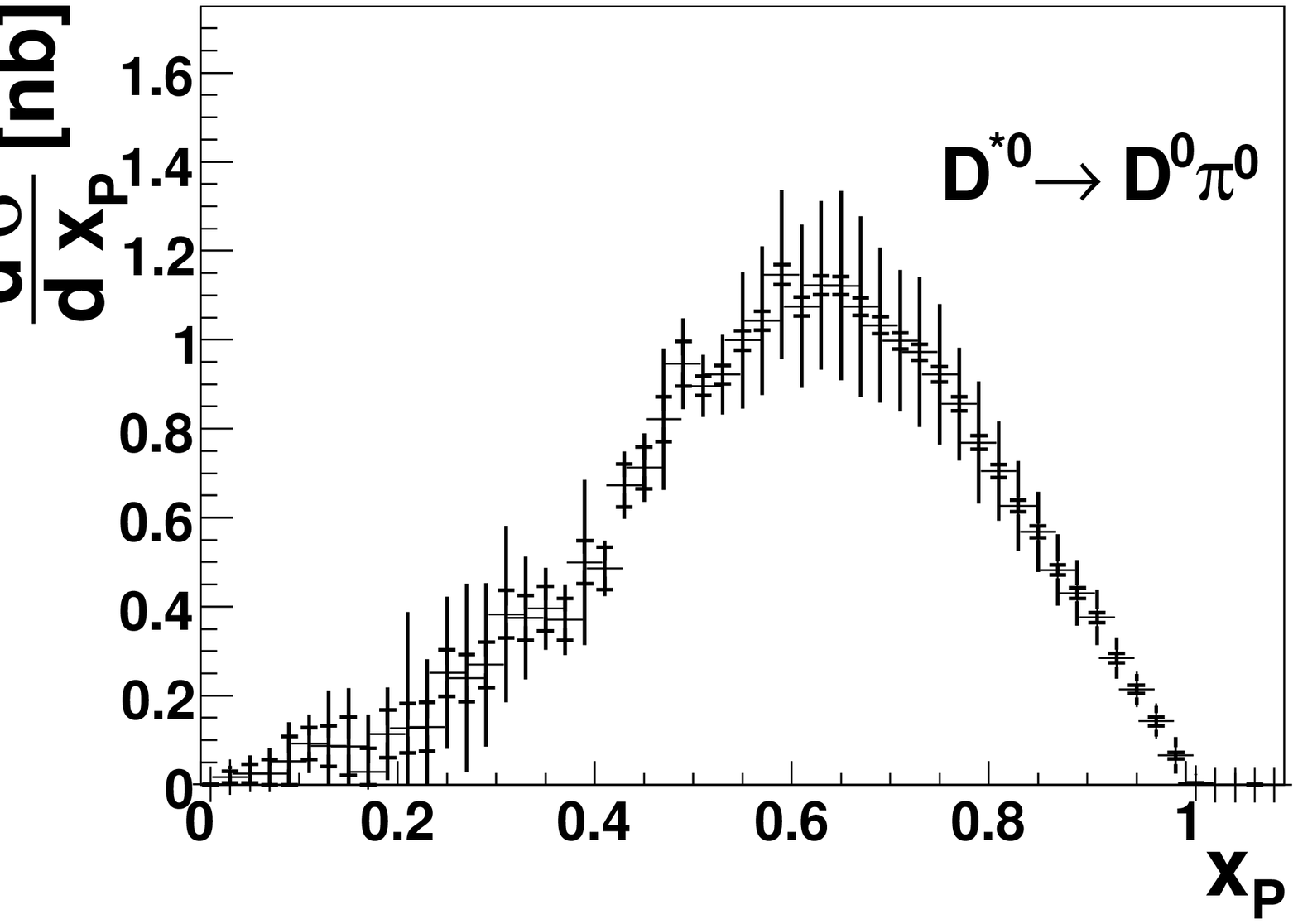}
    \caption{\label{xp-distribc}Efficiency corrected momentum distributions
      for the charmed hadrons produced in \epem-annihilation events,
      i.e. from fragmentation of charm quarks.
      The order of the particles is the same as in
      Fig.\ \ref{mass-xp-fita}. For $\xP>0.5$, the on-resonance and
      continuum data have been combined by a weighted average.
      The inner error bars show the statistical, the outer error bars
      the total uncertainties.
    }
  \end{figure}
 \end{center}

 Fig.\ \ref{xp-distribc} shows the efficiency-corrected \xP\
 distributions for the different particles for \epem\ annihilation
 events, {\it i.e.} spectra of hadrons formed in the fragmentation of charm
 quarks.
 Above $\xP>0.5$, the differential \xP\ distributions of the
 on-resonance sample and the continuum sample have been combined by a
 weighted average, where the inverse of the squared statistical
 uncertainty was used as the weight. As the systematic uncertainties
 for both samples are highly correlated, the larger uncertainty of
 the on-resonance and the continuum samples was used for the combined
 sample.
 
 As most efficiencies do not depend strongly upon \xP, the
 shapes of the efficiency-corrected distributions are similar to those
 of the uncorrected distributions. All distributions peak around
 $\xP\sim0.6-0.7$.
 To determine the peak position, a direct fit of the data to the 
 Peterson fragmentation function was tried. 
 The shape of the data agreed very poorly as this model does not include
 gluon radiation or decays from higher resonances.
 Therefore, a Gaussian function was used to determine the peak position.
 The fit ranges were chosen from $\xP=0.4$--$0.8$. The
 results of the fits are listed in Table \ref{peak}, together with the
 statistical and systematic uncertainties. The statistical uncertainty
 was determined by the RMS of the distribution divided by $\sqrt{N}$.
 
 \begin{table}[h]
  \caption{\label{peak}The peak positions of all hadrons, fitted with
    a Gaussian near the peak position. The fit range was
    $0.4<\xP<0.8$. Above $\xP>0.5$, the continuum sample and the
    on-resonance sample have been combined.
    The given errors are the statistical and systematic uncertainties,
    respectively. }
  \begin{center}
  \begin{ruledtabular}
   \begin{tabular}{Bcccc}
     & $\xP {}^{\mathrm{PEAK}}$ \\ \hline
    \DZ\to B \KM\PP\    & $0.587 \pm 0.001 \pm 0.002$ \\
    \DP\to B \KM\PP\PP\ & $0.600 \pm 0.001 \pm 0.001$ \\
    \Ds\to B \PHI\PP\   & $0.681 \pm 0.002 \pm 0.003$ \\
    \LC\to B \pr\KM\PP\ & $0.612 \pm 0.001 \pm 0.004$ \\
    \DSP\to B \DZ\PP\   & $0.631 \pm 0.001 \pm 0.002$ \\
        \to B \DP\PZ\   & $0.618 \pm 0.011 \pm 0.023$ \\
    \DSZ\to B \DZ\PZ\   & $0.631 \pm 0.001 \pm 0.003$ \\
   \end{tabular}
  \end{ruledtabular}
  \end{center}
 \end{table}

 \subsection{Average Number of Charmed Hadrons per $B$ Decay}
\begin{center}
  \begin{figure}[h]
    \includegraphics[width=0.45\textwidth]{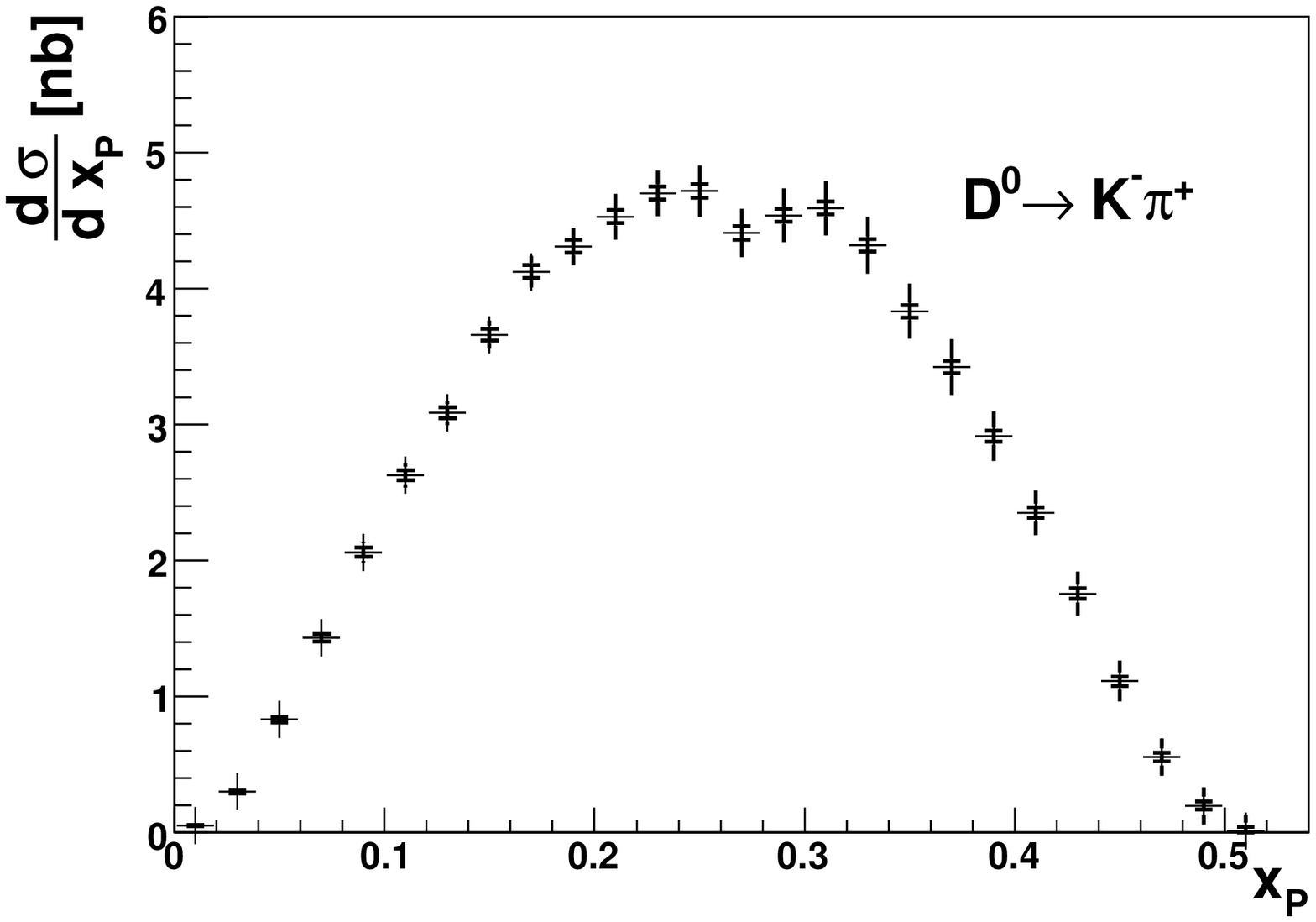}
    \includegraphics[width=0.45\textwidth]{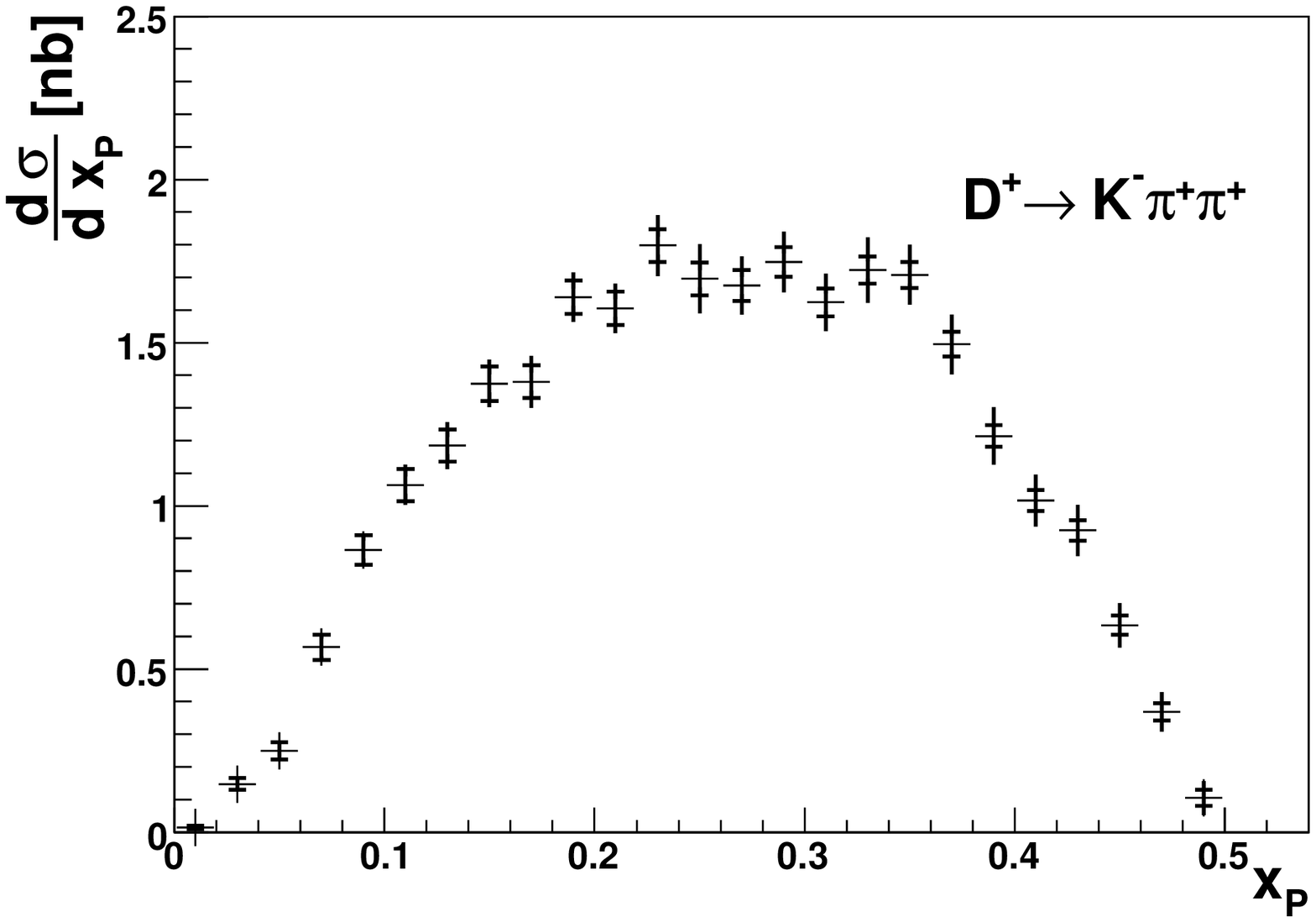} \\
    \includegraphics[width=0.45\textwidth]{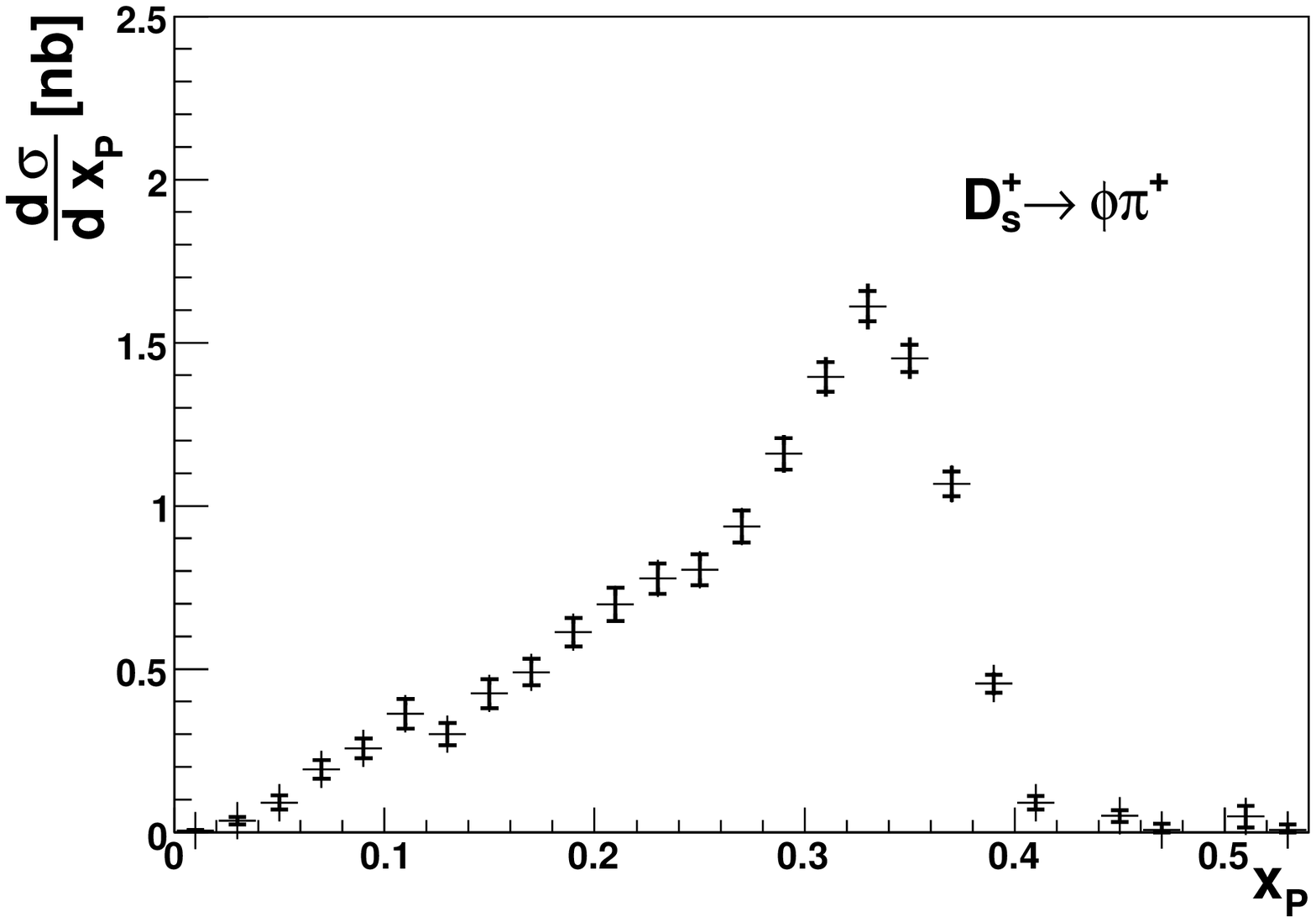}
    \includegraphics[width=0.45\textwidth]{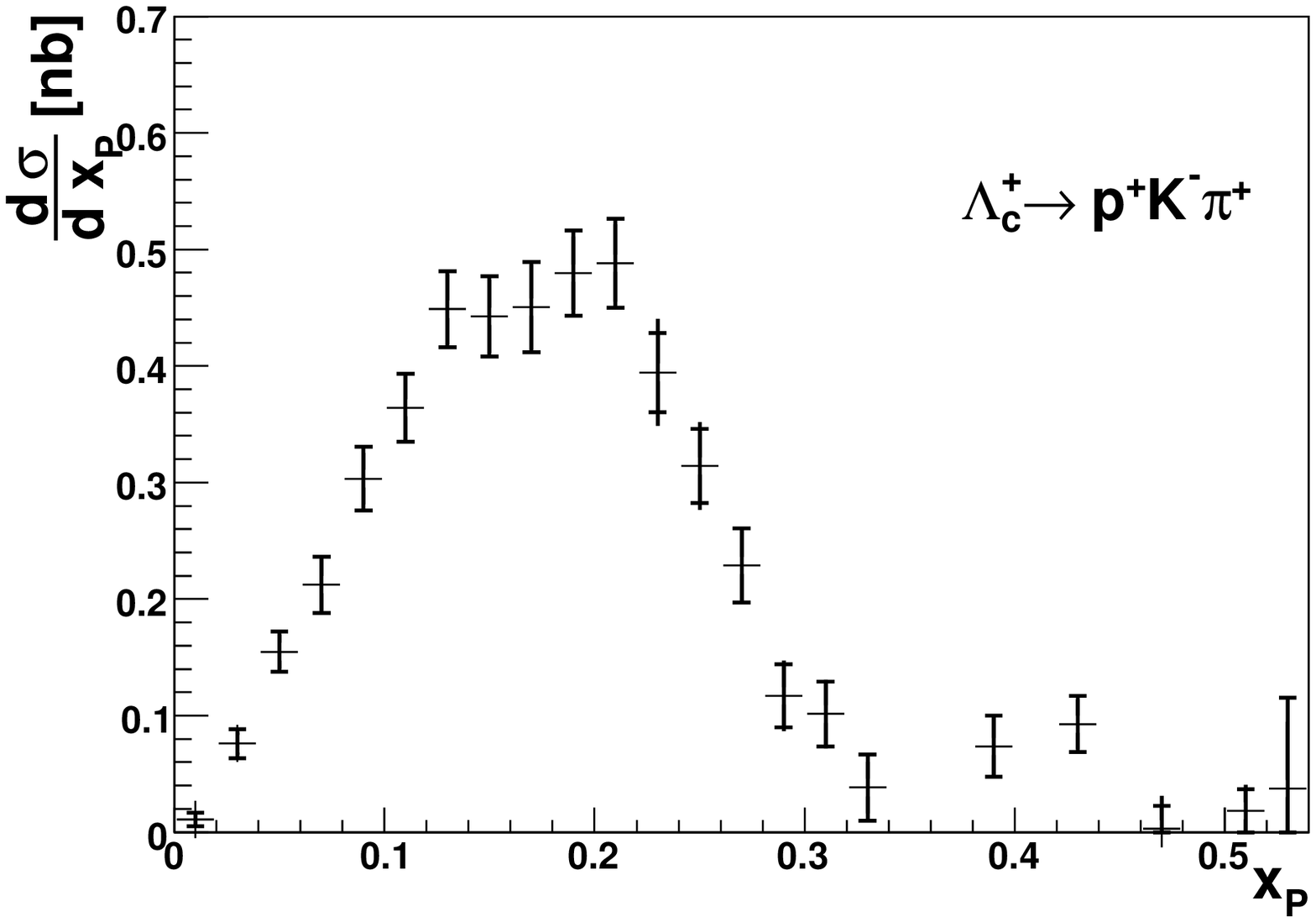} \\
    \includegraphics[width=0.3\textwidth]{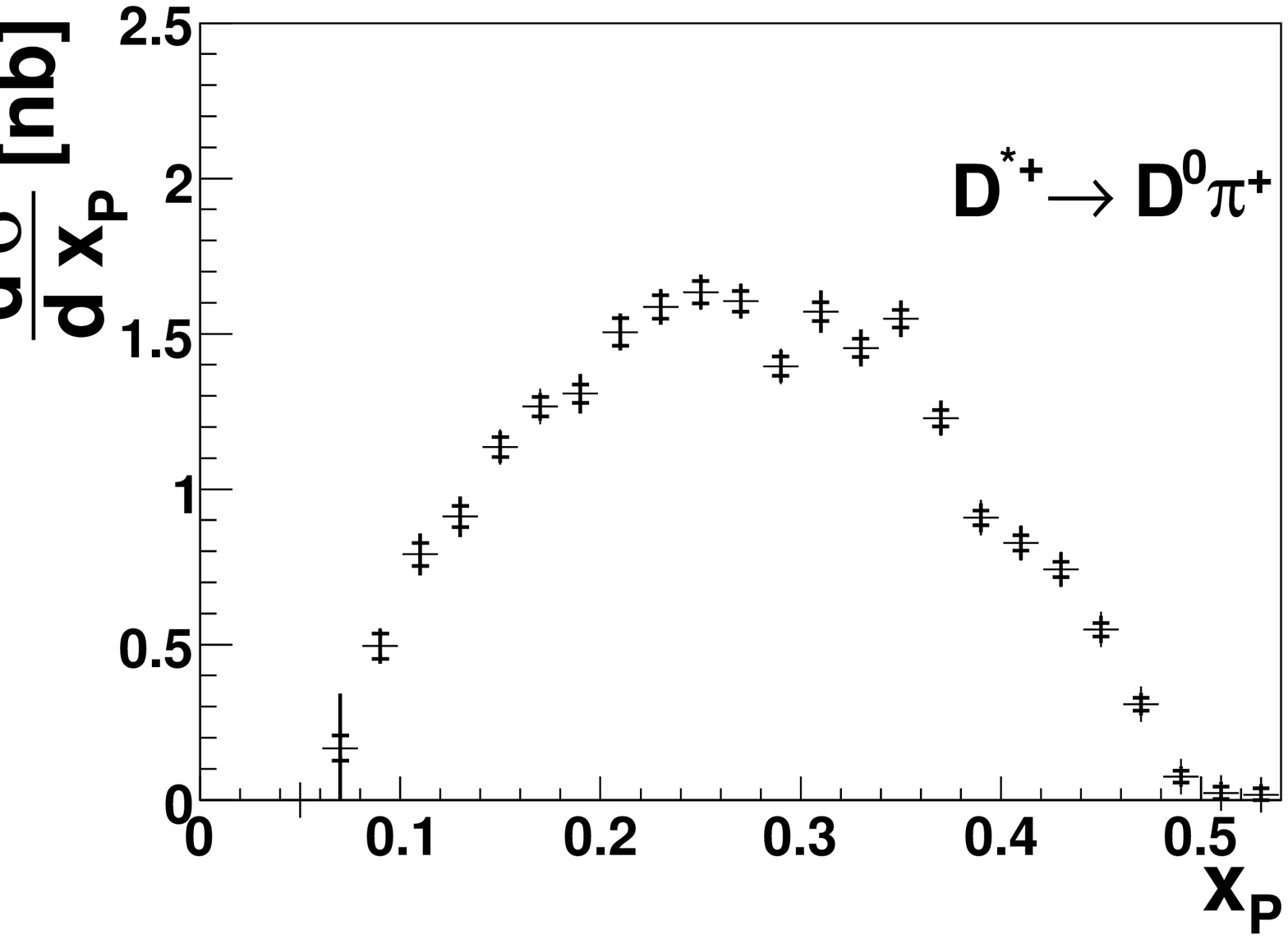}
    \includegraphics[width=0.3\textwidth]{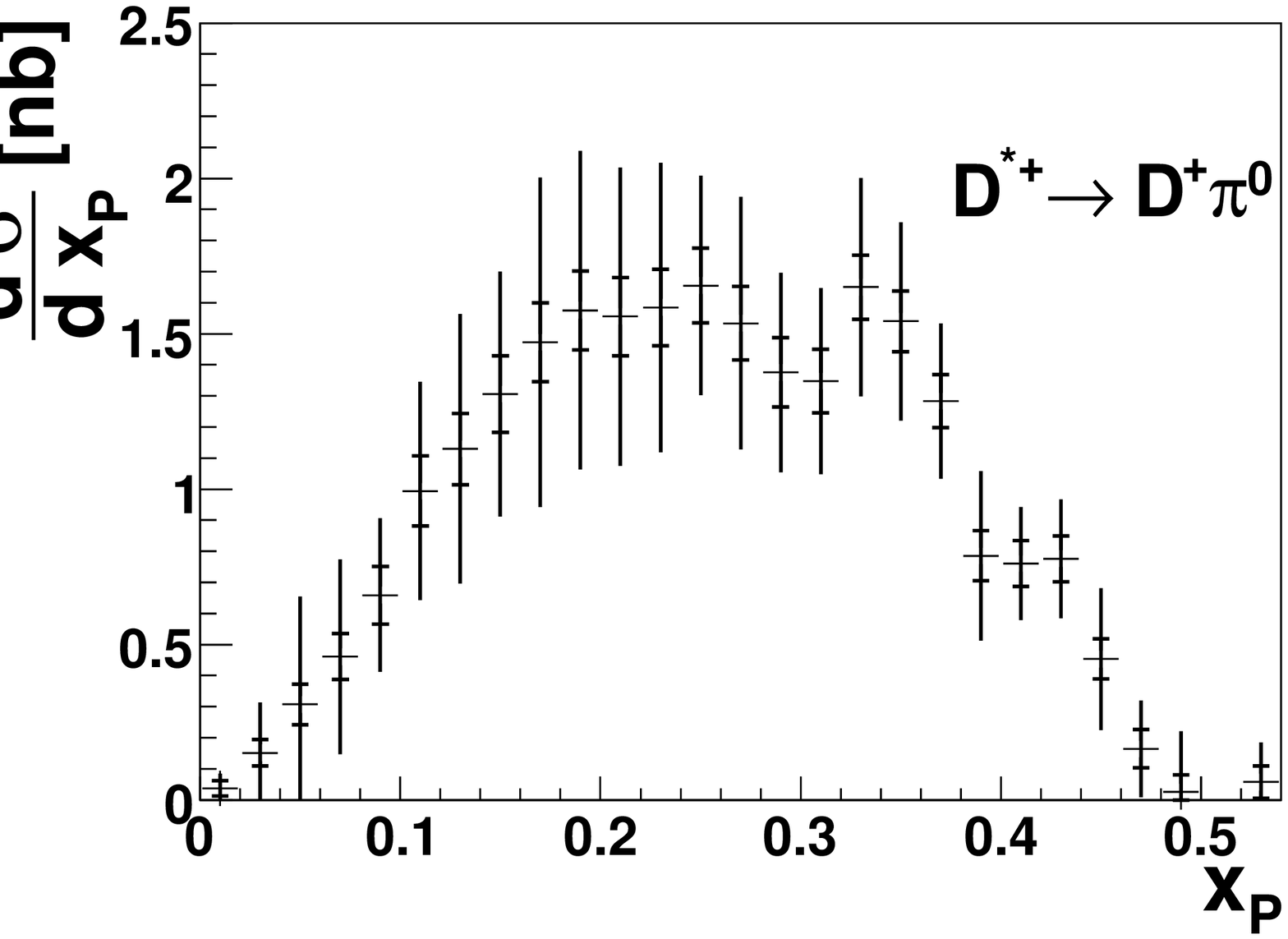}
    \includegraphics[width=0.3\textwidth]{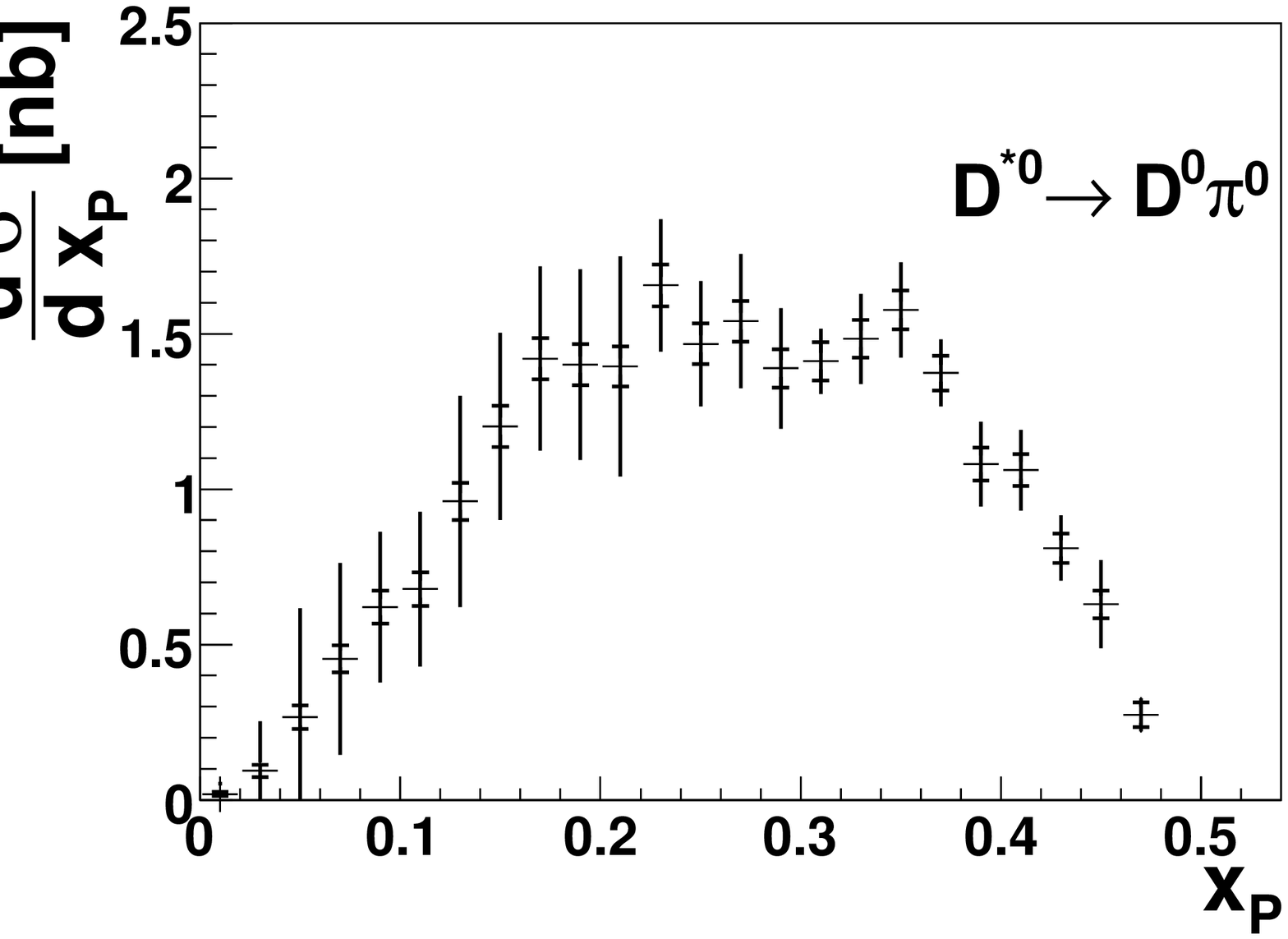}
    \caption{\label{xp-distribb}Efficiency corrected and continuum
      subtracted momentum distributions for the charmed hadrons from $B$
      decays used in this analysis.
      The \xP\ range is restricted to $\xP<0.55$.
      The order of the particles is the same as in Fig.\ \ref{mass-xp-fita}.
      The inner error bars show the statistical, the outer error bars
      the total uncertainties.}
  \end{figure}
 \end{center}

 The \xP\ distributions of the on-resonance and continuum samples
 differ in the contribution from $B$ decays for $\xP<0.5$.
 Fig.\ \ref{xp-distribb} shows this difference: the differential \xP\
 distribution of the continuum sample was subtracted from that of the
 on-resonance sample. Thus, up to statistical fluctuations it contains
 only contributions from decays of $B$ mesons.
 
 Table \ref{average-decay} lists the average number of charmed
 hadrons per $B$ meson decay together with the present
 world average \cite{PDG2004}. 
 In order to determine the average number of charmed hadrons produced
 per $B$ decay, we take the difference between the production rate in the
 on-resonance and the continuum sample and normalise by the
 $B$\ meson production cross section, which is estimated to be 
 \mbox{$(1.073\pm0.019)~\nb$} based on the measured luminosity and the
 measured number of $B\overline{B}$ pairs in this sample.
 Note that this visible production cross-section depends strongly upon
 the energy spread of the accelerator.
 The uncertainties in Table \ref{average-decay} are from the limited
 statistics (first), the systematics as discussed
 in Section \ref{systematics}
 (second), and the luminosity measurement and the
 uncertainties on the branching fractions (third). Note that the
 luminosity measurement and the determination of the number of
 $B\bar{B}$ are strongly correlated.
 Both values agree well within one standard deviation with each other,
 only the average number of produced \DSZ's here is lower
 by about one standard deviation and is closer to that of \DSP's.

 The small bump seen in the \xP\ distributions of the charmed mesons
 except the \Ds\ at $\xP=0.35$ is due to two body decays of the $B$
 mesons such as $B\TO\DZ D^{(*)}$ in case of the \DZ.
 \begin{table}[h]
  \caption{\label{average-decay}The average number
    $N_{B \rightarrow c}$
    of charmed
    hadrons per $B$ meson decay, corrected for acceptance and
    reconstruction efficiencies. The listed uncertainties are
    statistical, systematic, and the one due to the
    uncertainties on the branching fractions of the decays involved as well
    as on the luminosity, respectively.}
  \begin{center}
  \begin{ruledtabular}
   \begin{tabular}{Bcc}
           & $N_{B \rightarrow c}$ & PDG(2004) \\ \hline
    \DZ\to B\KM\PP    & $0.644 \pm 0.003 \pm 0.024 \pm 0.021$ & $0.640 \pm 0.030$ \\
    \DP\to B\KM\PP\PP & $0.248 \pm 0.004 \pm 0.033 \pm 0.020$ & $0.235 \pm 0.019$ \\
    \Ds\to B\PHI\PP   & $0.122 \pm 0.015 \pm 0.033 \pm 0.030$ & $0.105 \pm 0.026$ \\
    \LC\to B\pr\KM\PP & $0.042 \pm 0.011 \pm 0.033 \pm 0.018$ & $0.064 \pm 0.011$ \\
    \DSZ\to B\DZ\PZ   & $0.217 \pm 0.014 \pm 0.020 \pm 0.018$ & $0.260 \pm 0.027$ \\
    \DSP\to B\DZ\PP   & $0.218 \pm 0.007 \pm 0.020 \pm 0.015$ & $0.225 \pm 0.025$ \\
        \to B\DP\PZ   & $0.202 \pm 0.014 \pm 0.022 \pm 0.018$ & $0.225 \pm 0.025$ \\
    average  B \DSP   & $0.215 \pm 0.006 \pm 0.022 \pm 0.016$ & $0.225 \pm 0.025$ \\
   \end{tabular}
  \end{ruledtabular}
  \end{center}
 \end{table}
 
 \subsection{High \xP\ Region}
 \label{highxp}
\begin{center}
  \begin{figure}[h]
   \includegraphics[width=0.45\textwidth]{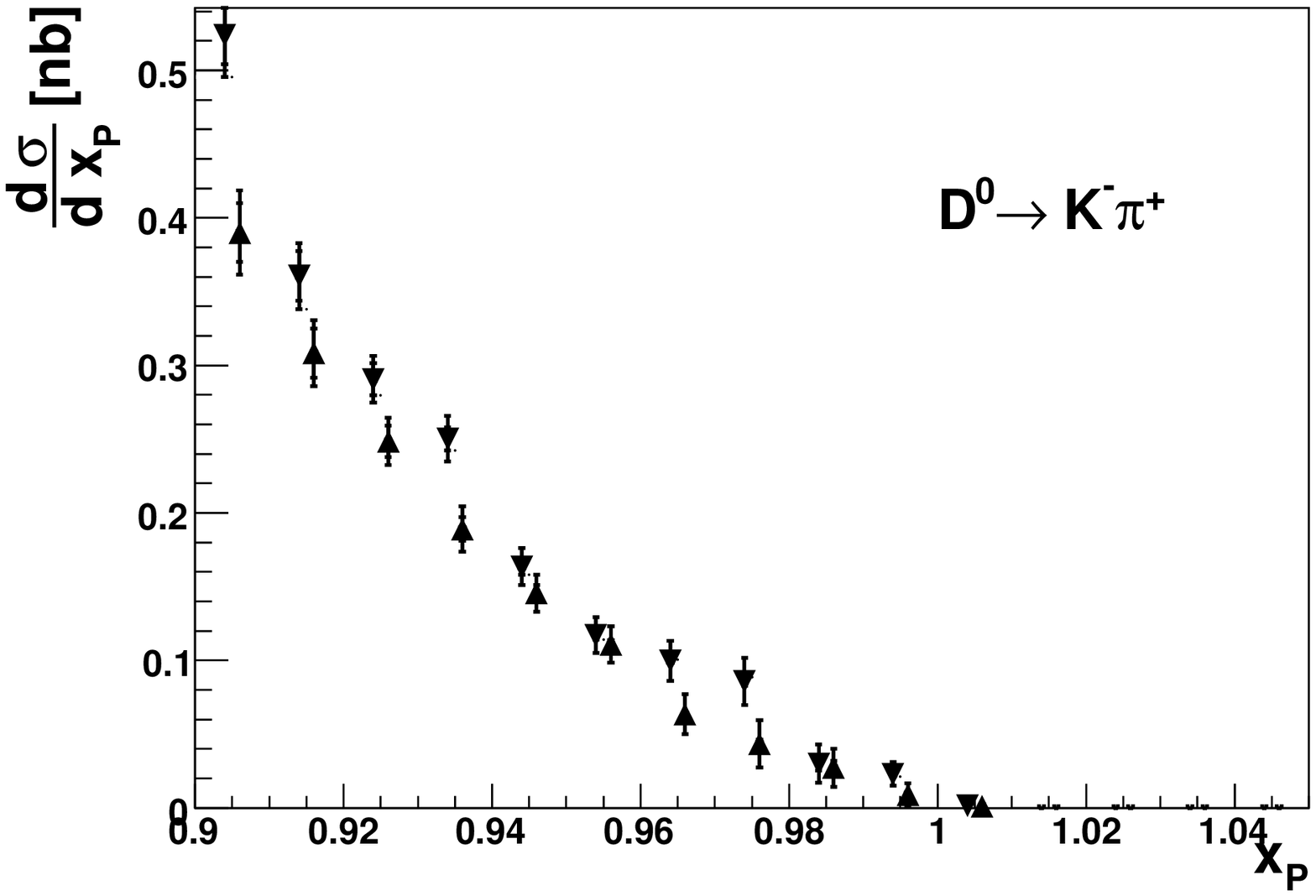}
   \includegraphics[width=0.45\textwidth]{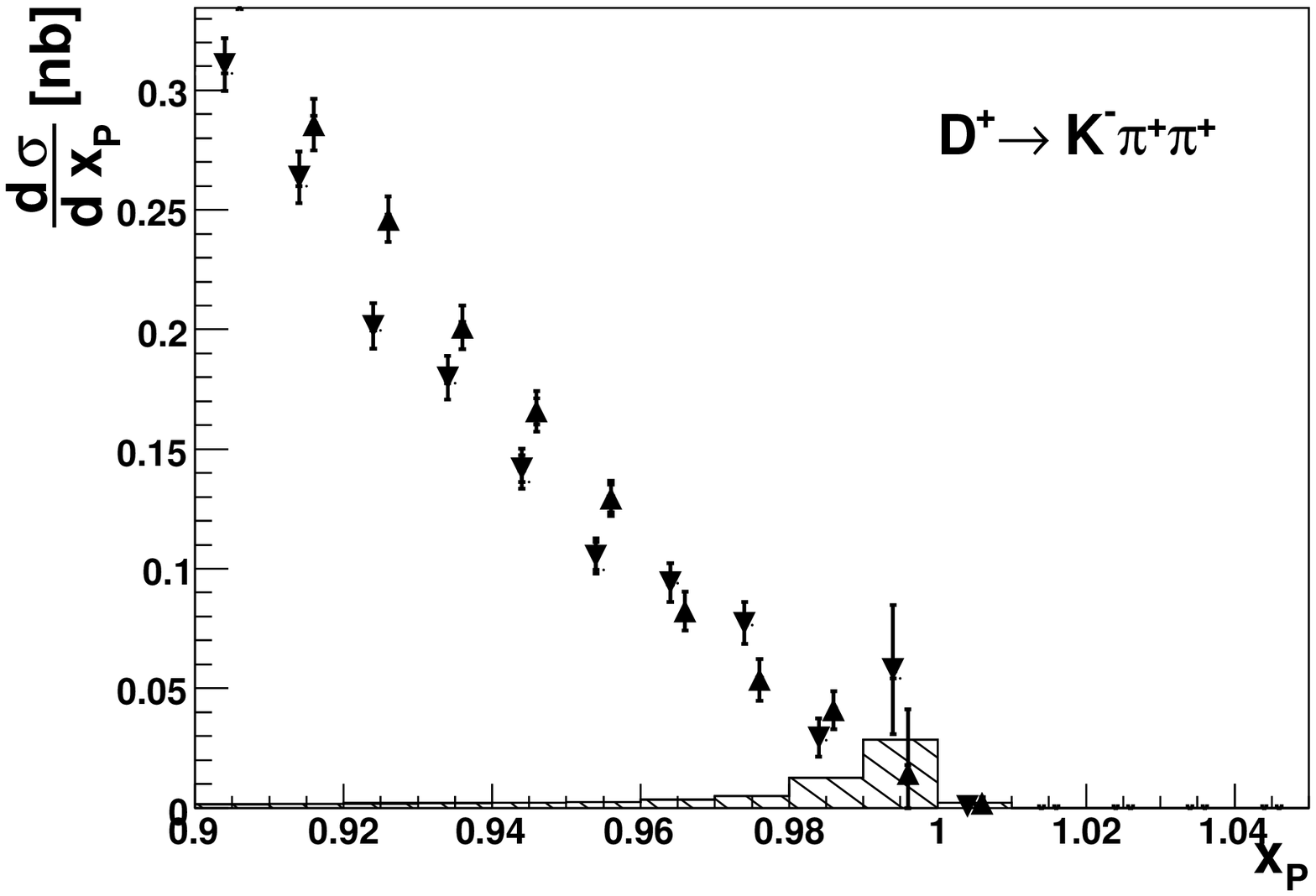}\\
   
   \includegraphics[width=0.45\textwidth]{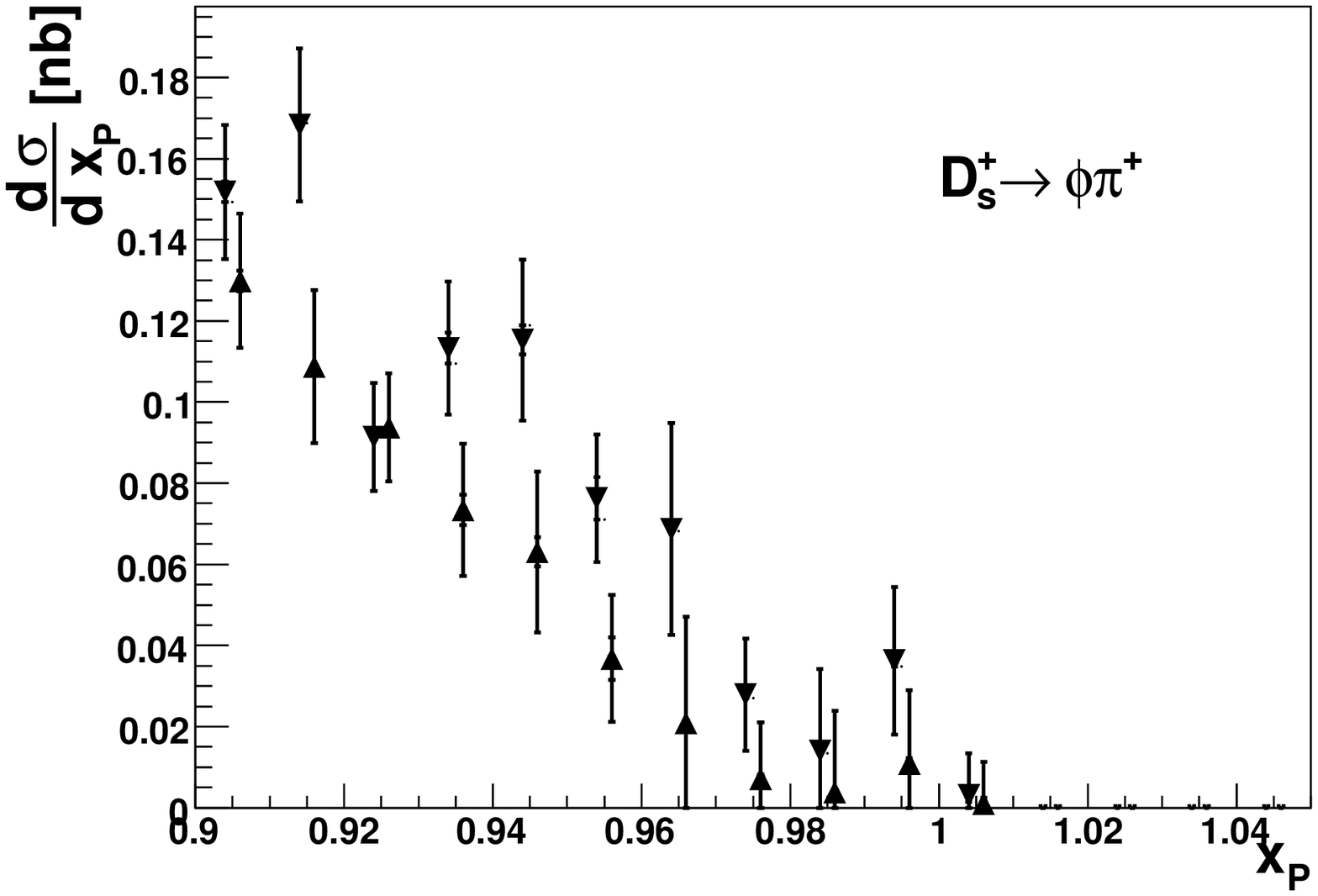}
   \includegraphics[width=0.45\textwidth]{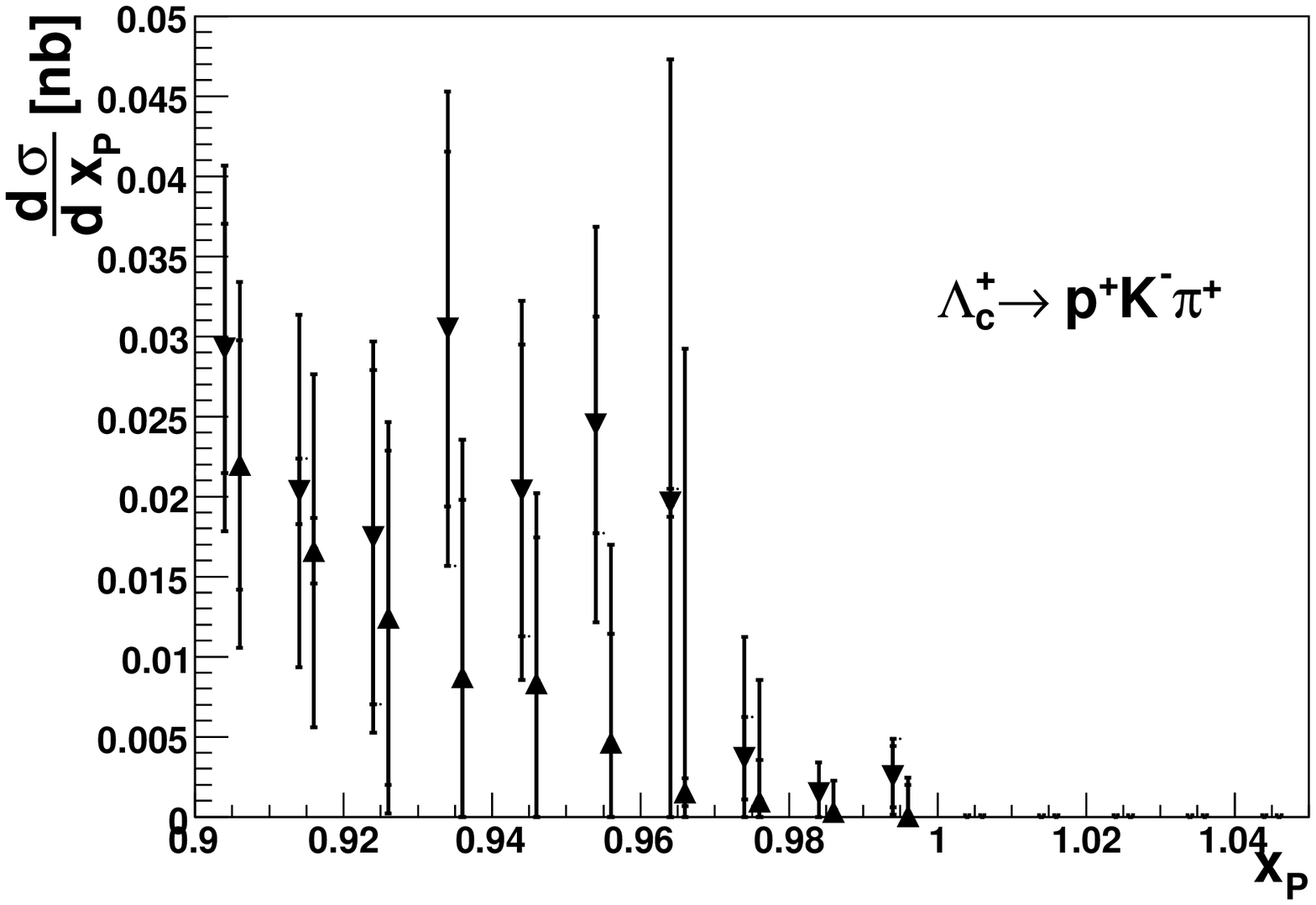}\\
   
   \includegraphics[width=0.3\textwidth]{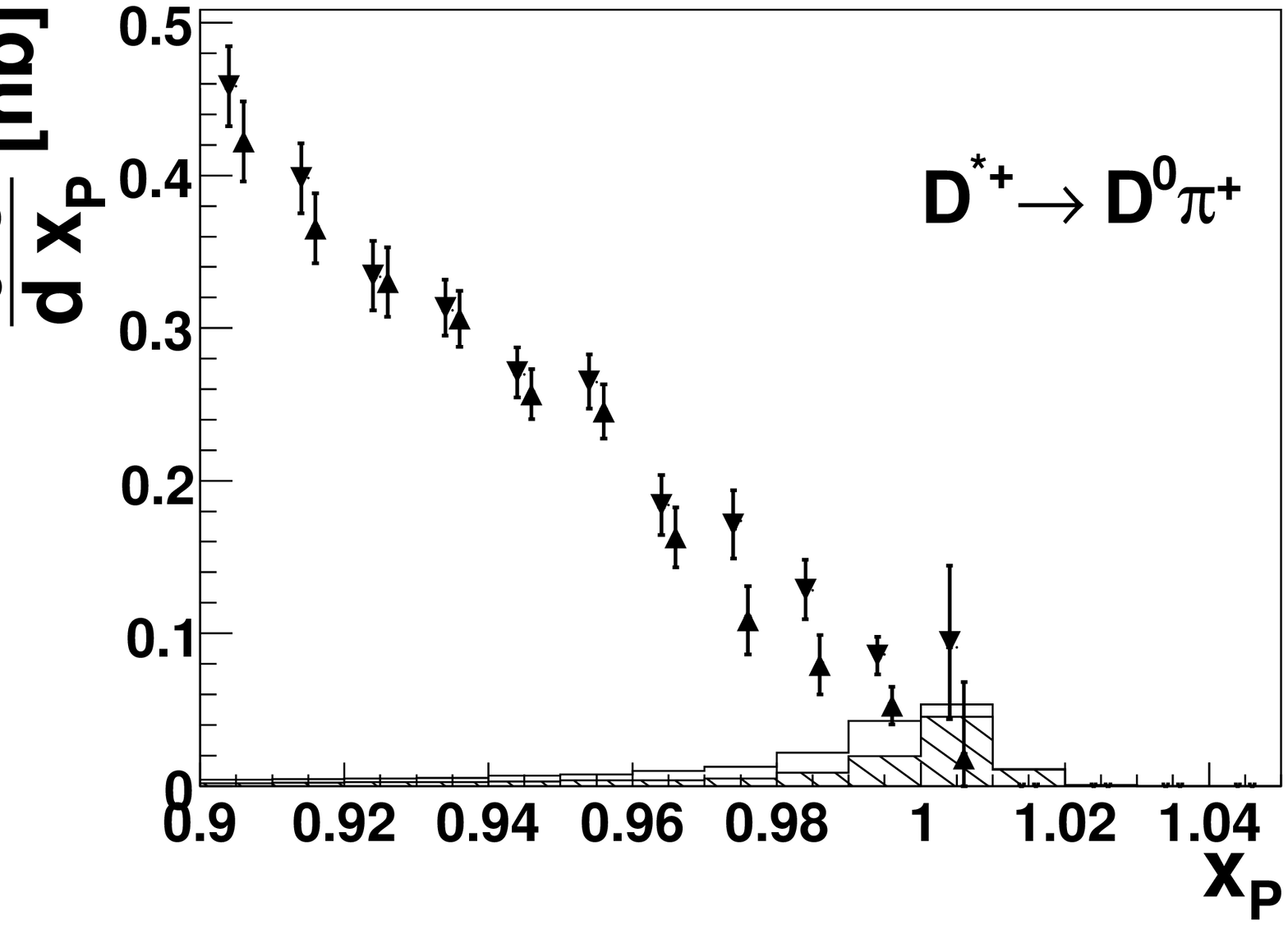}
   \includegraphics[width=0.3\textwidth]{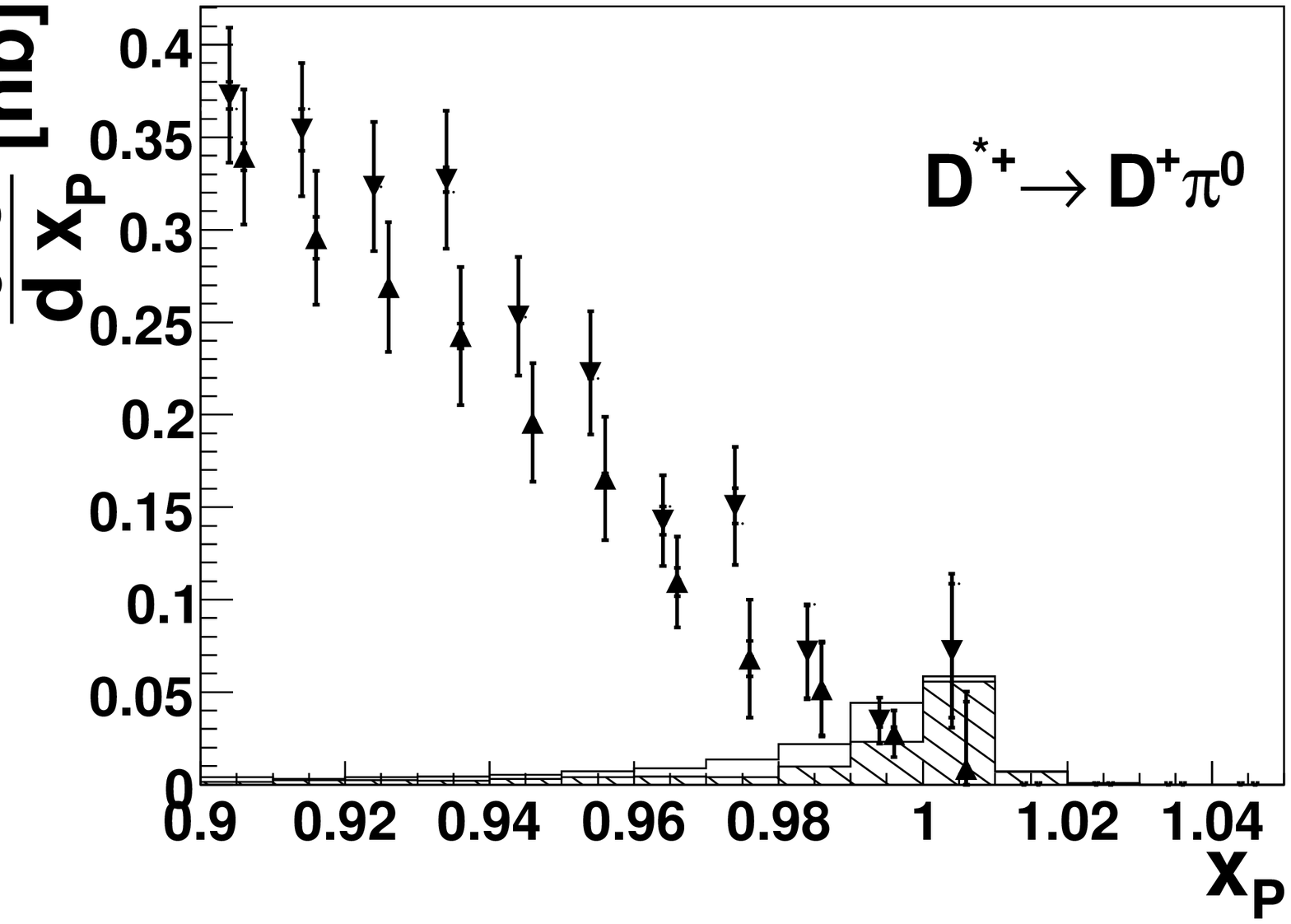}
   \includegraphics[width=0.3\textwidth]{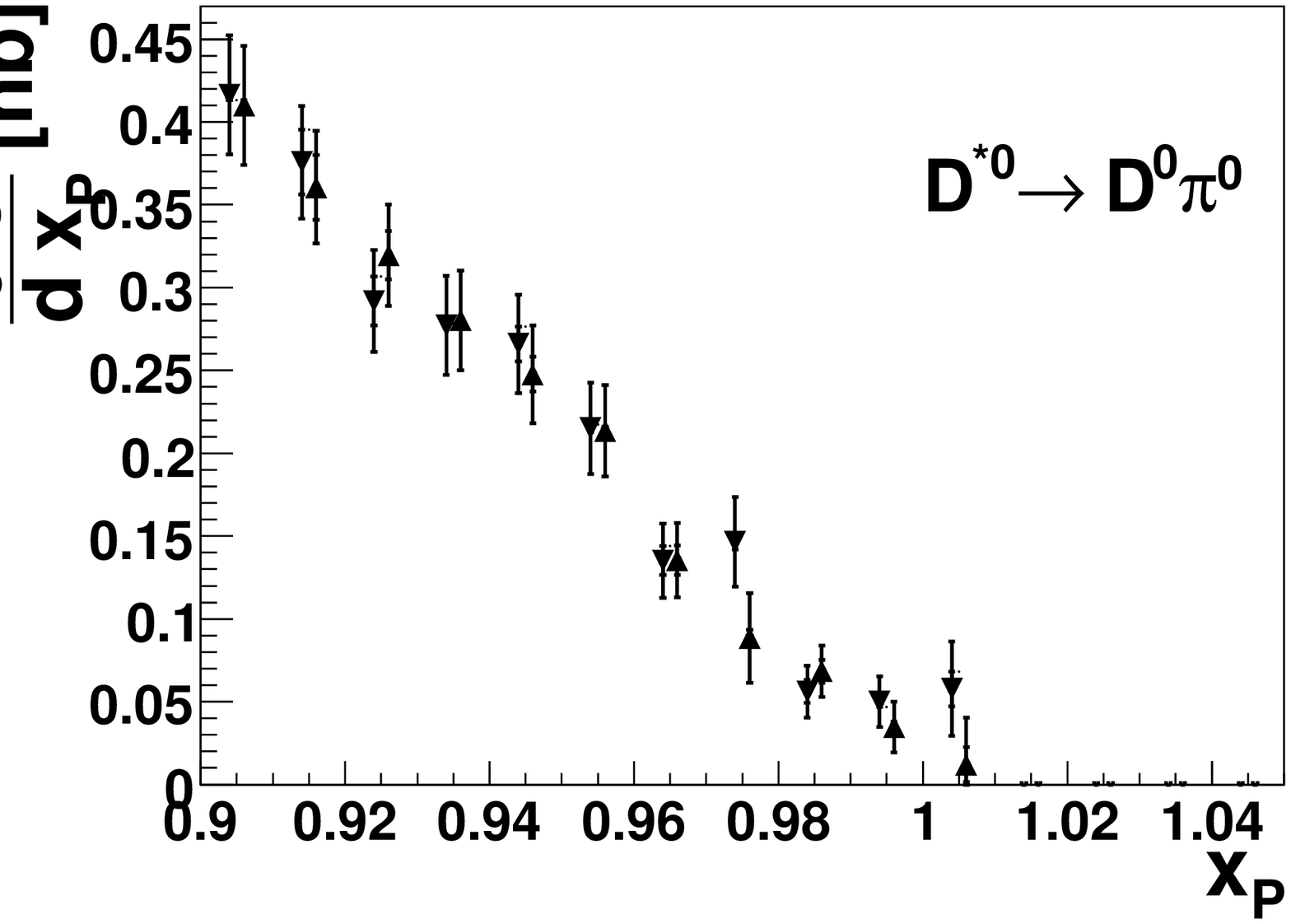}

   \caption{\label{high-xp}An expanded view of the high \xP\ region. 
     The downward (upward) triangles show the efficiency-corrected (unfolded)
     histograms.  Events above the naive limit of \xP = 1 can be produced via
     $e^+e^- \to D^{*+}D^{(*)}$; this is shown in the $D^{*+}$ \xP\
     distributions
     as a hatched (open) histogram for $e^+e^- \to D^{*+}D^-$ (sum of
     $e^+e^- \to D^{*+}D^-$ and $e^+e^- \to D^{*+}D^{*-}$).
     The order of the particles is the same as in Fig.\
     \ref{mass-xp-fita}.
     The inner (outer) error bars show the statistical (total) uncertainties.}
  \end{figure}
 \end{center}

 An expanded view of the high \xP\ region is shown in Fig.\
 \ref{high-xp}. The downward triangles show the efficiency-corrected data;
 the upward triangles show the corrected and unfolded data.
 
 Unfolding was done using the singular-value-decomposition (SVD)
 method \cite{SVD}.
 From \MC, we determined the response matrix of the detector for
 producing for a certain true input value of $\xP_{,true}$ a 
 measured value of $\xP_{,measured}$. 
 This matrix was decomposed using the SVD into two orthogonal and one
 diagonal matrices which can easily be inverted.
 Inverting the diagonal matrix was limited by a criteria defined
 in \cite{SVD} to contain only elements, which are of statistical significance.
 
 The hatched histogram show the only process
 \mbox{$\epem\to\DSP D^{-}$}, the open histogram shows the sum of 
 the previous process and \mbox{$\epem\to\DSP D^{(*)-}$}.
 
 The \xP\ distributions for the ground states \DZ, \DP, \Ds\ and \LC\
 extend up to the na\"{\i}ve kinematic endpoint $\xP=1$ and no
 significant number of events are present for $\xP>1$.
 
 All three \xP\ distributions for the excited $D$ mesons, however, show an
 enhancement at $\xP>1$. These events
 above $\xP=1$ correspond to events of the processes \mbox{$\epem\to\DSP
 D^{(*)-}$} or \mbox{$\epem\to\DSZ D^{(*)0}$} and are in 
 good agreement with the measured cross sections \cite{Uglov} of
 $0.55\pm0.03\pm0.05~\pbc$ for \mbox{$\epem\to\DSP D^{-}$}
 and $0.62\pm0.03\pm0.06~\pbc$ for \mbox{$\epem\to\DSP D^{*-}$}.
 Note that these events populate $\xP>1$ 
 only because of the use of the simplified upper limit
 $|\mathrm{\vec{p}^{MAX}_{candidate}}|$, for producing two $\mathrm{D^*}$ mesons.
 A background fluctuation producing an artificial peak is unlikely for
 two reasons. First, at high \xP, the background is negligible, and
 second, the unfolding procedure tends to identify signals at the edge
 of a distribution as statistical fluctuations rather than real
 signals, thus decreasing the significance of the signals.
 
 \subsection{Total Production Cross-Section}
 The total production cross-section is given by the integral of the
 \xP\ distribution.
 This integral was determined for the continuum sample using the
 current value of the world average product branching fraction
 of each particle, see Table \ref{pdgvalues} and \cite{PDG2004}. 
 The results are listed in Table \ref{production}, where the third
 error component reflects the uncertainty on the product branching
 fraction.
 
 The results by CLEO \cite{CLEOprel}
 given in the last column used their own branching
 fractions, which differ slightly from the world averages used here.
 The results, however, agree well with each other.
 Another measurement by BaBar \cite{BaBar_Ds} is given in the same column.
 The total production cross-section for the \DSZ\ differs only
 slightly from that of the \DSP. This can be understood as resulting
 from different feed-down contributions from higher resonances.
 
 \begin{table}[h]
  \caption{\label{production}The total production cross-sections
    $\epem\to D X$ (or $\LC X$),
   which have been corrected using the current world average of the
   respective product branching fractions.
   The listed uncertainties are statistical, systematic and the
    uncertainty due to the uncertainty on the branching ratios.
   Other measurements of production cross-sections are listed.
   The third column shows measurements by CLEO \cite{CLEOprel},
   BaBar \cite{BaBar_Ds} (marked ${}^{(1)}$) and
   an older CLEO measurement \cite{CLEOc} (marked ${}^{(2)}$).
  }
  \begin{center}
  \begin{ruledtabular}
   \begin{tabular}{Bcc}
    B X & $\sigma_\mathrm{PROD}$ [pb]  
                                   & $ \sigma_{PROD(CLEO'04/BaBar)}$ [pb] \\ \hline
   \DZ\to B\KM\PP & $ 1449 \pm     2 \pm    64 \pm    38$ 
                                   & $1521 \pm    16 \pm    62 \pm    36$ \\
   \DP\to B\KM\PP\PP  & $  654 \pm     1 \pm    36 \pm    46$ 
                                   & $ 640 \pm    14 \pm    35 \pm    43$ \\
   \Ds\to B\PHI\PP  & $  231 \pm     2 \pm    92 \pm    77$ 
                                   & $ 210 \pm     6 \pm     9 \pm    52^{(1)}$ \\
   \LC\to B\pr\KM\PP  & $  189 \pm     1 \pm    66 \pm    66$ & $ 270 \pm  90 \pm 70^{(2)}$
                                   \\
   \DSZ\to B\DZ\PZ & $  510 \pm     3 \pm    84 \pm    39$ 
                                   & $ 559 \pm    24 \pm    35 \pm    39$ \\
   \DSP\to B\DZ\PP & $  598 \pm     2 \pm    77 \pm    20$ 
                                   & $ 583 \pm     8 \pm    33 \pm    14$ \\
   \DSP\to B\DP\PZ & $  590 \pm     5 \pm    78 \pm    53$ 
                                   & - \\
   average B \DSP\ & $  597 \pm     2 \pm    78 \pm    25$ & - 
   \end{tabular}
  \end{ruledtabular}
  \end{center}
 \end{table}
 
 \subsection{Mean Values for \xP\ and Moments}
 In addition to the peak position for the seven \xP\ distributions,
 the mean and higher moments of these distributions were
 determined from distributions in $(\xP)^n$ with a bin width of 0.02
 in \xP\ and a bin-by-bin efficiency correction was applied.
 The $n^{th}$ moment was determined by the mean of the efficiency
 corrected distributions in $(\xP)^n$, and its statistical uncertainty
 was determined by $\mathrm{\sigma}/\sqrt{N_0}$, where $N_0$ is the number of
 entries in the uncorrected $(\xP)^n$ distribution. Tables \ref{moments1}
 and \ref{moments2} show the moments for the different decay modes.
 
 \begin{table}[h]
  \caption{\label{moments1}The first three moments of the \xP\
   distribution for the seven particles/decay modes used in this
   analysis. The listed uncertainties are statistical and systematic
   uncertainties.}
  \begin{center}
  \begin{ruledtabular}
   \begin{tabular}{Bcccc}
     & $\langle\xP\rangle \times 1000$ &
    $\langle\xP^2\rangle \times 1000$ &
    $\langle\xP^3\rangle \times 1000$ \\ \hline
    \DZ\TO B \KM\ \PP\       & 
    $570.33 \pm 0.18 \pm 2.23$ &
    $353.98 \pm 0.29 \pm 2.50$ &
    $233.85 \pm 0.28 \pm 2.54$ \\
    \DP\TO B \KM\ \PP\ \PP\  & 
    $578.03 \pm 0.18 \pm 1.47$ &
    $363.42 \pm 0.29 \pm 1.58$ &
    $243.58 \pm 0.27 \pm 1.58$ \\
    \Ds\TO B \PHI\ \PP\      & 
    $635.34 \pm 0.47 \pm 4.22$ &
    $442.52 \pm 0.81 \pm 8.64$ &
    $323.52 \pm 0.83 \pm11.14$ \\
    \LC\TO B \pr\ \KM\ \PP\  & 
    $582.45 \pm 0.39 \pm 2.53$ &
    $364.94 \pm 0.63 \pm 3.40$ &
    $239.59 \pm 0.59 \pm 2.24$ \\
    \DSP\TO B \DZ\ \PP\       & 
    $612.17 \pm 0.36 \pm 1.43$ &
    $407.96 \pm 0.61 \pm 2.01$ &
    $286.97 \pm 0.60 \pm 3.38$ \\
    \TO B \DP\ \PZ\       & 
    $586.06 \pm 0.37 \pm16.10$ &
    $380.99 \pm 0.64 \pm17.89$ &
    $266.49 \pm 0.62 \pm17.05$ \\
    \DSZ\TO B \DZ\ \PZ\       & 
    $607.63 \pm 0.42 \pm 6.07$ &
    $401.98 \pm 0.69 \pm 5.60$ &
    $282.65 \pm 0.68 \pm 5.90$ \\
   \end{tabular}
  \end{ruledtabular}
  \end{center}
 \end{table}
 
 \begin{table}[h]
  \caption{\label{moments2}The fourth through the sixth moments of the \xP\
    distribution for the seven particles/decay modes used in this
    analysis. The listed uncertainties are statistical and systematic
    uncertainties.}
  \begin{center}
  \begin{ruledtabular}
   \begin{tabular}{Bcccc}
     & $\langle\xP^4\rangle \times 1000$ &
    $\langle\xP^5\rangle\times 1000$ &
    $\langle\xP^6\rangle\times 1000$ \\ \hline
    \DZ\TO B\KM\PP\       & 
    $161.83 \pm 0.25 \pm 2.19$ &
    $116.97 \pm 0.22 \pm 1.90$ &
    $ 88.08 \pm 0.19 \pm 4.77$ \\
    \DP\TO B\KM\PP\PP\  & 
    $171.54 \pm 0.24 \pm 1.28$ &
    $125.62 \pm 0.22 \pm 1.16$ &
    $ 95.52 \pm 0.19 \pm 1.11$ \\
    \Ds\TO B\PHI\PP\      & 
    $244.69 \pm 0.81 \pm12.06$ &
    $188.72 \pm 0.76 \pm10.64$ &
    $150.59 \pm 0.72 \pm 9.75$ \\
    \LC\TO B\pr\KM\PP\  & 
    $163.04 \pm 0.52 \pm 5.11$ &
    $115.07 \pm 0.46 \pm 1.94$ &
    $ 85.06 \pm 0.41 \pm 2.31$ \\
    \DSP\TO B\DZ\PP\       & 
    $211.55 \pm 0.57 \pm 5.36$ &
    $162.26 \pm 0.53 \pm 7.05$ &
    $128.24 \pm 0.49 \pm 8.17$ \\
    \TO B\DP\PZ\       & 
    $196.24 \pm 0.58 \pm16.67$ &
    $150.28 \pm 0.53 \pm15.30$ &
    $118.85 \pm 0.49 \pm13.89$ \\
    \DSZ\TO B\DZ\PZ\       & 
    $215.63 \pm 0.68 \pm12.47$ &
    $160.05 \pm 0.59 \pm 7.98$ &
    $126.87 \pm 0.54 \pm 8.72$ \\
    
   \end{tabular}
  \end{ruledtabular}
  \end{center}
 \end{table}

 \subsection{Production Angle}
 Taking the interference between the exchange of virtual photons and
 \Zn\ bosons into account, the differential cross section for
 $\epem\to\ccbar$ is modified from the $1+\cos^2\theta$ form of
 the Born amplitude for pure photon exchange:
 \begin{equation}
   \frac{d \sigma}{d \cos\theta}=
   \frac{3}{8}(1+\cos^2\theta)\sigma_T+
   \frac{3}{4}\sin^2\theta\sigma_L+
   \frac{3}{4}\cos\theta\sigma_A
 \end{equation}
 Here, $\theta$ describes the angle between the incoming electron beam
 and the outgoing hadron containing the charmed quark, as measured in
 the CM frame.
 The term $\sigma_T$ describes the contribution of pair
 production of spin-1/2 particles from transverse polarised vector
 bosons, the term $\sigma_L$ the contribution from longitudinal
 polarised vector bosons and the term $\sigma_A$ denotes the parity
 violating asymmetry due to the interference between \Zn\ bosons and 
 virtual photons.
 
 The ${\cal KK}$ \MC\ generator \cite{KK} was used to predict the
 production angle distributions for the different charmed hadrons.
 This \MC\ generator includes interference between initial and final
 state radiation (ISR and FSR) as well as electro-weak corrections.
 $10^8$ \ccbar\ events were generated with ${\cal KK}$ and
 hadronised with the PYTHIA generator.

\begin{center}
  \begin{figure}[h]
   \includegraphics[width=0.45\textwidth]{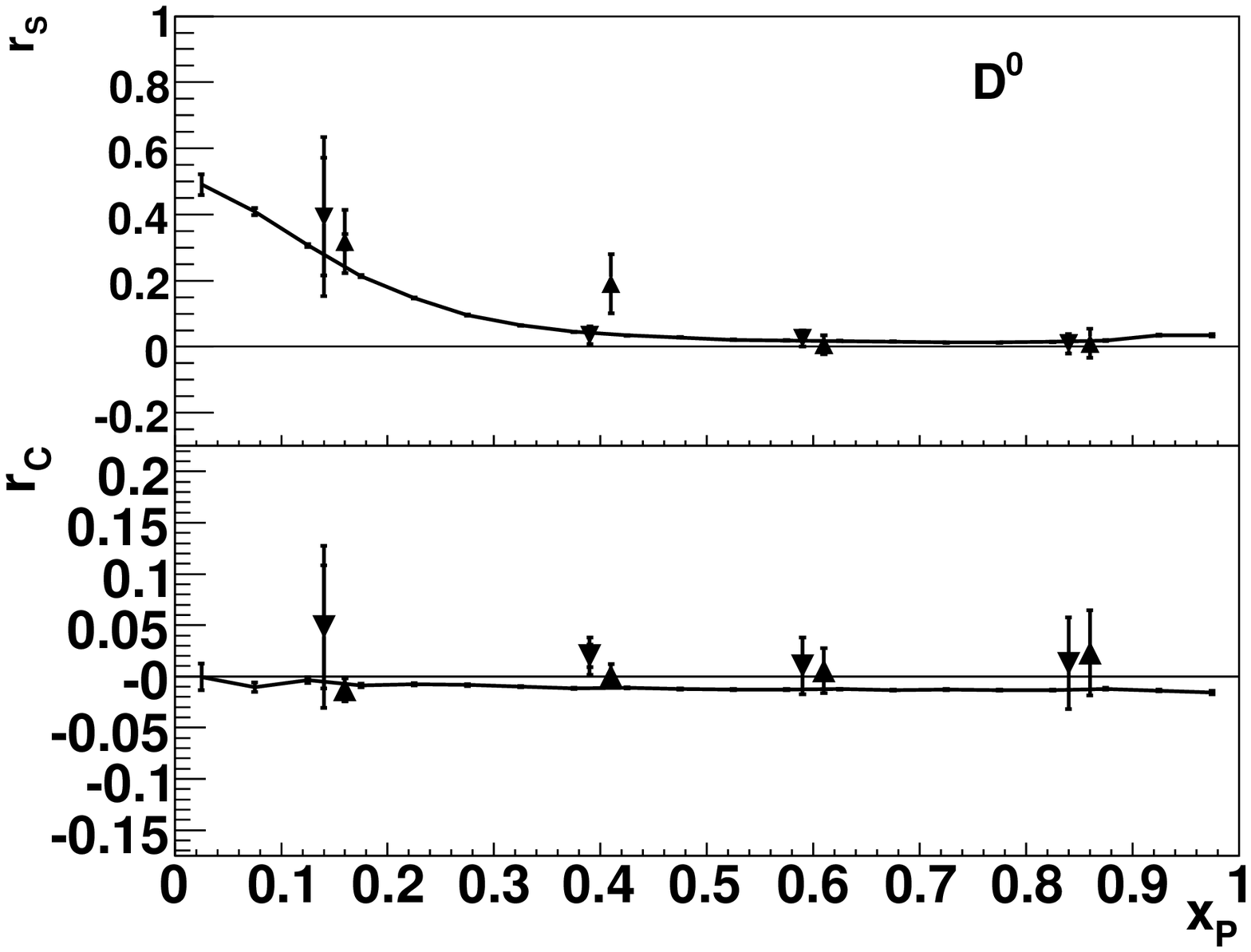}
   \includegraphics[width=0.45\textwidth]{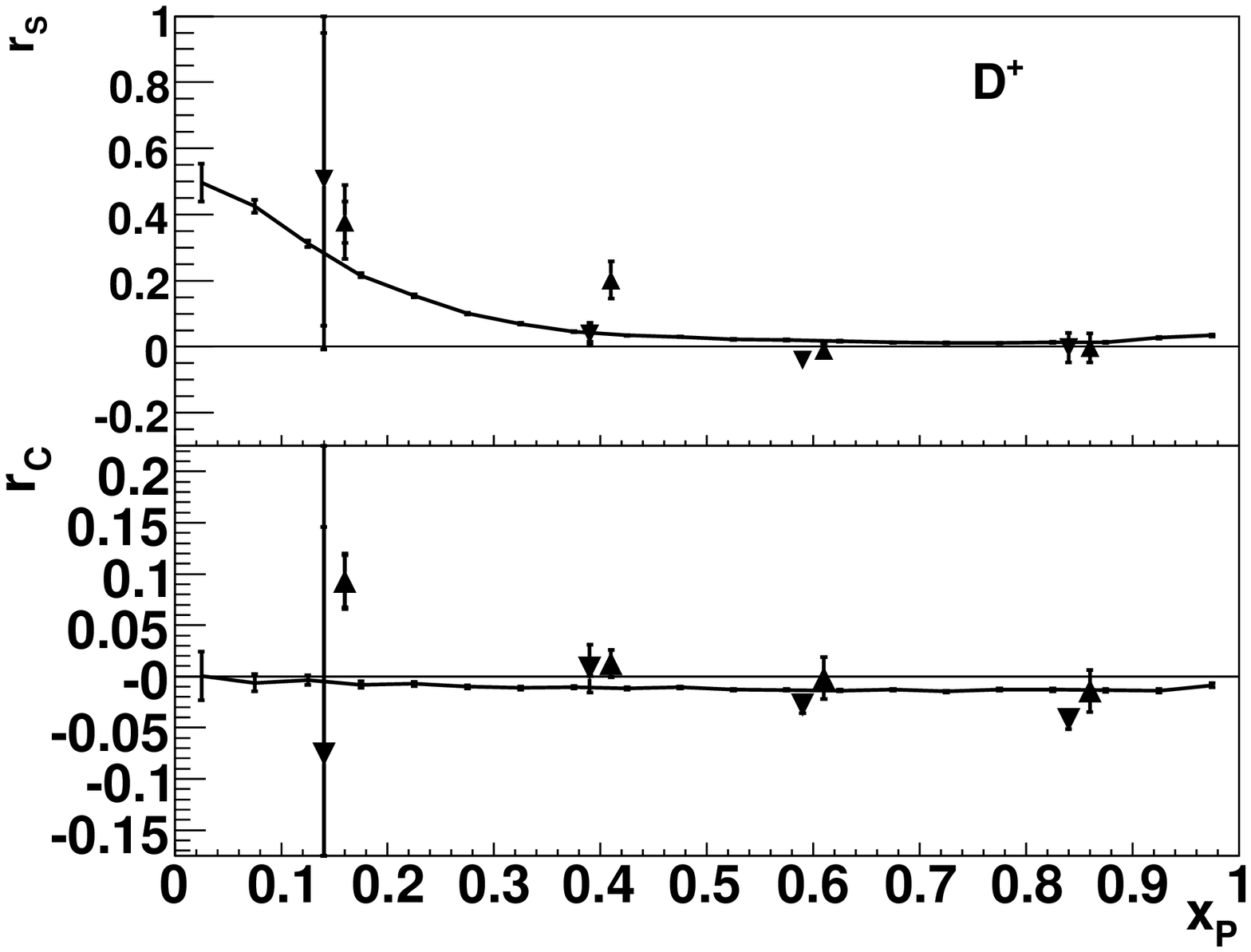}\\
   \includegraphics[width=0.45\textwidth]{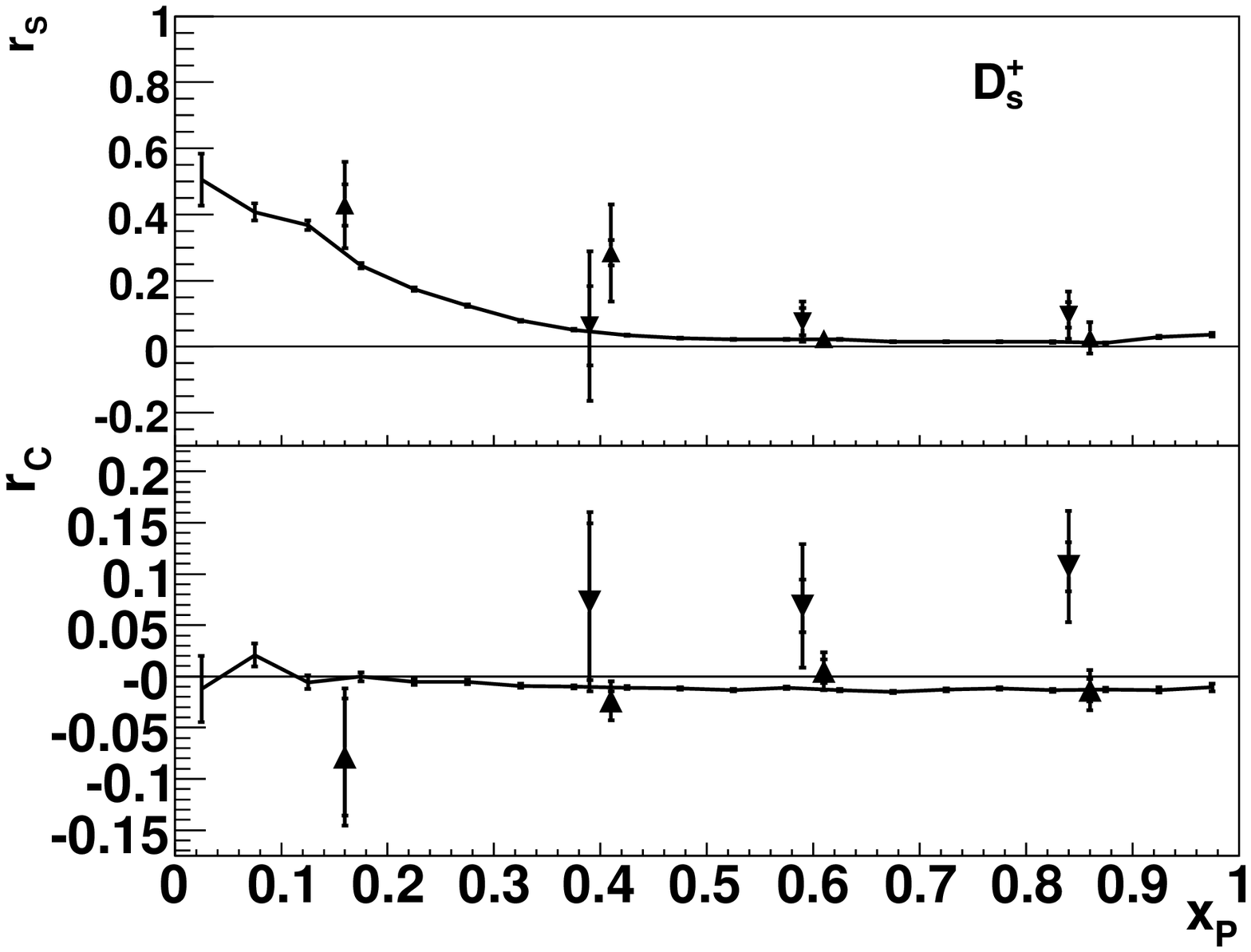}
   \includegraphics[width=0.45\textwidth]{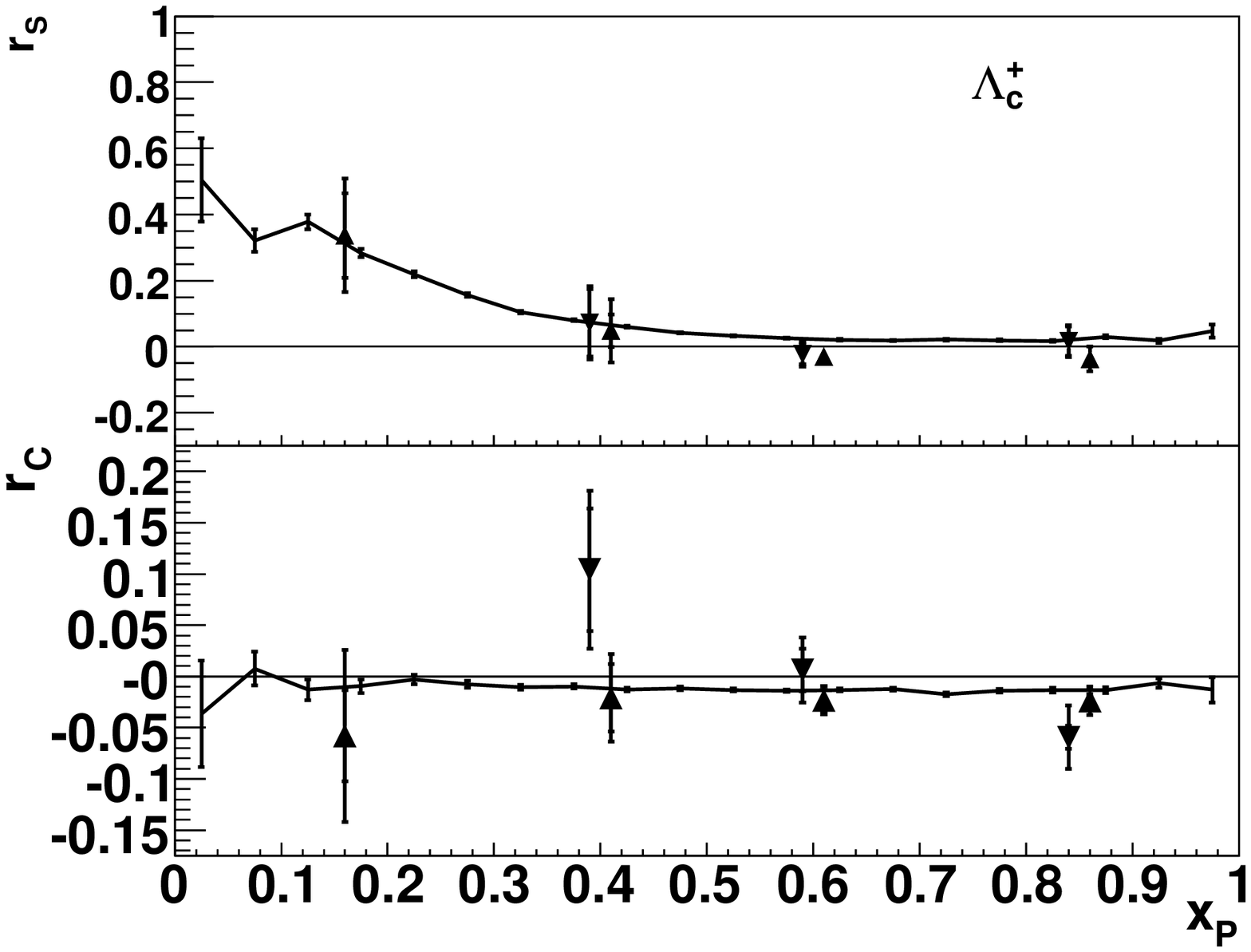}\\
   \includegraphics[width=0.45\textwidth]{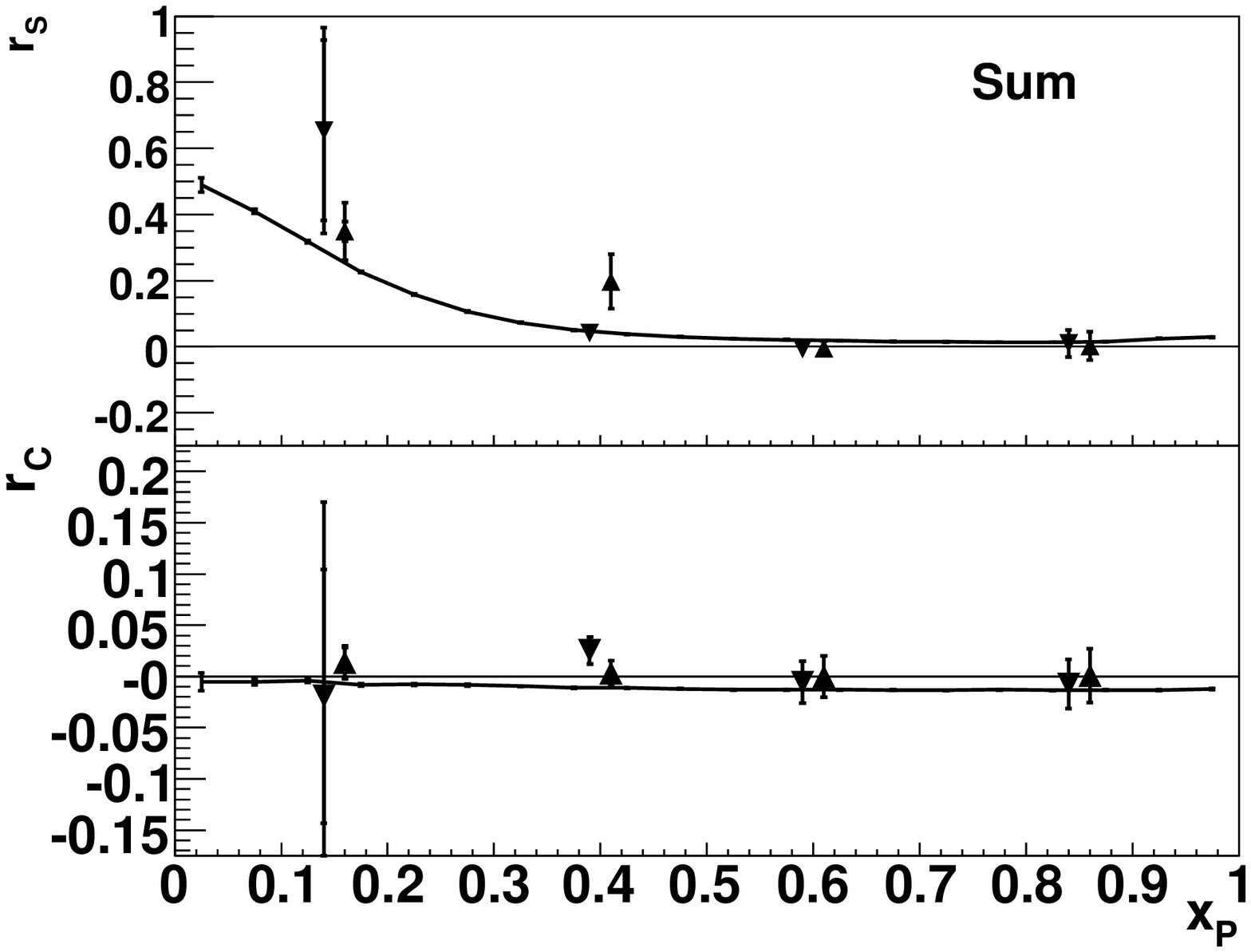}
   \includegraphics[width=0.45\textwidth]{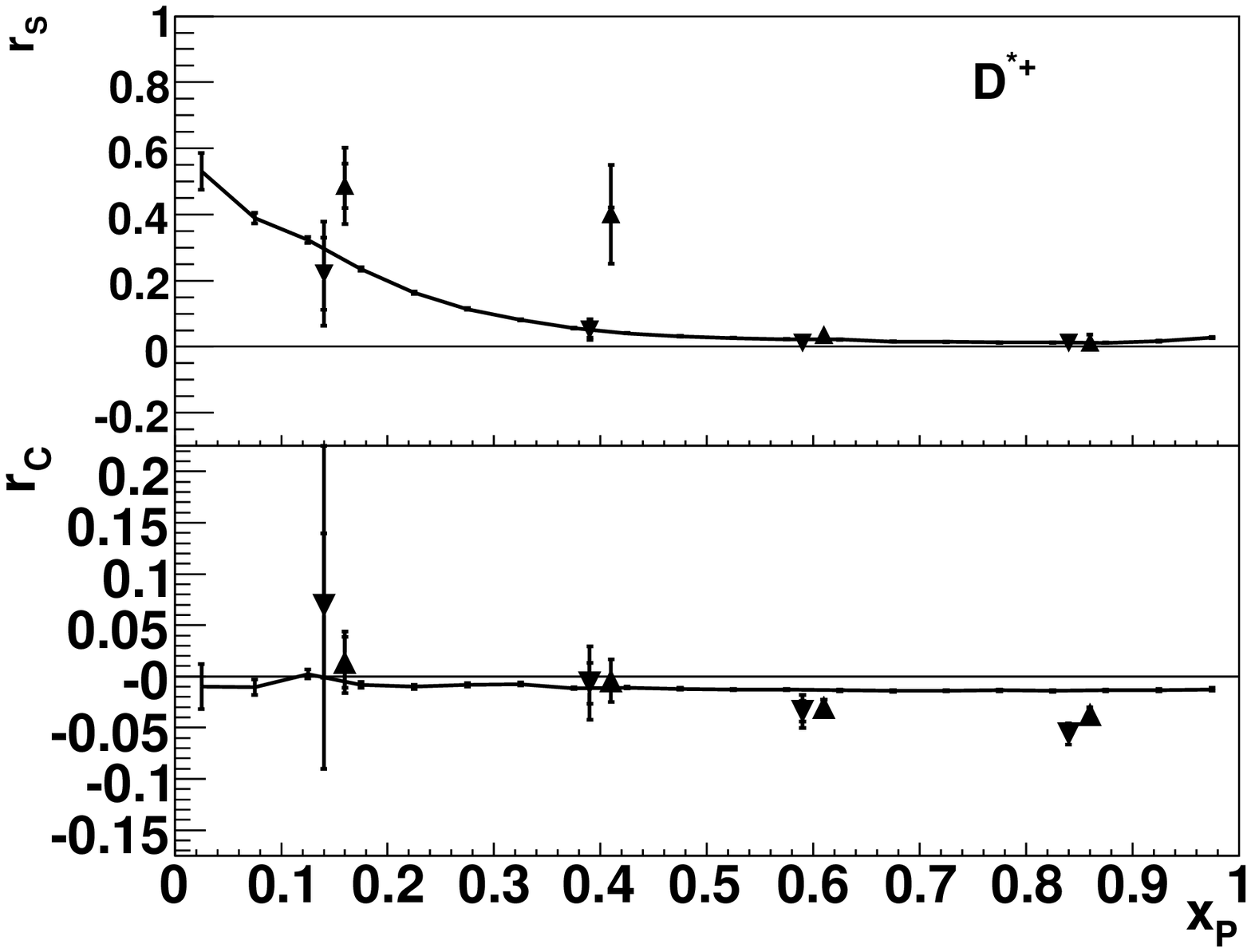}
   \caption{\label{prodAngle}
     The production angle coefficients $r_S$\ (upper distributions) and
     $r_C$\ (lower distributions) for the four ground state charmed hadrons:
     $D^0$\ and $D^+$\ at the top, $D^+_s$\ and $\Lambda^+_c$\ in the
     middle.
     The sum of all ground states after efficiency correction
     and an excited $D$\ state ($D^{*+} \to D^0 \pi^+$) are shown at
     the bottom.
     The upwards (downwards) triangles show on-resonance (continuum) data.
     The connected points show the results from the $\mathcal{KK}$
     generator.  For display purposes, the points are slightly separated
     in \xP. }
  \end{figure}
 \end{center}

 For the generated events, \xP\ was divided into 
 20 bins of equal width and a three-parameter fit to the
 production angle was performed:
 \begin{equation}
   \label{equation}
   f(\theta,\xP)=a_0(\xP)\left[
     \frac{3}{8}
     (1+\cos^2\theta)+
     \frac{3}{4}
     r_S(\xP)\sin^2\theta+
     \frac{3}{4}
     r_C(\xP)\cos\theta\right]
 \end{equation}
 where $a_0$ is the normalisation and $r_S$ and $r_C$ are the relative
 contributions for the $\sin^2\theta$ and the $\cos\theta$ terms,
 respectively. The results of these fits are shown in
 Fig.\ \ref{prodAngle} as the solid line, together with the measured
 data points.
 
 \begin{table}[h]
  \caption{\label{frf23}
    The coefficients in front of the sine-squared term ($r_S$) and the
    cosine term ($r_C$) for different \xP\ bins in the continuum sample.}
  {\small
  \begin{center}
  \begin{ruledtabular}
   \begin{tabular}{ccrr}
    particle & range in \xP\ & \multicolumn{1}{c}{$r_S$} & \multicolumn{1}{c}{$r_C$}  \\ \hline
    $D^0$ & 0.0 - 0.3 & 
    $ 0.393 \pm 0.178 \pm 0.166 $ & $ 0.048 \pm 0.079 \pm 0.008 $ \\ &  0.3 - 0.5 &
    $ 0.035 \pm 0.016 \pm 0.007 $ & $ 0.020 \pm 0.011 \pm 0.003 $ \\ &  0.5 - 0.7 &
    $ 0.025 \pm 0.008 \pm 0.014 $ & $ 0.010 \pm 0.006 \pm 0.001 $ \\ &  0.7 - 1.0 &
    $ 0.009 \pm 0.010 \pm 0.004 $ & $ 0.013 \pm 0.007 \pm 0.009 $ \\ \hline
    $D^+$ & 0.0 - 0.3 & 
    $ 0.507 \pm 0.443 \pm 0.349 $ & $ -0.076 \pm 0.222 \pm 0.005 $ \\ &  0.3 - 0.5 &
    $ 0.039 \pm 0.034 \pm 0.008 $ & $ 0.008 \pm 0.024 \pm 0.002 $ \\ &  0.5 - 0.7 &
    $ -0.040 \pm 0.011 \pm 0.006 $ & $ -0.028 \pm 0.008 \pm 0.004 $ \\ &  0.7 - 1.0 &
    $ -0.003 \pm 0.012 \pm 0.019 $ & $ -0.042 \pm 0.008 \pm 0.010 $ \\ \hline
    $D^+_s$ &  0.3 - 0.5 &
    $ 0.063 \pm 0.120 \pm 0.013 $ & $ 0.073 \pm 0.077 \pm 0.029 $ \\ &  0.5 - 0.7 &
    $ 0.076 \pm 0.041 \pm 0.004 $ & $ 0.069 \pm 0.026 \pm 0.007 $ \\ &  0.7 - 1.0 &
    $ 0.096 \pm 0.038 \pm 0.027 $ & $ 0.107 \pm 0.024 \pm 0.017 $ \\ \hline
    $\Lambda^+_c$ &  0.3 - 0.5 &
    $ 0.072 \pm 0.112 \pm 0.027 $ & $ 0.104 \pm 0.077 \pm 0.066 $ \\ &  0.5 - 0.7 &
    $ -0.023 \pm 0.029 \pm 0.042 $ & $ 0.006 \pm 0.021 \pm 0.004 $ \\ &  0.7 - 1.0 &
    $ 0.016 \pm 0.048 \pm 0.037 $ & $ -0.059 \pm 0.031 \pm 0.010 $ \\ \hline
    $\mathrm{Sum}$ & 0.0 - 0.3 & 
    $ 0.654 \pm 0.312 \pm 0.142 $ & $ -0.019 \pm 0.124 \pm 0.039 $ \\ of &  0.3 - 0.5 &
    $ 0.042 \pm 0.019 \pm 0.007 $ & $ 0.025 \pm 0.013 \pm 0.008 $ \\ ground &  0.5 - 0.7 &
    $ -0.007 \pm 0.007 \pm 0.011 $ & $ -0.006 \pm 0.005 \pm 0.002 $ \\ states &  0.7 - 1.0 &
    $ 0.010 \pm 0.008 \pm 0.007 $ & $ -0.007 \pm 0.005 \pm 0.008 $ \\ \hline
    $D^{*+}$ & 0.0 - 0.3 & 
    $ 0.221 \pm 0.157 \pm 0.779 $ & $ 0.069 \pm 0.070 \pm 0.333 $ \\ &  0.3 - 0.5 &
    $ 0.051 \pm 0.031 \pm 0.035 $ & $ -0.007 \pm 0.020 \pm 0.017 $ \\ &  0.5 - 0.7 &
    $ 0.011 \pm 0.014 \pm 0.013 $ & $ -0.034 \pm 0.010 \pm 0.005 $ \\ &  0.7 - 1.0 &
    $ 0.011 \pm 0.015 \pm 0.012 $ & $ -0.056 \pm 0.010 \pm 0.004 $ \\
   \end{tabular}
  \end{ruledtabular}
  \end{center}}
 \end{table}
 
 \begin{table}[h]
   \caption{\label{frf22}
     The same as in Table \ref{frf23}, but now for the on-resonance sample.}
  {\small
   \begin{center}
  \begin{ruledtabular}
   \begin{tabular}{ccrr}
    particle & range in \xP\ & \multicolumn{1}{c}{$r_S$} & \multicolumn{1}{c}{$r_C$}  \\ \hline
    $D^0$ & 0.0 - 0.3 & 
    $ 0.318 \pm 0.023 \pm 0.022 $ & $ -0.013 \pm 0.011 \pm 0.008 $ \\ &  0.3 - 0.5 &
    $ 0.191 \pm 0.007 \pm 0.016 $ & $ 0.001 \pm 0.004 \pm 0.004 $ \\ &  0.5 - 0.7 &
    $ 0.005 \pm 0.004 \pm 0.004 $ & $ 0.006 \pm 0.003 \pm 0.002 $ \\ &  0.7 - 1.0 &
    $ 0.010 \pm 0.005 \pm 0.010 $ & $ 0.023 \pm 0.004 \pm 0.006 $ \\ \hline
    $D^+$ & 0.0 - 0.3 & 
    $ 0.377 \pm 0.063 \pm 0.026 $ & $ 0.093 \pm 0.027 \pm 0.016 $ \\ &  0.3 - 0.5 &
    $ 0.202 \pm 0.015 \pm 0.008 $ & $ 0.012 \pm 0.008 \pm 0.001 $ \\ &  0.5 - 0.7 &
    $ -0.011 \pm 0.006 \pm 0.005 $ & $ -0.002 \pm 0.004 \pm 0.001 $ \\ &  0.7 - 1.0 &
    $ -0.003 \pm 0.006 \pm 0.007 $ & $ -0.014 \pm 0.004 \pm 0.001 $ \\ \hline
    $D^+_s$ & 0.0 - 0.3 & 
    $ 0.428 \pm 0.130 \pm 0.032 $ & $ -0.079 \pm 0.057 \pm 0.030 $ \\ &  0.3 - 0.5 &
    $ 0.284 \pm 0.038 \pm 0.011 $ & $ -0.024 \pm 0.019 \pm 0.005 $ \\ &  0.5 - 0.7 &
    $ 0.025 \pm 0.017 \pm 0.008 $ & $ 0.005 \pm 0.011 \pm 0.001 $ \\ &  0.7 - 1.0 &
    $ 0.026 \pm 0.017 \pm 0.008 $ & $ -0.013 \pm 0.011 \pm 0.008 $ \\ \hline
    $\Lambda^+_c$ & 0.0 - 0.3 & 
    $ 0.336 \pm 0.172 \pm 0.044 $ & $ -0.058 \pm 0.084 \pm 0.009 $ \\ &  0.3 - 0.5 &
    $ 0.048 \pm 0.049 \pm 0.027 $ & $ -0.021 \pm 0.033 \pm 0.020 $ \\ &  0.5 - 0.7 &
    $ -0.030 \pm 0.014 \pm 0.015 $ & $ -0.023 \pm 0.010 \pm 0.003 $ \\ &  0.7 - 1.0 &
    $ -0.037 \pm 0.020 \pm 0.002 $ & $ -0.024 \pm 0.014 \pm 0.004 $ \\ \hline
    $\mathrm{Sum}$ & 0.0 - 0.3 & 
    $ 0.349 \pm 0.030 \pm 0.001 $ & $ 0.014 \pm 0.014 \pm 0.012 $ \\ of &  0.3 - 0.5 &
    $ 0.197 \pm 0.008 \pm 0.013 $ & $ 0.004 \pm 0.005 \pm 0.001 $ \\ ground &  0.5 - 0.7 &
    $ -0.004 \pm 0.004 \pm 0.001 $ & $ -0.000 \pm 0.002 \pm 0.001 $ \\ states &  0.7 - 1.0 &
    $ 0.002 \pm 0.004 \pm 0.006 $ & $ 0.001 \pm 0.003 \pm 0.003 $ \\ \hline
    $D^{*+}$ & 0.0 - 0.3 & 
    $ 0.487 \pm 0.067 \pm 0.049 $ & $ 0.014 \pm 0.030 \pm 0.012 $ \\ &  0.3 - 0.5 &
    $ 0.401 \pm 0.021 \pm 0.004 $ & $ -0.004 \pm 0.010 \pm 0.008 $ \\ &  0.5 - 0.7 &
    $ 0.037 \pm 0.008 \pm 0.002 $ & $ -0.030 \pm 0.005 \pm 0.001 $ \\ &  0.7 - 1.0 &
    $ 0.014 \pm 0.008 \pm 0.001 $ & $ -0.037 \pm 0.005 \pm 0.004 $ \\
   \end{tabular}
  \end{ruledtabular}
   \end{center}}
 \end{table}
 
 For data, the signal yield was determined in bins of \xP\ and
 $\cos\theta$, where $\theta$ is the production angle of the
 charmed hadron. It should be noted that here, the efficiency correction
 depends on \xP\ and $\cos\theta$.
 
 \xP\ was divided into only 4 bins:
 $\xP<0.3$, $0.3<\xP<0.5$, $0.5<\xP<0.7$ and $0.7<\xP$. 
 The boundaries were chosen in order to roughly equalise the number of
 candidates per bin. Two bins each were chosen below and above $\xP=0.5$,
 which is the upper kinematic limit for hadrons from $B$ decays.
 
 $\cos\theta$ was divided into 20 bins.
 In each bin of \xP\ and $\cos\theta$, the efficiency corrected signal
 yield in the mass or mass difference distributions was fitted
 separately for the on-resonance and continuum samples. The same
 signal and background functions were used as in the fits which
 depended only on \xP.
 
 In a second step, a three-parameter fit (similar to Eq.\ \ref{equation})
 to the signal yields was performed in bins of \xP.
 The fit range was restricted to $-0.8<\cos{\theta}<0.8$. As a
 systematic check, the range was tightened to $-0.7<\cos{\theta}<0.7$.
 
 No significant deviation of the $r_S$ and the $r_C$ parameters from
 their expectation was found for the continuum sample. The expectation
 from the ${\cal KK}$ generator was of the same order as the
 statistical uncertainties except for the $r_S$ term in the lowest
 \xP\ bin, where gluon radiation introduces a longitudinal momentum
 component and therefore smears out the initial distribution of the
 production angle. This smearing introduces a significant
 $\sin^2\theta$ term,
 to which the measured values in this regime agree well.
 As the number of entries in the low \xP\ bins also diminish,
 the statistical uncertainties increase to roughly the same size as
 the expected effect. 
 The fitted values for $r_S$ and $r_C$ in the lowest \xP\ bin in the
 continuum samples suffer from very low statistics (see
 Fig.\ \ref{signal-yield}).
 For the \Ds\ and the \LC, the lowest bins in the continuum samples
 have been neglected as the numbers of entries in these bins were too low
 to perform an angular analysis.
 
 For the on-resonance sample with higher statistics, the $r_S$ terms
 significantly deviate from zero for $\xP<0.5$, as $B$ decays with
 different production angle distributions also contribute.
 The $r_C$ term is again consistent with both zero and the expectation
 from the \MC\ generator. Tables \ref{frf23} and \ref{frf22} list the
 measured values for the $r_S$ and $r_C$ values for the continuum
 and the on-resonance data. The systematic uncertainties here include
 the uncertainties discussed in the standard analysis as well as the
 uncertainty due to the restricted fit range.
 
 \section{Interpretation}
 \subsection{Contributions from Higher Resonances}
 Contributions from excited states have been considered only in the
 \xP\ distributions of the \DZ\ and the \DP\, and for these, only
 contributions from \DSZ\ and \DSP\ were considered.
 For higher resonances, both production cross-section and
 branching fractions of {\it e.g.} $D^{**}$ have large uncertainties and have
 been neglected.
 In order to reduce the statistical uncertainty, a \MC-based
 correction was applied:
 Three large samples of several million \MC\ events were 
 generated without a detailed detector simulation. These samples
 were required to contain the decay modes \DSP\TO\DZ\PP,
 \DSP\TO\DP\PZ\ and \DSZ\TO\DZ\PZ, respectively.
 For these events, the \xP\ of the parent \DSP/\DSZ\ vs. the \xP\ of
 the daughter \DZ/\DP\ were stored in a two dimensional matrix.
 The measured and efficiency corrected \DSZ\ and \DSP\ \xP\
 distributions were multiplied with this matrix in order to
 estimate the \xP\ distribution for all \DZ's and \DP's coming from
 \DSP/\DSZ\ decays.
 
\begin{center}
  \begin{figure}[h]
   \includegraphics[width=0.45\textwidth]{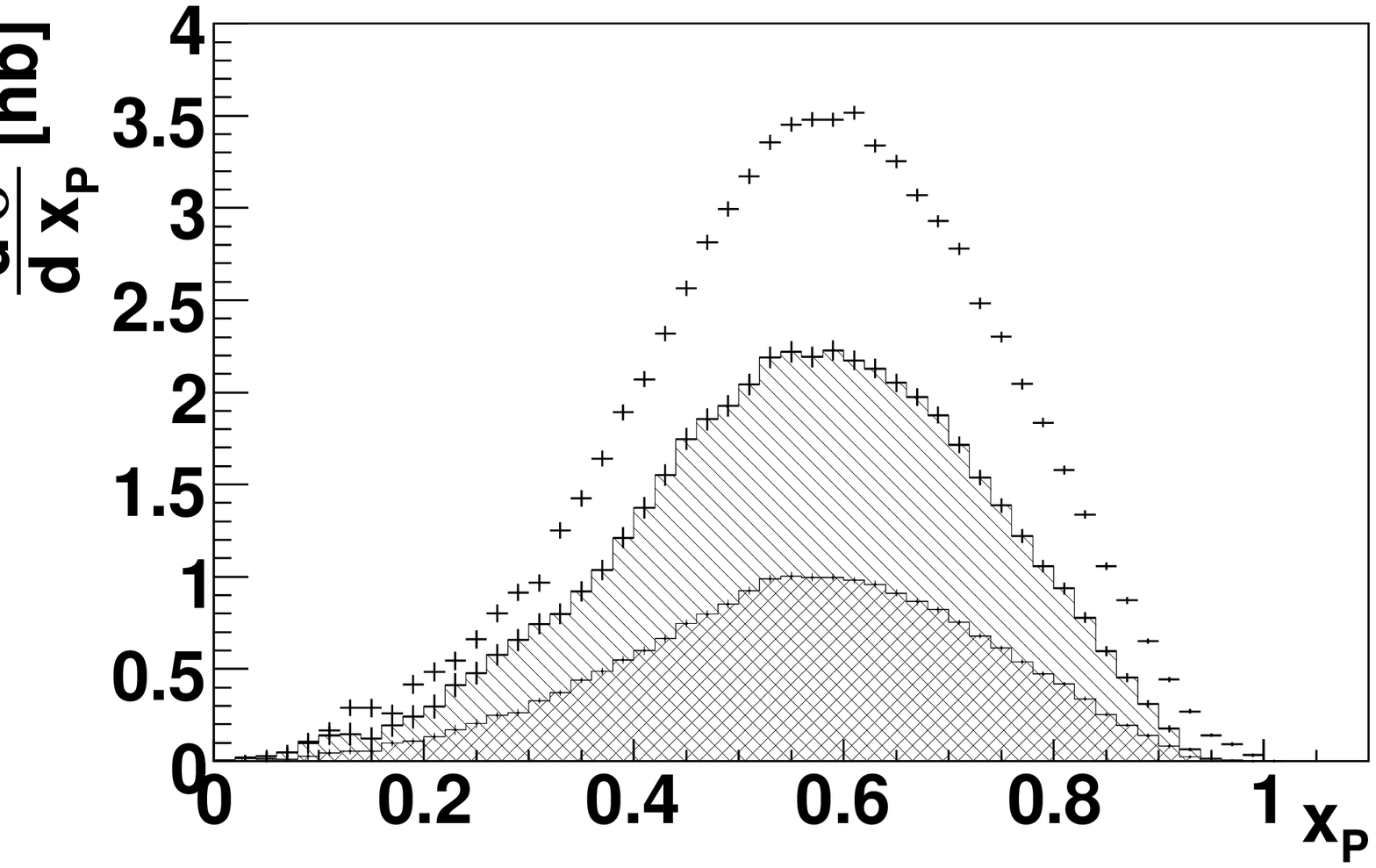}
   \includegraphics[width=0.45\textwidth]{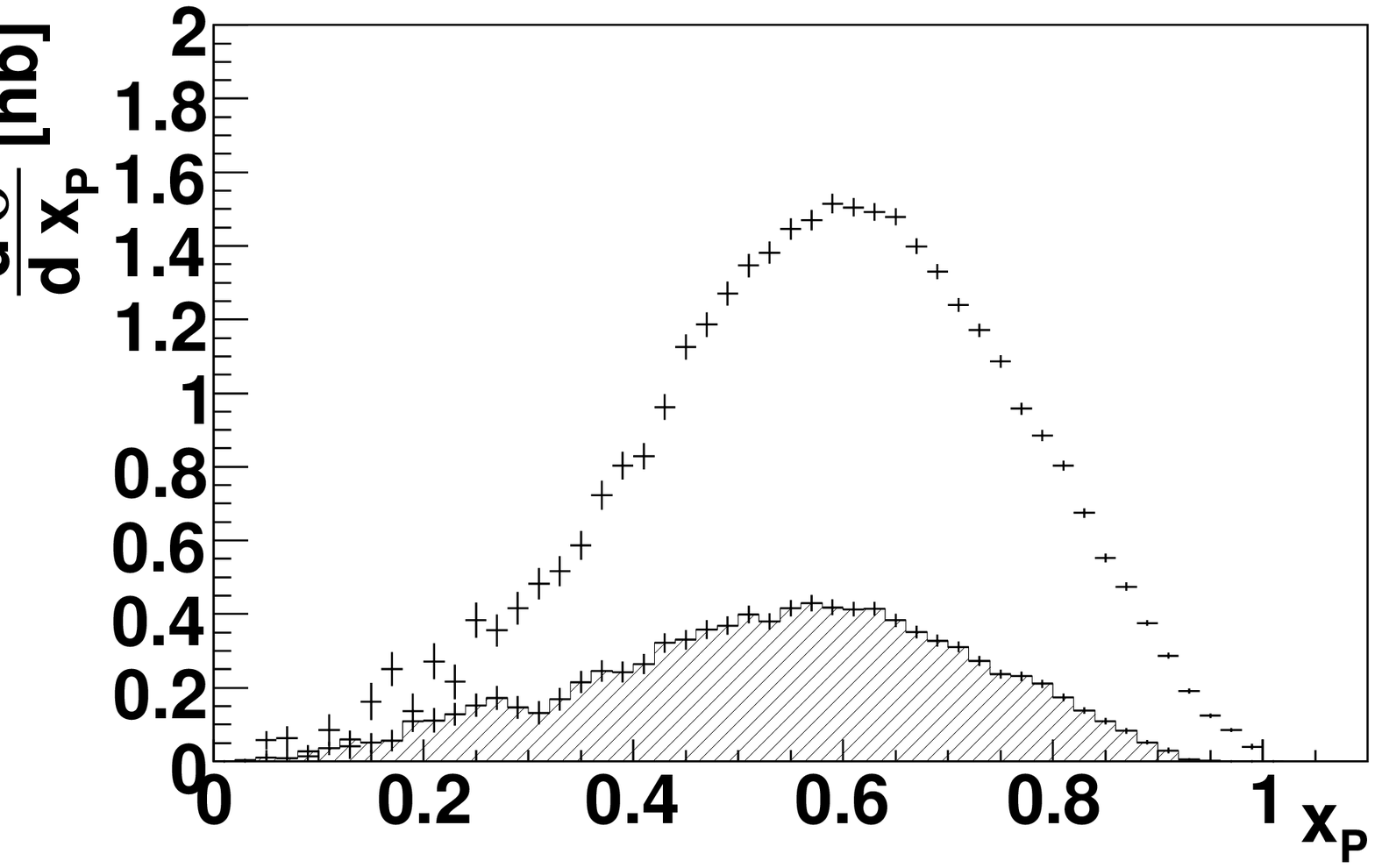}\\
   \includegraphics[width=0.85\textwidth]{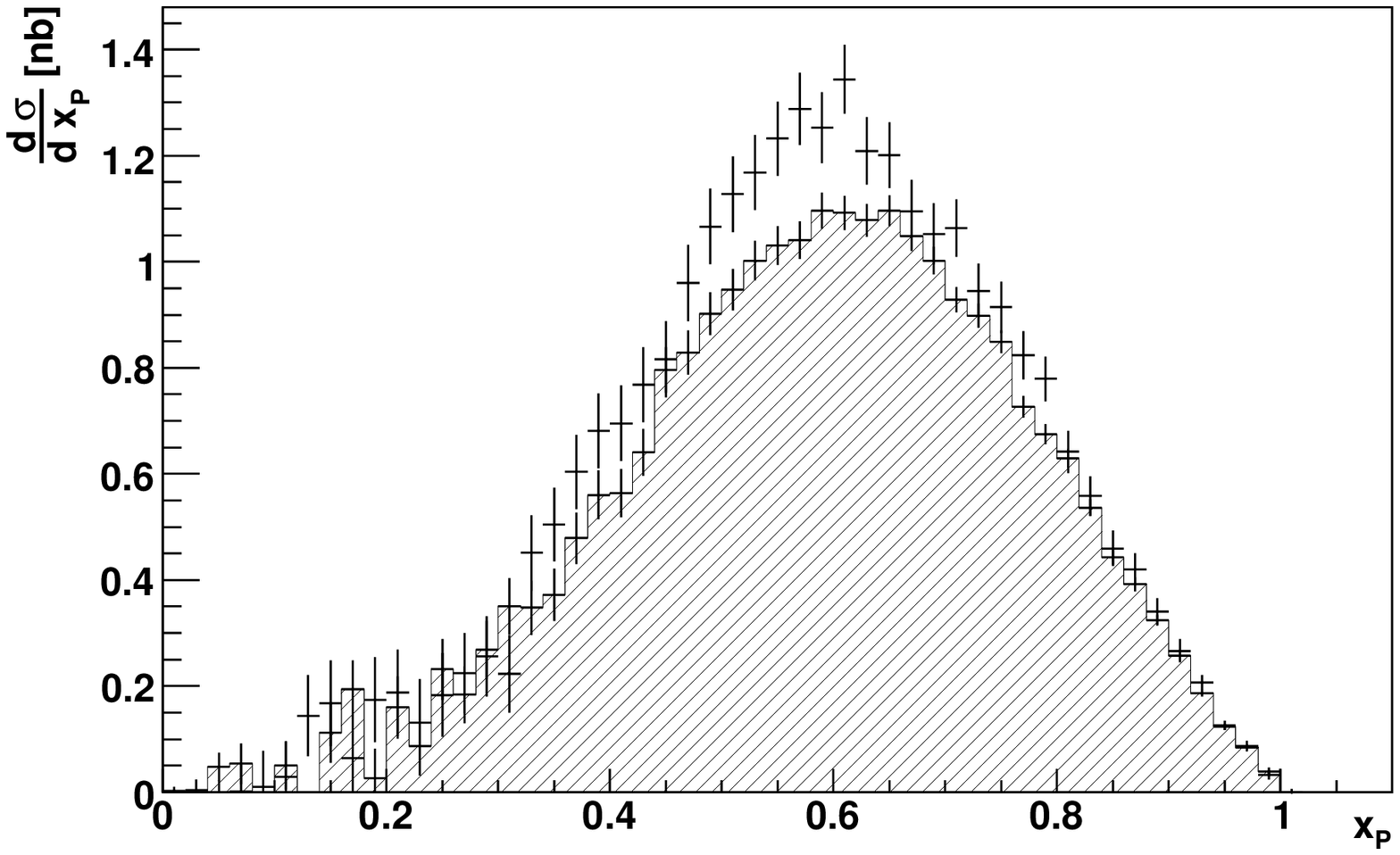}

   \caption{\label{primary}
     Upper plots: The contributions from $D^{*+}$\ (up-right hatching)
     and $D^{*0}$\ (up-left hatching) decays to 
     the $D^0$\ (left) and $D^+$\ (right) \xP\ distributions.
     These plots are not corrected for the branching fraction of the $D$ decay.
     Lower plot:
     The \xP\ distributions for primarily produced $D^0$\ (error bars)
     and $D^+$\ (up-right hatching) mesons.
     Note that only contributions from $D^{*+}$\ and $D^{*0}$\ decays have
     been considered; the exclusion of higher resonance contributions can
     partially account for the differences in the \xP\
     distributions of the $D^0$\ and $D^+$.
     Only statistical uncertainties are shown. }
  \end{figure}
 \end{center}

 The two plots at the top of Fig.\ \ref{primary} show the
 contributions of \DSP\ and \DSZ\ decays to the \DZ\ fragmentation
 function (left), and the contribution of \DSP\ decays to the \DP\
 \frf\ (right). These plots are not corrected for the branching
 fraction of the $D$ decay.
 The bottom plot in Fig.\ \ref{primary} shows primary \DZ\ and \DP\
 fragmentation spectra: the total \DZ\ (\DP) spectrum minus the
 contribution from \DSZ\ and \DSP\ (\DSP only) decays.
 The difference of 13\% between the sum of primary produced \DZ\ and
 \DP\ should be compared to the 6.5\% relative uncertainty in the
 \DP\TO\KM\PP\PP\ branching fraction.
 Also, as only the contribution from $D^*$ decays has been
 considered,
 the remaining difference may be due to the contribution of
 other resonances. In the generic \MC\ sample, where primary \DZ's and
 \DP's were produced in equal amounts, there was an excess of 6\% in
 the production of \DZ\ (compared to \DP) mesons in the decay of
 resonances other than \DSP.
 
\subsection{Ratios}
\label{secratios}
Comparisons of production rates for various particles are useful for
understanding the dynamics of fragmentation,
as systematic errors cancel in the ratio.
In this section we present ratios of both integrated cross-sections
and cross-sections as a function of \xP, to characterise general properties
of fragmentation and to test the agreement between \MC\ simulation and
data.

Table~\ref{tabratio} presents three ratios of total production cross sections.
Since the production of $D^{*+}$ and $D^{*0}$ is included in
the total $D^+$ and $D^0$ production rate 
(all $D^{*+}$ decay to either $D^+$ or $D^0$, and all $D^{*0}$ to $D^0$),
the ratio of $D^*$ to $D$ production measures $V/(V+P)$,
the probability of producing a vector charmed meson.
(Here we write $V$ for vector and $P$ for pseudo-scalar meson production rates.)
A correction is necessary to account for higher resonances decaying
directly to $D^{+,0}$. For example, based on the measured production rates
of the $D_1(2420)$ and $D_2^*(2460)$~\cite{CLEO_Dstst} mesons,
known branching fractions~\cite{PDG2004}, and isospin relations,
we find a correction of $-(3.7 \pm 3.3)\%$ to the first ratio.
In principle, further corrections due to decays of broad $D^{**}$ states and
charmed-strange mesons are also required.
However, no corrections have been applied to the values presented in Table~\ref{tabratio}.

Similarly, the second ratio measures the production rate of charmed-strange
mesons as a fraction of all charmed mesons, up to corrections for 
$D_{s1}(2536)$ and $D_{s2}^*(2573)$ decays. The third ratio measures
the production rate of charmed baryons relative to that of charmed mesons,
excluding the charmed-strange baryon states.
For comparison, see~\cite{Gladilin}. 

\begin{center}
  \begin{figure}[h]
   \includegraphics[width=0.45\textwidth]{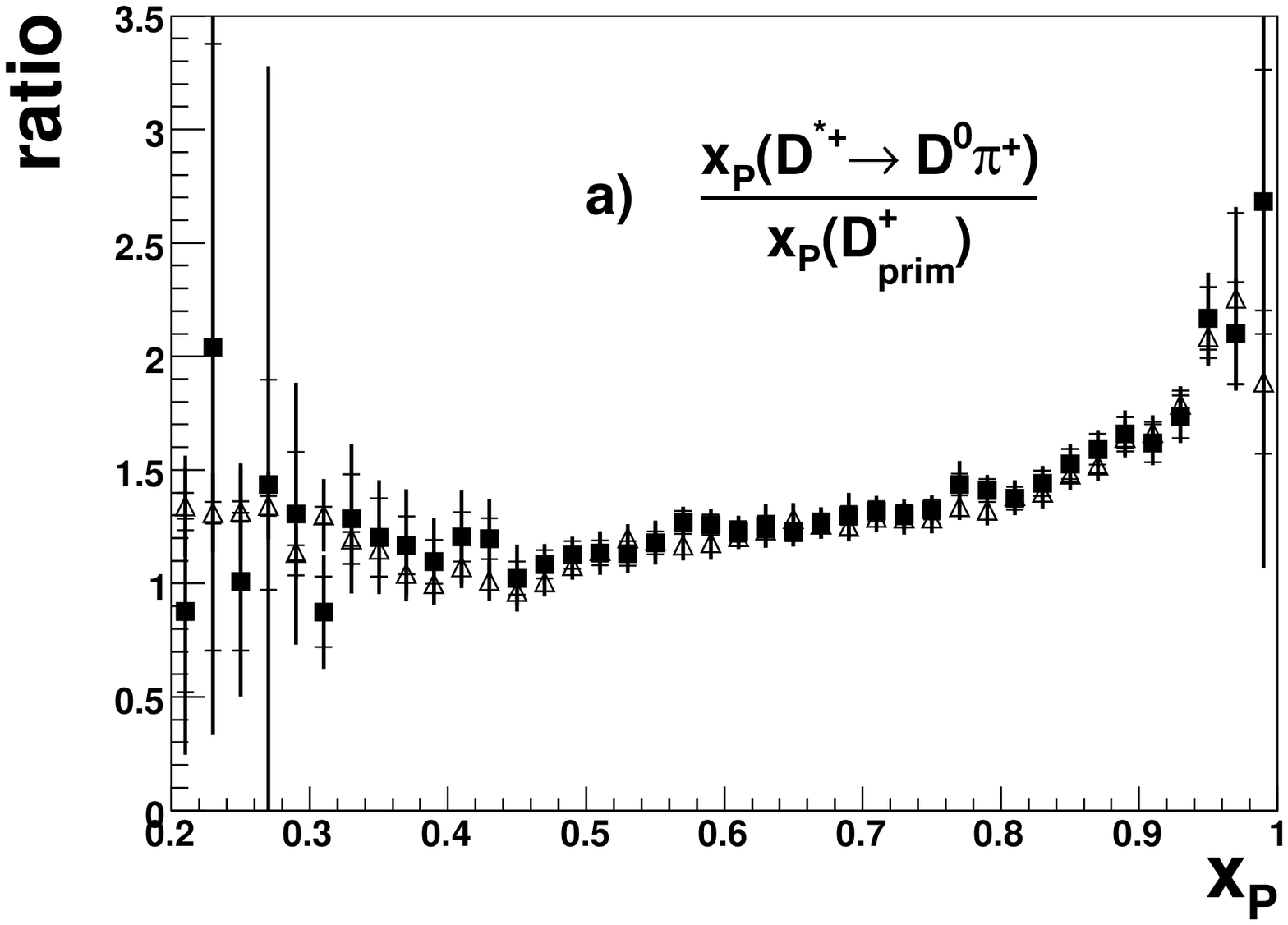}
   \includegraphics[width=0.45\textwidth]{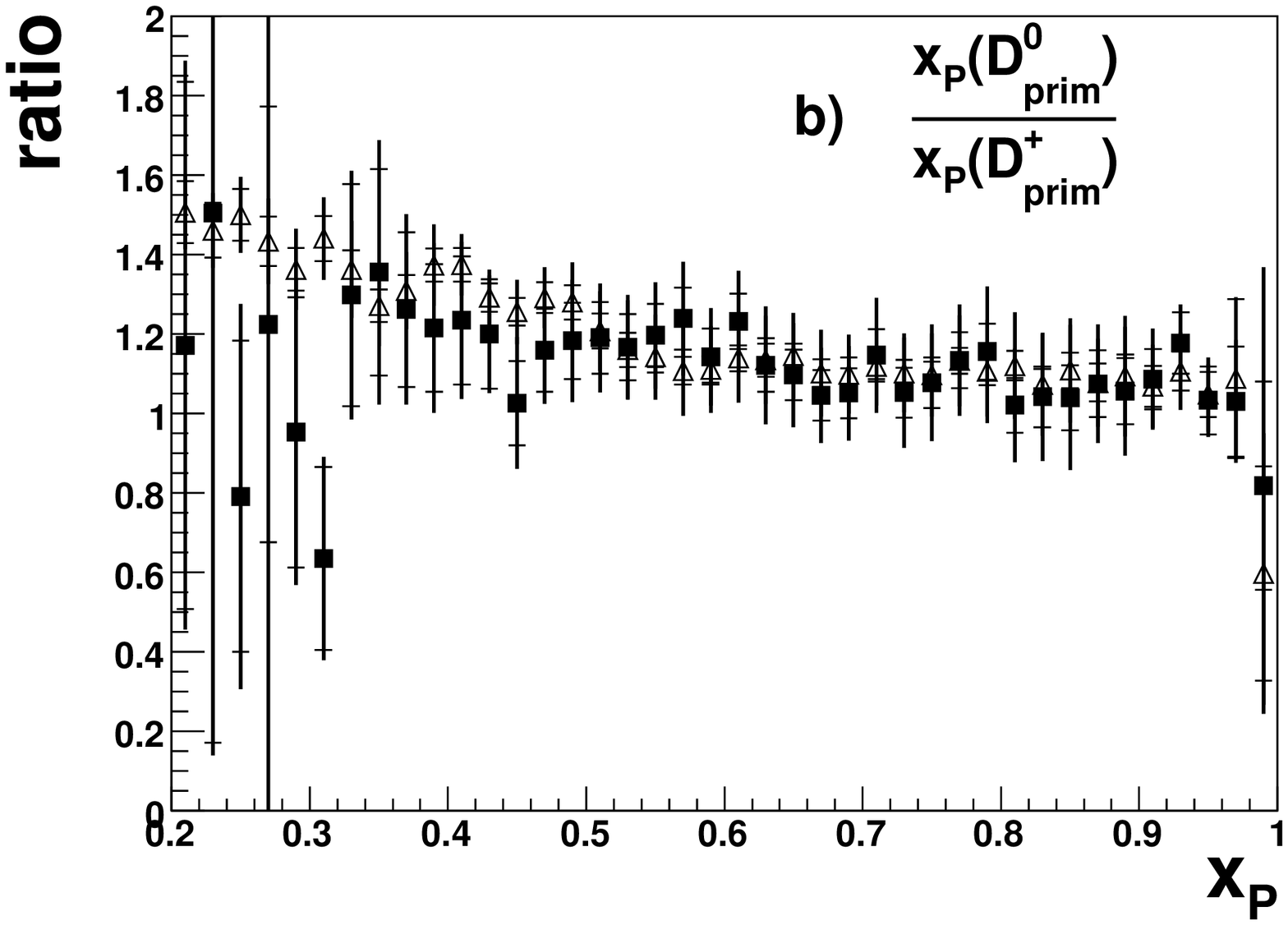}\\
   \includegraphics[width=0.45\textwidth]{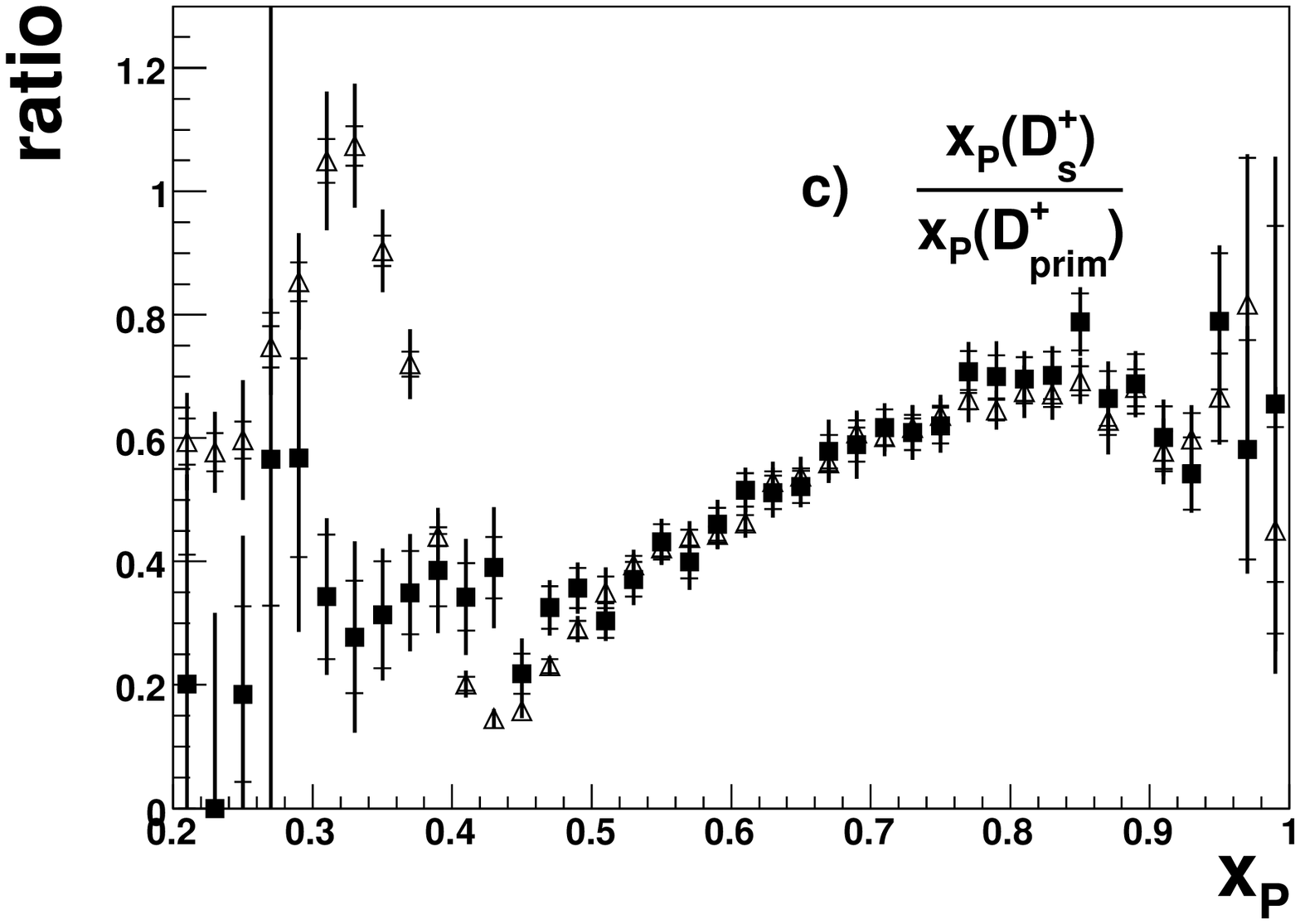}
   \includegraphics[width=0.45\textwidth]{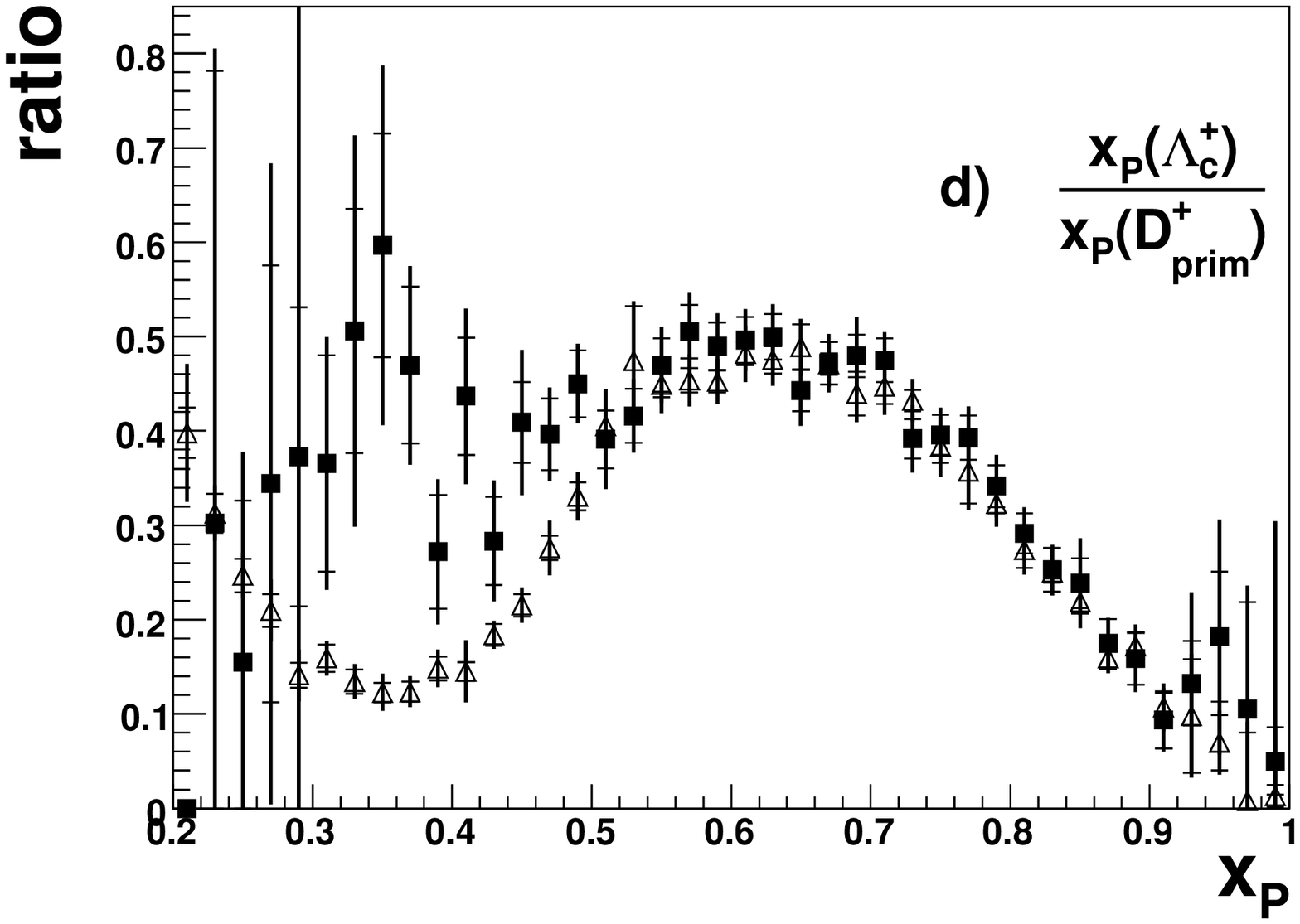}\\
   \includegraphics[width=0.45\textwidth]{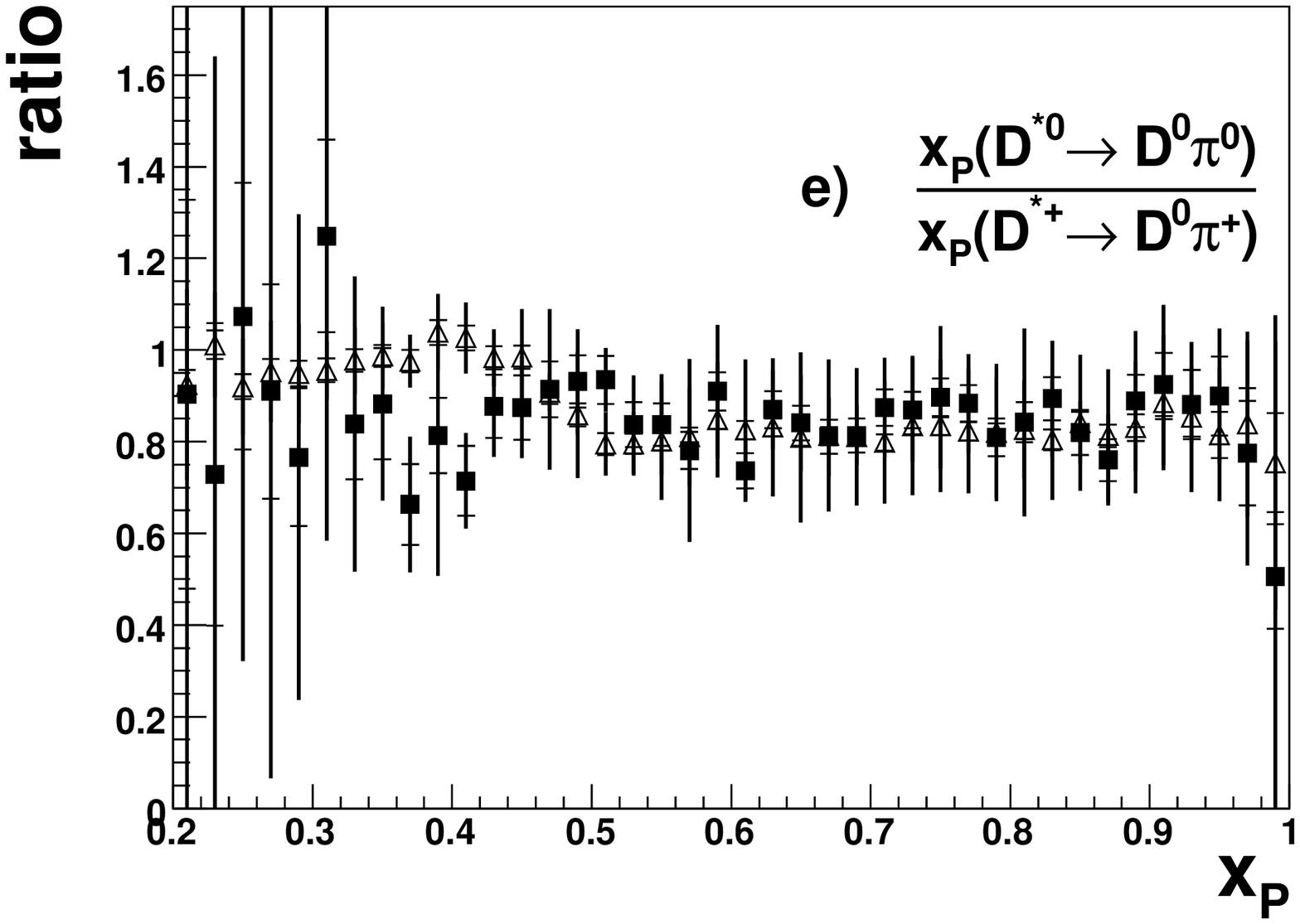}
   \caption{\label{ratios}
     The ratios \xP($D^{*+}$)/\xP($D^+_{prim}$) and \xP($D^0_{prim}$)/\xP($D^+_{prim}$) at
     the top, 
     \xP($D^+_s$)/\xP($D^+_{prim}$) and \xP($\Lambda^+_c$)/\xP($D^+_{prim}$) in the middle
     and \xP($D^{*0}$)/\xP($D^{*+}$) at the bottom.  The
     open upward triangles represent on-resonance data, and the full squares with
     error bars represent continuum data. The inner (outer) error bars show the
     statistical (total) uncertainties.}
  \end{figure}
 \end{center}

Ratios of production rates as a function of \xP\ allow momentum-dependent
effects in fragmentation to be studied, although contributions from decays
of higher states also appear.
Fig.~\ref{ratios} shows the following five ratios as a function of \xP,
for both on-resonance and continuum data:
\begin{itemize}
  \item[(a)]	$\xP(D^{*+})/\xP(\dpprim)$,
		sensitive to the production rate of vector relative to
		pseudo-scalar mesons;
  \item[(b)]	$\xP(\dzprim)/\xP(\dpprim)$,
		sensitive to charged relative to neutral pseudo-scalar production;

  \item[(c)]	$\xP(D^+_s)/\xP(\dpprim)$,
		sensitive to the production of strange quarks;
  \item[(d)]	$\xP(\Lambda_c^+)/\xP(\dpprim)$,
		sensitive to the production of baryons relative to mesons;
  \item[(e)]	$\xP(D^{*0})/\xP(D^{*+})$,
		the relative production rate of the vector mesons.
\end{itemize}
The suffix ``prim'' denotes \xP\ distributions corrected for the contributions
from $D^*$ decays; $D^+$ production has been chosen as the denominator
in (a)--(d), as this correction is smaller than that for $D^0$.
No other corrections for excited states have been applied.

The production ratios in Fig.~\ref{ratios}(a) and (b) are similar for
on-resonance and continuum data.
In Fig.~\ref{ratios}(c), the contribution of $B$ meson decays to $D_s^+$
production can be clearly seen in the low-\xP\ region.
In Fig.~\ref{ratios}(d), baryon production in $B$ decays
is seen to be suppressed below $\xP \approx 0.4$.
As \xP\ approaches unity, the $\Lambda_c^+ / D^+$ production ratio goes to zero,
consistent with the conservation of baryon number.

\begin{center}
  \begin{figure}[h]
   \includegraphics[width=0.45\textwidth]{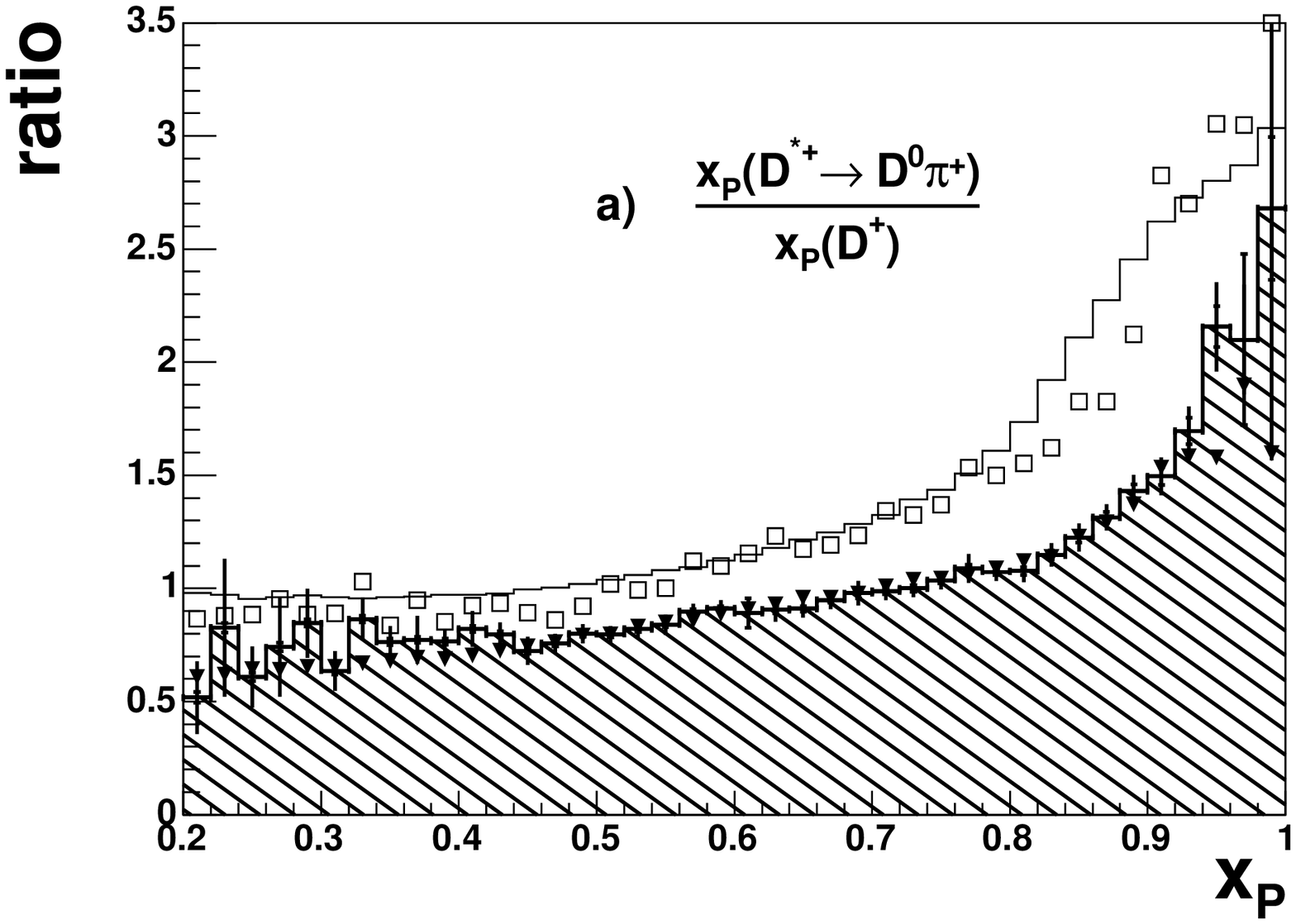}
   \includegraphics[width=0.45\textwidth]{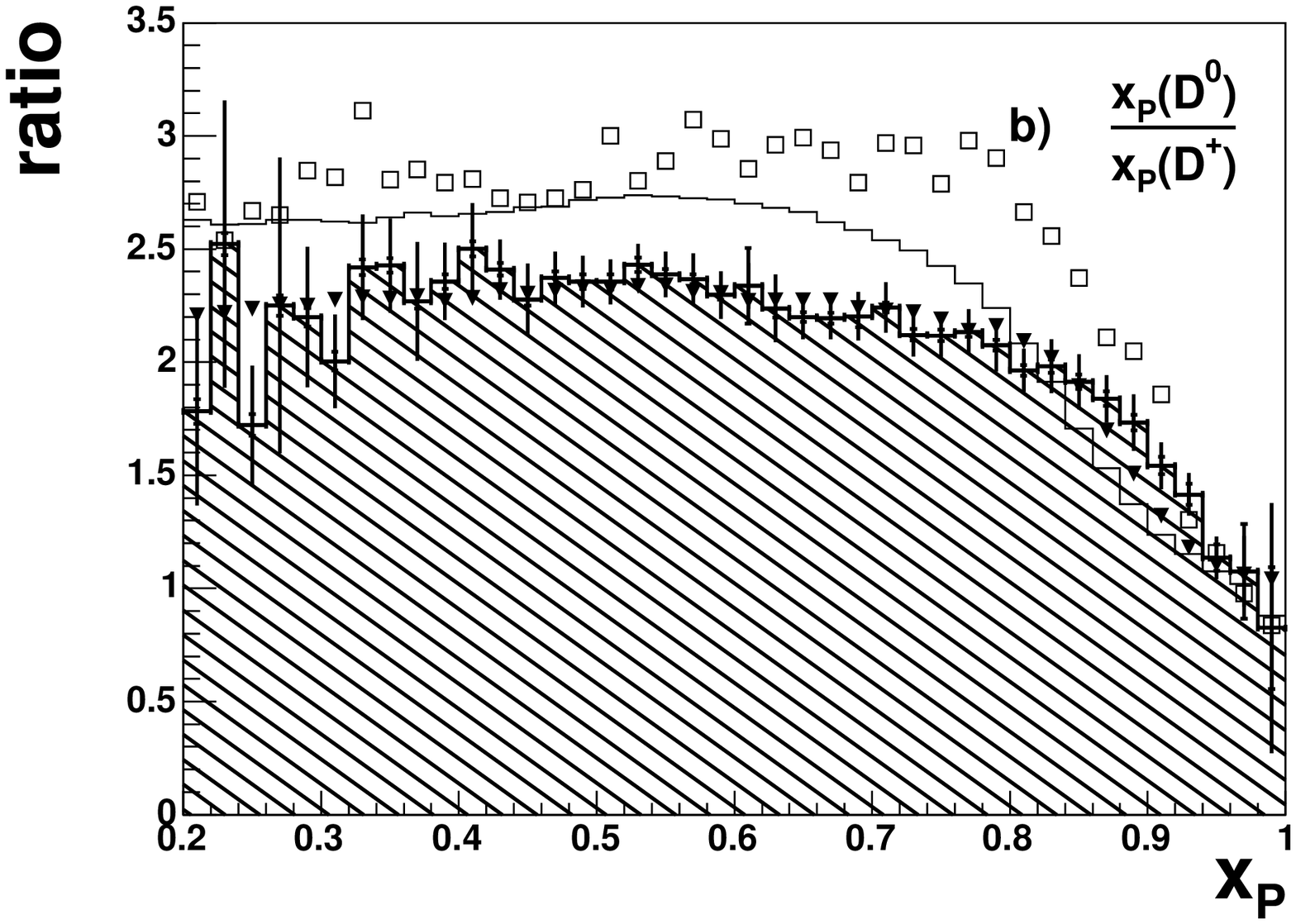}   
   \includegraphics[width=0.45\textwidth]{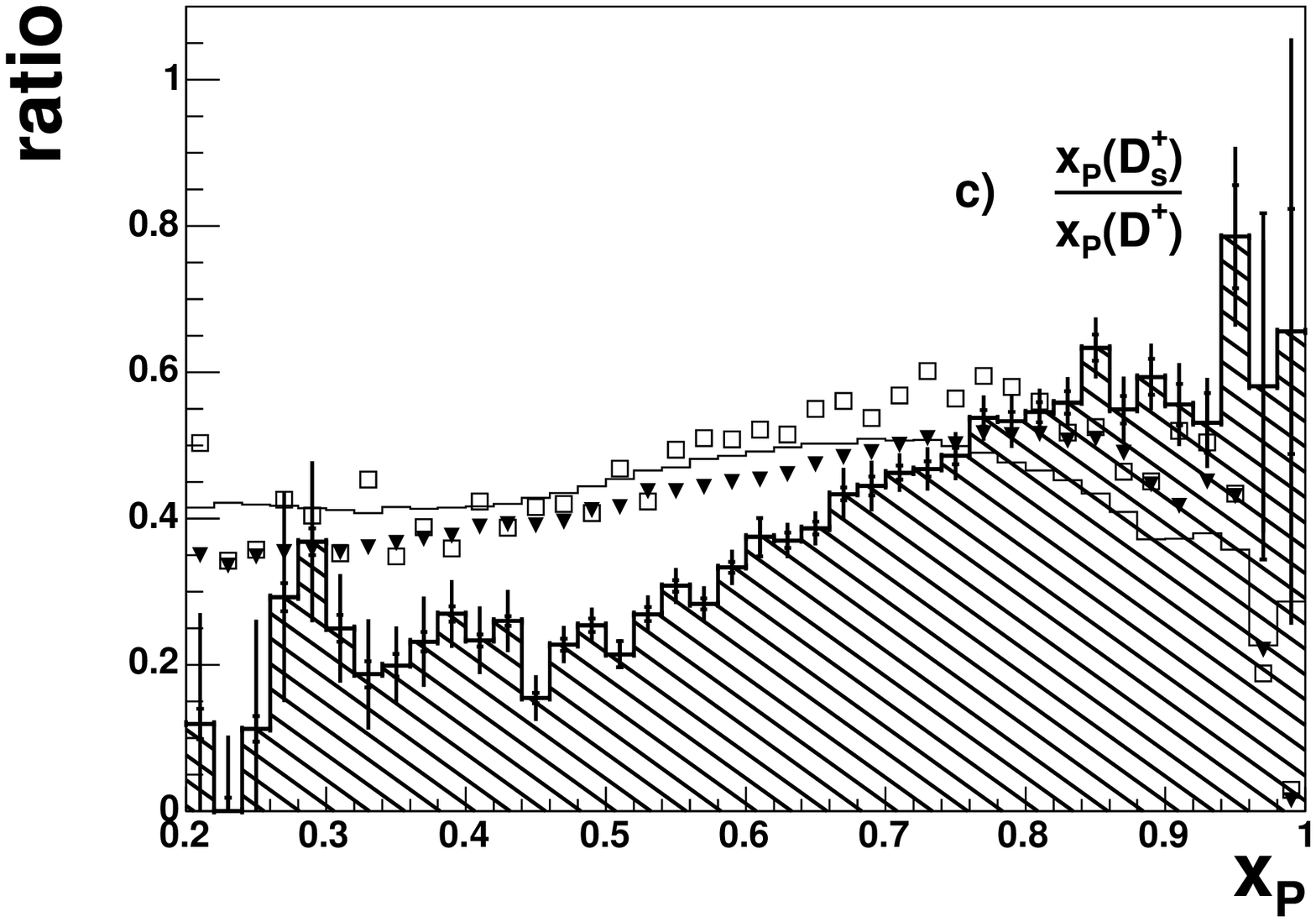}
   \includegraphics[width=0.45\textwidth]{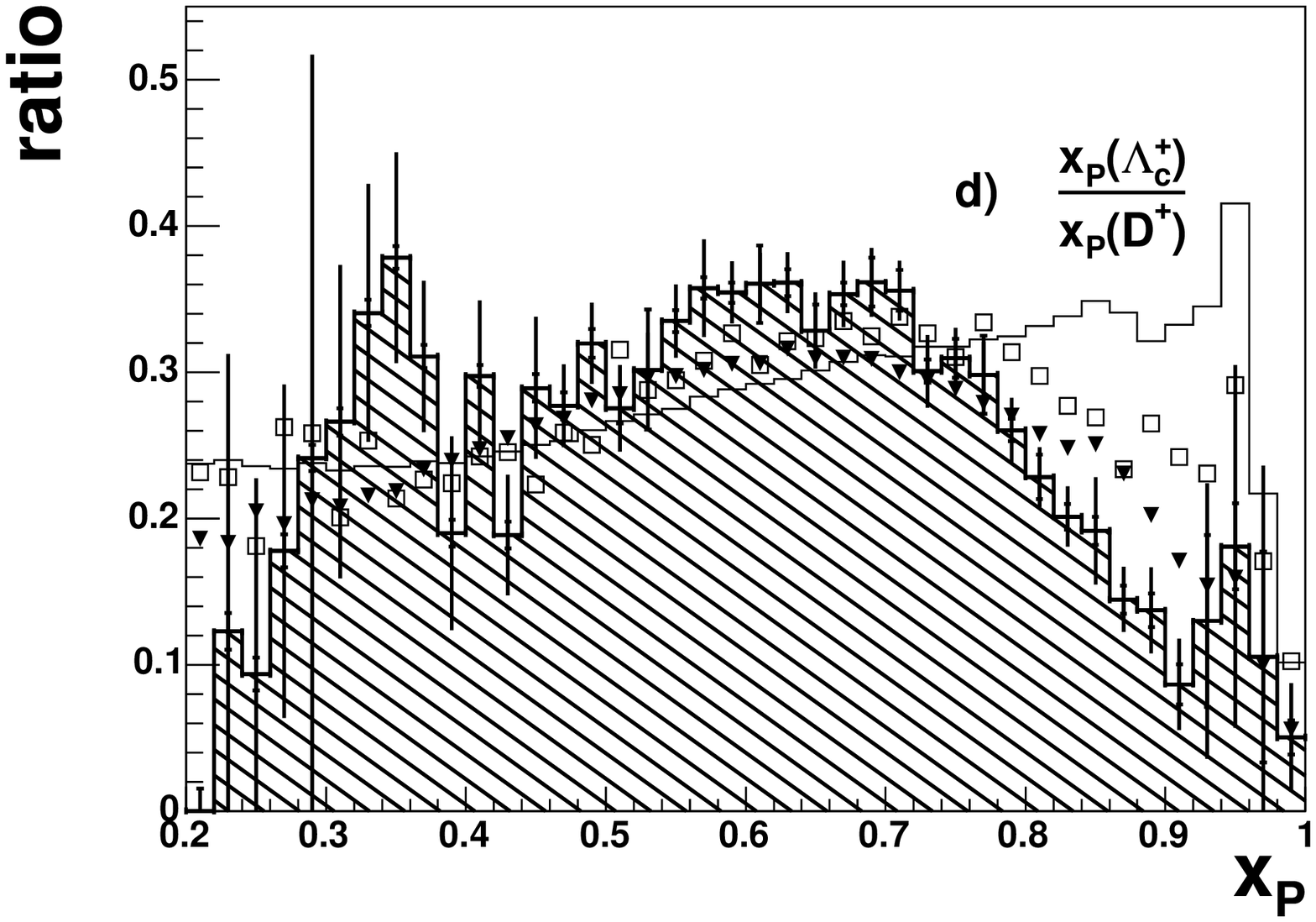}\\   
   \caption{\label{MCratios}The ratios
     \xP(\DSP)/\xP(\DP) in a), \xP(\DZ)/\xP(\DP) in b),
     \xP(\Ds)/\xP(\DP) in c) and \xP(\LC)/\xP(\DP) in d).
     The hatched histograms represent the ratio for continuum data, the
     open histogram shows the distribution for the corresponding
     ratios from generic \MC, the open squares show the predictions
     using the Bowler fragmentation function with default parameters.
     The full triangles show the predictions 
     using the Bowler fragmentation function with a tuned value
     for the probability of producing a charmed meson with
     spin=1: $\mathrm{\bf PARJ(13)}=0.59$ . }
  \end{figure}
 \end{center}

Four similar ratios are shown in Fig.~\ref{MCratios}(a)--(d) for both
continuum data (full squares) and MC simulations, to test the performance
of the MC for various fragmentation function parameters. 
In these plots, the total $D^+$ production rate, without $D^*$ subtraction,
is used in the denominator of the ratios.
The open histograms show the generic MC sample, which agrees with the
data only for the highest values of \xP\ of the distributions in
Fig.~\ref{MCratios}(a) and (b), but fails to describe the data
distributions at lower values.
The open squares show a second MC sample generated with the Bowler
fragmentation function, which shows a similar behaviour.

Noting that the parameter \parj\ in PYTHIA gives the probability
for a charmed hadron produced in fragmentation to have spin one,
50 MC samples of $10^7$ events each were generated, 
with \parj\ values ranging from $0.25$ to the default value $0.75$ given
by spin counting.
These samples were generated in addition to the ``reweighted samples''
used for more refined \MC\ comparisons as described in the next section.
A reduced chi-squared $\tilde{\chi}^2$ was calculated
for these samples and the measured and corrected ratios, and a fourth-order
polynomial in \parj\ was fitted to the results. 
The minimum $\tilde{\chi}^2$ was found to occur at
$\parj = 0.592 \pm 0.021$ for the $\xP(D^{*+})/\xP(D^+)$ ratio,
and     $0.592 \pm 0.046$ for the $\xP(D^0)/\xP(D^+)$ ratio,
where the uncertainties denote the $1\sigma$ range in the fitted polynomial.
We note that models of hadron production more sophisticated than spin counting
predict values for \parj\ below $0.75$;
see~\cite{VoverPV} and references therein. 

In Fig.~\ref{MCratios}(a)--(d) a third MC sample generated with the
Bowler fragmentation function, and $\parj = 0.59$,
is shown with closed triangles.
This sample and the data agree well within the error bars over almost
the entire range in Fig.~\ref{MCratios}(a) and~(b).
All three MC samples fail to describe the ratios
in Fig.~\ref{MCratios}(c) and~(d): both the endpoints and the shape 
disagree. The difference between the MC samples is small compared to the 
discrepancy with data for $D_s^+/D^+$ production in Fig.~\ref{MCratios}(c);
while Bowler fragmentation (open squares and triangles) gives an improved
description of $\Lambda_c^+/D^+$ production in Fig.~\ref{MCratios}(d),  
the agreement is still poor. There are no obvious parameters in the MC which
can affect these two ratios in such a way as to improve the agreement
between data and MC. 

\subsection{Comparison of \xP\ distributions with predictions from MC generators}
\label{MC}
The models used by MC generators are based on simplified assumptions and 
require input from experiment: this is reflected in the models' input
parameters.
 \begin{table}[h]
   \caption{\label{tabratio}Three ratios of total cross sections, each
    of the form $\sigma(\epem\to A X)/\sigma(\epem\to B Y)$
    for the continuum sample. 
    The denominators of all ratios contain the contribution from
    $D^* \to D$ and other decays.
    For comparison, see \cite{Gladilin}}
   \begin{center}
     \begin{ruledtabular}
       \begin{tabular}{ccc}
	 A & B & ratio \\ \hline
	 \DSZ\ + \DSP\ & \DP\ + \DZ\ &
           $0.527 \pm 0.013 \pm 0.024$ \\
	 \Ds\ & \Ds\ + \DP\ + \DZ\ &
           $0.099 \pm 0.003 \pm 0.002$ \\
	 \LC\ & \Ds\ + \DP\ + \DZ\ &
           $0.081 \pm 0.002 \pm 0.003$ 
       \end{tabular}
     \end{ruledtabular}
   \end{center}
 \end{table}
 The commonly used JETSET/PYTHIA generators are based on the Lund
 or string model, in which a coloured string is expanded between two
 emerging partons. The energy stored in the string increases with
 increasing distance, and eventually allows the production of a new
 quark anti-quark pair. The quark (anti-quark) then produces a meson
 together with the initial anti-quark (quark). The energy distribution
 of the new quark or anti-quark is described by a fragmentation
 function. Various fragmentation function models have been
 published; see Table \ref{TabFF} for a summary.
 These models depend upon up to two independent
 variables, these are the transverse mass
 $m_{\perp}=\sqrt{m^2+p_{\perp}^2}$ of the newly created hadron, and $z$,
 the fraction of the longitudinal energy $E+p_{\parallel}$ which the
 meson inherits from the initial quark.
 
 Not all models listed in Table \ref{TabFF} are implemented in the
 JETSET/PYTHIA generator. In order to be able to compare all models to
 data, a reweighting technique has been applied. Here, several million
 events referred to as ``reweighted samples'' have been generated,
 allowing a more elaborate comparison than described in Section \ref{secratios}. For
 these events, $z$ and $p_{\perp}$ were stored together with the
 event. This allowed each event to be reweighted in order to mimic any
 other fragmentation function.
 Scans through the parameter space of the five listed fragmentation
 functions have been performed on these special samples. 
 This analysis was performed on five different hadrons;
 \DSZ\TO\DZ\PZ\ and \DSP\TO\DP\PZ\ have been omitted because of the
 large systematic uncertainty due to the detection efficiency of the
 slow neutral pion. 
 
 \begin{table}[h]
   \caption{\label{TabFF}The functional form of the \frf s used in
     this analysis. The normalisation $N$ is different for all
     functions. The models by Collins and Spiller and by
     Kartvelishvili are not included in the JETSET/PYTHIA
     generator.
   }
  \begin{center}
  \begin{ruledtabular}
   \begin{tabular}{lcr}
     fragm. function & functional form & comment \\ \hline
     Bowler &
     $N {1 \over z^{1+bm^2}} (1-z)^a \exp \left(-{b m_\perp^2 \over z}\right)$
      & $a$, $b$ identical for all quarks \\
     Lund   &
     $N {1 \over z} (1-z)^a \exp \left(-{b m_\perp^2 \over z}\right)$
      & $a$, $b$ identical for all quarks \\
     Kartvelishvili &
     $Nz^{\alpha_{c}}(1-z)$  & \\
     Collins-Spiller &
     $N\left({1-z \over z} + {(2-z)\varepsilon_{c}' \over 1-z}\right)
     (1+z^2) \left(1-{1 \over z}-{\varepsilon_{c}' \over
     1-z}\right)^{-2}$ \\
     Peterson  &
     $N {1 \over z}\left(1-{1\over z}-{\varepsilon_{c} \over
     1-z}\right)^{-2}$  & widely used \\
   \end{tabular}
  \end{ruledtabular}
  \end{center}
 \end{table}

 \begin{table}[h]
  \caption{\label{frf}
    The minimum of the chi-squared distribution, $\chi^2_{min}$,
    for MC samples reweighted to represent the fragmentation functions
    shown, varying their respective parameters. The number of degrees
    of freedom (d.o.f.) is also shown for each case.}
  \begin{center}
  \begin{ruledtabular}
   \begin{tabular}{cccccc}
     & $D^0$ & 
     $D^+$         & 
     $D^+_s$       & 
     $\Lambda^+_c$ & 
     $D^{*+}$      \\ \hline
     &
     $\chi_{min}^2/d.o.f.$ &
     $\chi_{min}^2/d.o.f.$ &
     $\chi_{min}^2/d.o.f.$ & 
     $\chi_{min}^2/d.o.f.$ &
     $\chi_{min}^2/d.o.f.$ \\ \hline
     Bowler &
     1327.0 / 59 &
      188.4 / 60 &
      730.7 / 55 &
      269.1 / 60 &
      541.8 / 55 \\
      Lund &
     1500.5 / 59 &
      527.1 / 60 &
      513.2 / 55 &
      266.6 / 60 &
      965.6 / 55 \\
      Collins and Spiller &
     3032.1 / 58 &
      948.0 / 60 &
     1412.5 / 55 &
     2836.7 / 59 &
     1540.7 / 54 \\
     Kartvelishvili &
     3210.4 / 59 &
      861.4 / 60 &
      735.3 / 55 &
      390.7 / 60 &
     1271.1 / 54 \\
     Peterson & 
     5070.2 / 59 &
     2229.6 / 60 &
      829.6 / 55 &
     1345.0 / 59 &
     3003.0 / 54 \\
   \end{tabular}
  \end{ruledtabular}
  \end{center}
 \end{table}
 \begin{table}[h]
  \caption{\label{frf-par}The parameters of the fragmentation functions at
  the minimum of the $\chi^2/d.o.f.$ distributions .}
  \begin{center}
  \begin{ruledtabular}
   \begin{tabular}{ccccccc}
     & & $D^0$ & 
     $D^+$         & 
     $D^+_s$       & 
     $\Lambda^+_c$ & 
     $D^{*+}$      \\ \hline
     & parameter
     & par$s$ at min.
     & par$s$ at min.
     & par$s$ at min.
     & par$s$ at min.
     & par$s$ at min. \\ \hline
     Bowler & $a|b$ &
      0.12 $|$ 0.74 &
      0.12 $|$ 0.58 &
      0.12 $|$ 0.68 &
      0.34 $|$ 0.74 &
      0.22 $|$ 0.56 \\
      Lund & $a$ &
      0.26 &
      0.45 &
      0.2 &
      0.55 &
      0.58 \\
      Collins and Spiller & $\varepsilon_{c}'$ &
     0.04 &
     0.055 &
     0.04 &
     0.04 &
     0.075 \\
     Kartvelishvili & $\alpha_{c}$ &
     4.6 &
     4 &
     5.6 &
     3.6 &
     5.6 \\
     Peterson & $\varepsilon_{c}$ & 
     0.028 &
     0.039 &
     0.008 &
     0.011 &
     0.054 \\
   \end{tabular}
  \end{ruledtabular}
  \end{center}
 \end{table}
 
 For data and \MC, the \xP\ distributions were compared using uncorrected
 data and the reweighted special \MC\ samples after full detector
 simulation. A $\chi^2$ was calculated based upon the distribution of
 the reweighted special sample and the measured data
 distribution. Only statistical uncertainties in each \xP\ bin were
 taken into account and only bins which contained entries in data or
 \MC\ were included. The number of bins minus the number of parameters
 of the \frf\ was used as the number of degrees of freedom ($d.o.f.$). The
 weights in the reweighting procedure were constructed in such a way
 that the number of events before and after reweighting stayed
 constant. This way the total value of the $\chi^2$ becomes dependent
 on the size of the data and \MC\ samples; the relative $\chi^2$
 values, however, allow a direct comparison between the different \frf
 s.
 
 Table \ref{frf} shows the $\chi_{min}^2/d.o.f.$ for all five
 particles and five \frf s.
 For all five particles a similar trend is visible. The Bowler model
 in general agrees best with the data. The Lund models shows a similar
 performance in describing the spectra, its $\chi_{min}^2/d.o.f.$
 being by factors of 2--3 better than the next best model. For
 \DP\ and \DSP\ the $\chi_{min}^2/d.o.f.$ is slightly worse than for the
 Bowler model. In the minimum of the $\chi^2/d.o.f.$ distributions the
 $a$ parameter deviates 
 strongly from the default for most of the particles. As the Lund
 model is employed for fragmentation of all flavour species, such a large
 change in the parameter would also change the particle spectrum of
 light mesons. Therefore, further tuning of the second parameter
 to the Lund fragmentation function has been omitted.
 
 The models by Collins and Spiller and by Kartvelishvili show a
 similar $\chi_{min}^2$ for all particles, about factors of two to
 three worse
 than that of the best models.  
 The last model, that of Peterson, shows the worst agreement with a
 reduced $\chi_{min}^2$ of 15 and well above, ruling out this 
 model for describing data at this CME.
 
 The input parameters for the \frf s at the minimum of the
 $\chi^2/d.o.f.$ distributions are listed in Table \ref{frf-par}.

 In summary, the Bowler model shows the best agreement between data
 and \MC, however, large differences are still present. These
 differences might be resolved by adjusting other parameters of the
 generators as well, but such a task is out of scope for this
 analysis. The Lund model shows the second best agreement. The models
 by Kartvelishvili and by Collins and Spiller show larger deviations and
 the commonly used model by Peterson shows the worst agreement between
 data and \MC.
 
 \section{Summary}
 A new determination of the charm fragmentation function at a CME 
 close to the \Ys\ resonance has been presented.
 The measured \xP\ spectra have been compared to those of five different
 parametrisations in \MC\ via a reweighting procedure, and the best
 input parameters have been found.
 The best agreement between data and \MC\ has been found for the
 Bowler model and the Lund model.
 Additionally, the peak positions and the first six moments of the \xP\
 distributions have been measured. These measurements will allow
 detailed comparisons between experiment and theory. The total
 production cross-section, as well as \xP\ dependent ratios of the
 \frf s, place stringent tests on existing \MC\ generators, which so
 far completely fail to describe the \xP\ dependent ratios of
 $\xP(\Ds)/\xP(\DP)$ and $\xP(\LC)/\xP(\DP)$.
 For the first time, the production rates of \DP\ and
 \DZ\ excluding the decay of $D^*$ mesons have been measured. They
 were found to agree reasonably well with each other. 
 
 The efficiency corrected data points will be made available via
 download in the Durham HEP REACTION DATA DataBase \cite{DurhamHEP}.
 It is presented in a different way as shown this article. Separate
 sets of the continuum and the on-resonance samples are given as
 $s B d\sigma/d\xP$, {\it i.e.} scaled by the nominal center-of-mass
 energies of 10.52 \GeV\ and 10.58 \GeV, respectively, and not
 corrected for the branching ratios.
 The on-resonance data includes the additional correction for ISR of
 $+0.27\%$, see Section \ref{anaproc} for details.
 
 \section*{Acknowledgements}
 We thank O.~Biebel and A.~Mitov for discussion.
 We thank the KEKB group for the excellent operation of the
 accelerator, the KEK cryogenics group for the efficient
 operation of the solenoid, and the KEK computer group and
 the National Institute of Informatics for valuable computing
 and Super-SINET network support. We acknowledge support from
 the Ministry of Education, Culture, Sports, Science, and
 Technology of Japan and the Japan Society for the Promotion
 of Science; the Australian Research Council and the
 Australian Department of Education, Science and Training;
 the National Science Foundation of China under contract
 No.~10175071; the Department of Science and Technology of
 India; the BK21 program of the Ministry of Education of
 Korea and the CHEP SRC program of the Korea Science and
 Engineering Foundation; the Polish State Committee for
 Scientific Research under contract No.~2P03B 01324; the
 Ministry of Science and Technology of the Russian
 Federation; the Ministry of Higher Education, Science and Technology
 of the Republic of Slovenia;  the Swiss National Science Foundation;
 the National Science Council and the Ministry of Education of Taiwan;
 and the U.S.\ Department of Energy.
 

\end{document}